\documentclass[12pt,preprint]{aastex}
\usepackage{amsmath,amssymb,latexsym,mathrsfs}
\usepackage{graphicx}
\usepackage[figuresright]{rotating}
\usepackage{subfigure}
\usepackage{epsfig}

\def\rmi{{\rm i}}

\begin{document}

\title{3D Structures of equatorial waves and the resulting superrotation in the atmosphere of a tidally locked hot Jupiter}
\author{Shang-Min Tsai\altaffilmark{1}, Ian Dobbs-Dixon\altaffilmark{2}, \& Pin-Gao Gu\altaffilmark{1}}
\altaffiltext{1}{Institute of Astronomy and Astrophysics, Academia Sinica, Taipei
10617, Taiwan}
\altaffiltext{2}{NYU Abu Dhabi PO Box 129188 Abu Dhabi, UAE}

\begin{abstract}

Three-dimensional equatorial trapped waves excited by stellar isolation and the resulting equatorial superrotating jet in a vertical stratified atmosphere of a tidally-locked hot Jupiter are investigated. Taking the hot Jupiter HD 189733b as a fiducial example, we analytically solve linear equations subject to stationary stellar heating with a uniform zonal-mean flow included. We also extract wave information in the final equilibrium state of the atmosphere from our radiative hydrodynamical simulation for HD 189733b.
Our analytic wave solutions are able to qualitatively explain the three-dimensional simulation results. Apart from previous wave studies, investigating the vertical structure of waves allows us to explore new wave features such as the wavefronts tilts related to the Rossby-wave resonance as well as dispersive equatorial waves. We also attempt to apply our linear wave analysis to explain some numerical features associated with the equatorial jet development seen in the general circulation model by Showman and Polvani.
During the spin-up phase of the equatorial jet, the acceleration of the jet can be in principle boosted by the Rossby-wave resonance. However, we also find that as the jet speed increases, the Rossby-wave structure shifts eastward, while the Kelvin-wave structure remains approximately stationary, leading to the decline of the acceleration rate. Our analytic model of jet evolution implies that there exists only one stable equilibrium state of the atmosphere, possibly implying that the final state of the atmosphere is independent of initial conditions in the linear regime. Limitations of our linear model and future improvements are also discussed.

\end{abstract}


\section{Introduction}

Equatorial super-rotation, occuring when the equatorial winds in the atmosphere blow faster than the self rotation of the planet, has been an interesting and peculiar phenomenon on tidally-locked hot Jupiters. The feature has been demonstrated in many numerical simulations \citep[e.g.,][]{Showman08,SF09,TC10,RM10,Dobb,Perna,Heng,3D climate}. The existence of super-rotating equatorial jets on tidally locked, highly irradiated extrasolar planets has been corroborated by phase-curve observations. Observations of HD 189733b (Knutson et al 2009, Agol et al 2009, Knutson et al 2012) show that the maximum flux occurs several hours before secondary eclipse, indicative of a strong super-rotating jet that advects energy downwind.
Apart from the tidaly-locked gas giants, superrotation also exists in our solar system, such as Venus, Jupiter, and Saturn. Because the atmosphere in the equatorial region has the maximum angular momentum per unit mass about the planetary rotation axis, there must be non-axisymmetric processes (wave or eddy motions) that pump momentum to the equatorial region \citep{Hide}. The mechanism that gives rise to the superrotating flow may vary between planets and is worthy of detailed study. Mechanisims that have been explored include hydrodynamic instabilities, eddy mixing, atmospheric tides, and meridional circulation (e.g., \citet{Gierasch}, \citet{FL74}, \citet{HG85}, \citet{Will03}, \citet{Zhu}, \citet{TM07}, \citet{Liu10}, \cite{Liu11}, and see \citet{SP11} for more references).

The importance of internal angular momentum transport via atmospheric tides induced by intense stellar heating (namely, thermal tides) has been discussed in
the context of the atmospheric dynamics of hot Jupiters \citep{Cho,GO09,WC10,GH11,SP11}.
\citet{GO09} and \citet{GH11} carried out a linear analysis on the diurnal thermal tides in the form
of Hough waves on a non-synchronized hot Jupiter. Their estimates suggested that
the resulting internal torques associated with the Rossby/Kelvin waves damped by radiative
loss may be comparable to the external torques by stellar gravity. Apart from global equatorial waves, \citet{WC10} conducted a local linear analysis in the presence of a
background superrotating equatorial jet on a tidally locked hot Jupiter. As the
thermally-driven internal gravity waves propagate upwards, their amplitudes grow
and become saturated. The authors found that the damping of the local internal
gravity waves has a non-negligible impact on the background jet.

\citet{SP11} took one step further to elucidate the relation between the steady day-night thermal forcing and the formation of the superrotating equatorial jets on a synchronously rotating planet. Following the work of Matsuno (1966) and Gill (1980), they first applied a linear analysis of a two-layer shallow-water model with mass source/sink simulating the heating/cooling via a day-night contrast. Accompanied by their 3-D radiative hydrodynamic simulation using the primitive equations, the authors concluded that on these tidally locked planets, a heat-induced standing wave pattern arising from a combination of the
equatorial
Rossby and Kelvin waves plays the role of equatorward (up-gradient) momentum transport from higher latitudes, which is ultimately balanced by vertical momentum transport and/or fluid drag leading to a fast eastward jet. The contribution from vertical momentum transport is investigated in \citet{SP11} by taking into account momentum exchange between two layers in addition to the usual wave momentum transport
in their shallow-water approach \citep[also see][]{Shell}.
This additional consideration
is crucial for the model to obey momentum conservation and has the effect of weakening the westward vertical momentum transport that cancels out the momentum transport by horizontal waves, thus allowing superrotation to arise. Albeit simple and clearly illuminating, the nature of the two-layer model has simplified the vertical motions to an upwelling flow from the day-side mass source and a downwelling flow on the night-side mass sink. Therefore, it does not render complete information about the vertical structure of the waves and the resulting vertical momentum transport in a stably stratified atmosphere on a hot Jupiter.

Based on the method of separating variables, there have been two approaches treating the linear forced response in the three-dimensional stably stratified atmosphere/ocean on the rotating Earth. One is as in classic atmospheric tidal theories, to solve the homogeneous meridional structure equation first for the meridional free modes, then solve the forced vertical structure equation \citep{Lindzen67}. The other approach commonly used in oceanography, or the shallow-water approach, is to expand the homogeneous vertical equation first for the vertical normal modes, then solve the forced meridional structure equation \citep{Geisler82,Demaria}. The separation constants in both approaches that link the meridional structure equation and the vertical equation are the so-called ``equivalent depths," which bear important information about wave structures in a vertically stratified atmosphere. However, since they are the eigenvalues corresponding to different equations, equivalent depths have different meanings: a mathematical constant in the tidal approach and the depth of the hypothetical fluid in the shallow-water approach.

As discussed in \citet{Wu2000}, the tidal approach has been quite successful in studying diurnal tides \citep{Lindzen67}, but it has some limitations when studying the forced problem with a large-scale, low-frequency heat source. When the wave period is comparable to the damping time scale, damping can not be neglected. In the presence of damping, the resulting meridional structure equation would contain complex coefficients. In this case, the completeness of the eigenfunctions is questionable. In addition, no wave solutions exist in the meridional structure equation when the forcing frequency actually goes to zero. Hence, this approach is not suitable for studying the dissipative wave motions on tidally-locked planets with synchronous (i.e. stationary) heating.

In \citet{Wu2000}, a method following the shallow-water approach is developed to investigate the thermally forced response of the tropical atmosphere in the high rainfall regions on Earth by vertical decomposition when utilizing spectral representation,
wherein the linear equations consist of a homogeneous vertical structure equation and a set of forced shallow-water equations.

By calculating the projections of a prescribed vertical heating profile onto the vertical eigenfunctions, the contribution from different vertical modes can be obtained. In this paper, we follow the methods by \citet{Wu2000}, using an idealized model in which the vertical structure equation can be expressed as a sum of vertical eigenfunctions. Then the horizontal motion forced by stationary heating can be further expressed by equatorial free waves, with each set of shallow-water equations corresponding to a vertical eigenfunction. By summing up all the vertical eigenfunctions with the horizontal solutions, we can obtain the linear solutions with vertical structures while keeping the equations in a simple shallow-water manner. This approach allows us to learn both the response from different waves (as in \cite{matsuno}) and from each vertical mode (as in \cite{Wu2000}) individually, and to study the effects of a zonal-mean flow, such as wave resonance.

The purpose of this paper is to construct the aforementioned 3D linear equatorial wave theory in a stably stratified atmosphere on a tidally locked hot Jupiter. We take the atmosphere of the hot Jupiter HD 189733b as a fiducial example to test our wave theory, which will be used to explain the results for the planet seen in our 3D radiative hydrodynamical simulations based on \citet{DA13} (hereafter DA) and the results from the general circulation model (GCM) by \citet{SP11} (hereafter SP). We first introduce our 3D linear wave theory in Section 2. The linear results applied to HD 189733b is presented and compared with the simulation results in Section 3. The implication of the linear model to the evolution of an equatorial jet is also explored in Section 3. Finally, we summarize the results and discuss a number of shortcomings and potential outlooks of our work in Section 4.

\section{Three-Dimensional Linear Wave Theory}
\subsection{Basic equations for equatorial waves}
\label{sec:basic}
As has been commonly adopted in the study of equatorial waves, we work on the thermal-tide problem by utilizing Cartesian geometry on an equatorial beta plane for a hot Jupiter with the radius $R_p$ and spin rate $\Omega$. Any fluid quantity $q$ in the atmosphere is linearized in terms of its zonal-mean $q_0$ and its perturbation associated with planetary-scale waves $q'$. In an atmosphere of a hot Jupiter without thermal inversion, it can be assumed without loss of generality that the zonal-mean state (i.e. the basic state) of the atmosphere is isothermal and hydrostatic \citep[e.g.,][]{MS09}, with a uniform zonal flow at a speed of $u_0$ to resemble the planetary-scale super-rotation flow. As a result of strong vertical stable stratification as well as the fact that phase velocities of planetary waves are much less than the sound speed, we can further assume the perturbations to be vertically hydrostatic \citep[e.g.][]{SCM11} and nearly incompressible such that the anelastic approximation is applicable to eliminate acoustic waves \citep{OP62}.
Hence, considering wave motions in an atmosphere of a tidally locked planet driven by stationary thermal forcing, the linearized equations on an equatorial beta plane are given by \citep[e.g.,][]{Holton92}
\begin{subequations}
\begin{equation}
\alpha u' + u_{0} \frac{\partial u'}{\partial x} - \beta y v' + \frac{\partial \phi'}{\partial x} =0,
\end{equation}

\begin{equation}
\alpha v' + u_{0} \frac{\partial v'}{\partial x} + \beta y u' + \frac{\partial \phi'}{\partial y}  =0,
\end{equation}

\begin{equation}
\frac{\partial u'}{\partial x} + \frac{\partial v'}{\partial y} + \frac{1}{\rho_{0}}\frac{\partial}{\partial z} (\rho_{0} w')  =0,
\end{equation}

\begin{equation}
\alpha \frac{\partial \phi'}{\partial z} + u_{0} \frac{\partial^{2} \phi'}{\partial x \partial z} + N^{2} w' = \frac{\kappa}{H} J,
\label{energy eq}
\end{equation}
\label{tot eq}%
\end{subequations}
where we have already assumed that the $\partial/\partial t$ terms all vanish for steady-state solutions (which are actually treated as quasi-steady for the evolution of $u_0$ on a longer timescale, as will be discussed later in Section \ref{sec: momentum flux}). In the above equations, $x$, $y$, $z$ are the zonal, meridional, and vertical coordinates, and $u'$, $v'$, $w'$ are the horizontal, meridional, vertical velocities, respectively. $\beta = 2 \Omega / R_{p}$ is the Coriolis parameter in the equatorial beta-plane approximation, which approximates the latitudinal variation of the Coriolis force from the equator. $\phi' = p'/ \rho_{0}$ is the perturbed geopotential with $p'$ being the perturbed thermal pressure and $\rho_0$ being the basic density given by
\begin{equation}
\rho_{0}(z) = \rho_{0}(z_b)e^{-(z-z_b)/H},
\label{log z}
\end{equation}
where $H$ is the pressure and density scale height in the isothermal atmosphere
and $z_b$ is the prescribed location of the bottom boundary. Equation(\ref{log z}) states that $z$ is equivalent to the logarithmic pressure coordinate, which will be applied in this work. $N$ is the buoyancy frequency. $\alpha$ is the linear momentum damping rate (a.k.a. Rayleigh friction) in the momentum equations and thermal damping rate (a.k.a. Newtonian cooling) in the energy equation. These two types of damping are set equal here for the advantage of the mode analysis shown later in this paper to better elucidate underlying physics. Furthermore, $J$ is the stellar heating rate per unit mass in the planetary atmosphere. $\kappa$ is the specific gas constant $R$ divided by the specific heat at constant pressure $c_{p}$.
We note that $u_{0}$ is assumed to be uniform over the entire atmosphere under consideration. Although it is not self-consistent to ignore the Coriolis effect and shear of the zonal-mean zonal flow, we add a uniform $u_{0}$ here to study the effect of zonal-mean zonal flow on the global-scale waves and their momentum transport, without making the equations too difficult to solve by separation of variables \citep{Gill87,Arnold}.

We first follow the shallow-water approach employed by \cite{Wu2000} in which the solution is separated into a vertical part and a horizontal part, where the vertical part of the solution is first expressed by the vertical eigenfunctions (the free vertical modes). For each of these vertical modes, the horizontal structure is governed by the stationary forced shallow-water equations but with a different equivalent depth respectively. The wave solutions to the forced shallow-water equations can then be expressed by free horizontal modes (i.e. equatorial waves).
The summation of each vertical mode in combination with the horizontal modes as the solutions to the corresponding forced shallow-water equations are the complete solutions in the three-dimensional problem.

In this study, we focus solely on the $s=1$ response ($s$ is the longitudinal wavenumber), which is expected to be the gravest response of the atmosphere in a tidally-locked heating scenario. Following \cite{Wu2000}, we eliminate $w'$ in Equation(\ref{tot eq}) for easier decomposition in shallow-water equations later. The solutions can then be written as the sum of vertical modes $Z_m(z)$, each of which is denoted by $m$:

\begin{equation}
\begin{pmatrix}
u'(x,y,z)\\
v'(x,y,z)\\
\phi' (x,y,z)
\end{pmatrix}
=
\sum_{m}
\begin{pmatrix}
u_{m}(y)\\
v_{m}(y)\\
\phi_{m}(y)
\end{pmatrix}
Z_m(z) e^{i k x } e^{z/2H},
\label{m mode}
\end{equation}
where $e^{z/2H}$ is the density weighted factor resulting from the vertical stratification and $k=s/R_p=1/R_p$ is the wavenumber in the $x$ direction for the $s=1$ modes. By defining the relation describing the vertical
eigenfunctions (i.e. free vertical modes)
as follows
\begin{equation}
\left( \frac{d^{2}}{d z^{2}}-\frac{1}{4H^{2}} \right)Z_{m}(z)=-\frac{N^{2}}{gh_{m}}Z_{m}(z),
\label{vertical eq}
\end{equation}
where $g$ is the gravitational acceleration on the planetary surface and $h_m$ is the eigenvalue known as the ``equivalent depth", Wu et al. showed that Equation(\ref{tot eq}) can be transformed to sets of stationary forced shallow-water equations for the coefficients $u_m$, $v_m$, and $\phi_m$ of the solutions in Equation(\ref{m mode}). That is, the set of shallow-water equations for each vertical mode denoted by $m$ reads
\begin{subequations}
\begin{equation}
(\alpha + i k u_{0}) u_{m} - \beta y v_{m} + i k \phi_{m} = 0,
\end{equation}
\begin{equation}
\beta yu_{m} + (\alpha + i k u_{0}) v_{m} + \frac{\partial \phi_{m}}{\partial y} = 0,
\end{equation}
\begin{equation}
i k u_{m} + \frac{\partial v_{m}}{\partial y} + (\alpha + i k u_{0}) \frac{\phi_{m}}{gh_{m}}= -F_{m},
\end{equation}
\label{shallow eqs}%
\end{subequations}
where $F_{m}$ satisfies
\begin{equation}
Q \equiv \frac{\kappa}{N^{2}H} (\frac{\partial}{\partial z} - \frac{1}{2H})J e^{-z/2H}
= \sum_{m} F_{m}(y)Z_{m}(z)e^{\rmi kz}.
\label{Q}
\end{equation}
Namely, $F_m$ is the projection of a quantity related to the heating function onto a vertical eigenmode $m$. The summation in Equation(\ref{m mode}) and (\ref{Q}) represents a sum over all discrete modes or an integration over all continuous modes.

The set of Equation(\ref{shallow eqs}) is identical to the shallow-water equation in \citet{matsuno} by replacing $\alpha$ and the ocean depth with $\alpha+\rmi ku_0$ and $h_m$, respectively. Thus, it can again be solved straightforwardly by expanding the solutions ($u_m, \, v_m, \, \phi_m$) and the heating function $F_m$ in terms of the meridional eigenfunctions, i.e. equatorial Kelvin waves, equatorial Rossby waves, and inertia-gravity waves, obtained from the free equatorial wave equations subject to the zero-velocity boundary condition as done in \cite{matsuno}.
It follows that
the characteristic speed, length, and time for each vertical $m$ mode in Equation(\ref{m mode}) are given by
$C_m=\sqrt{gh_{m}}$,
$L_m=(\sqrt{gh_{m}}/ \beta)^{1/2}$,
and
$T_m=(\sqrt{gh_{m}} \beta)^{-1/2}$, respectively. $L_m$ is the so-called equatorial radius of deformation describing the characteristic meridional scale of equatorial trapped waves and $T_m$ is the characteristic wave travel time over $L_m$ with the characteristic speed (i.e. horizontal phase speed) of gravity wave  $\sqrt{gh_m}$. We can see from these expressions that a vertical $m$ mode with a larger $h_{m}$ has a larger horizontal phase speed $\sim C_m$ and a larger horizontal scale $\sim L_m$.

The expansion procedure leads to an expression describing how the atmosphere response, designated by $a$, reacts linearly to the thermal forcing, represented by $b$, in terms of each vertical mode $(m)$ and each type of the corresponding meridional eigenfunction\footnote{We adopt the labeling convention in \citet{matsuno}: $n$ is a non-negative integer and stands for meridional modes, each of which contains a Rossby wave and inertio-gravity waves further labeled by $l$. Additionally, $n=-1$ gives a Kelvin wave.} ($n,l$) denoted together by $(m,n,l)$ in the presence of a linear damping and a zonal-mean flow. In other words, we have (cf. Equation(34) in \cite{matsuno})
\begin{equation}
a_{m,n,l} = \frac{1}{\alpha - i (\omega _{m,n,l} - \omega) } b_{m,n,l},
\label{eq amp}
\end{equation}
where $\omega$ is the stationary forcing frequency Doppler-shifted by the zonal-mean zonal flow (i.e., $\omega =  u_{0} k$) and $\omega_{m,n,l}$ is the eigenfrequency of the free-wave mode $(m,n,l)$. For sign convention in this study, we always choose $k$ to be positive. Consequently, positive (negative) $\omega$ and $\omega_{m,n,l}$ represent the horizontal phase velocity of the thermal forcing and free modes going westward (eastward). Note that the simple relation between the atmospheric response and thermal forcing in Equation(\ref{eq amp}) results from the same damping rate applied to both the Rayleigh friction and Newtonian cooling \citep{matsuno}. While this procedure limits the parameter space in which the Prandtl number equals just one (cf. SP), its advantage is that it is easy to understand how an atmosphere responds to thermal forcing in terms of different contributions from eigenmodes. Equation(\ref{eq amp}) neatly tells us that
the atmospheric response of each mode, which is proportional to the heating projection, is mainly determined by the relative magnitude of damping rate and the Doppler-shifted free-wave frequency as viewed by the uniform zonal-mean flow. Specifically, the response amplitude in Equation(\ref{eq amp}) can be rewritten in the polar form
\begin{equation}
a_{m,n,l} = r_{m,n,l} \, \exp (\rmi \theta_{m,n,l}),
\end{equation}
where
\begin{equation}
r_{m,n,l} =  \frac{b_{m,n,l}}{ \sqrt{\alpha ^2 + (\omega - \omega_{m,n,l})^2}}, \qquad \theta_{m,n,l} = \arctan \left( \frac{\omega_{m,n,l} - \omega}{\alpha} \right).
\end{equation}
The phase difference in $\theta_{m,n,l}$ is crucial in introducing the correlation of horizontal velocities. For example, the horizontal momentum flux contributed from the modes with the same $m$ and $n$ is given by
\begin{equation}
(\overline{u'v'})_{m,n} = {1\over 2}\mbox{Re} \sum_{l,l'} \left[ u_{m,n,l} v^*_{m,n,l'} \right] =
{1\over 2}r_{m,n,l}r_{m,n,l'}
\sum_{l,l'}\mbox{sin}(\theta_{m,n,l}-\theta_{m,n,l'}) |u_{m,n,l}||v_{m,n,l'}|,\label{eq:uv_flux}
\end{equation}
where the overbar means taking a zonal mean and * means taking a complex conjugate.
In deriving the above relation, we have used the same convention as \cite{matsuno} for free-wave eigenfunctions in which $u_{m,n,l}$ is real and $v_{m,n,l}$ is pure imaginary. It is evident from the above equation that the term ``sin" associated with the phase difference between two modes provides the correlation between $u'$ and $v'$, giving rise to the horizontal momentum flux. For instance, when $l = l'$, there is no correlation from one single horizontal mode and the momentum transfer is zero. Therefore, different horizontal modes are necessary in order to create the phase differences and thus the horizontal momentum flux, as the roles played by the combined equatorial Kelvin and Rossby waves in SP.

Finally, we stress that even though the phase speed of Kelvin waves is not a function of $k$, it is not constant because $h_m$ changes with $m$. In this respect, baroclinic Kelvin waves with vertical propagation are ``dispersive" in a vertically stratified atmosphere. Likewise, baroclinic Rossby waves are not only dispersive in the horizontal direction but are also ``dispersive" in the same way as baroclinic Kelvin waves due to different vertical modes denoted by $m$. We shall discuss the dispersive effect due to vertical stratification in Section 3.2.3.

\subsection{Vertical eigenfunctions}
\label{sec:vertical eigen}

To solve Equation(\ref{vertical eq}) for eigenfunctions and eigenvalues, boundary conditions at the top and bottom boundary of the atmosphere of interest need to be specified. Since we shall compare our 3D linear analysis with our numerical results based on the 3D radiative hydrodynamical simulation in DA and the 3D GCM in SP, we employ the free-slip bottom boundary condition as adopted by DA, namely, the derivatives in the $z$ direction of $u'$ and $v'$ at the bottom boundary are equal to zero, which gives one constraint
\begin{equation}
\frac{d Z_{m}(z)}{d z} + \frac{Z_{m}(z)}{2H} = 0, \mbox{ at } z=z_{b}.
\label{bottom bc}
\end{equation}
This bottom boundary condition is similar to the impermeable condition used by SP for their circulation simulation.

Moreover, we use the outgoing radiation condition as $ z \rightarrow \infty $. It should be noted that the free-slip bottom boundary in our linear model reflects waves rather than allowing waves to travel downward according to Equation(\ref{energy eq}). It is not reasonable unless the waves propagating to $z=z_b$ have been strongly damped.

The eigenfunctions satisfying the boundary conditions consist of a single barotropic mode
\begin{equation}
Z_{m}(z) = e^{-z/2H}
\end{equation}
and the continuous baroclinic modes \citep[cf.][]{Wu2000}
\begin{equation}
Z_{m}(z) = \frac{A \mbox{sin} mz + B \mbox{cos} mz}{ \sqrt{A^{2}+B^{2}}},
\end{equation}
\\
where
\begin{equation*}
\begin{aligned}
A = 2 H m \sin (m z_b) - \cos (m z_b), \\
B = 2 H m \cos (m z_b) + \sin (m z_b),
\end{aligned}
\end{equation*}
with $m \equiv \sqrt{N^2/gh_m - 1/4H^2}$ being any real positive number, representing the vertical wavenumber for the continuous baroclinic mode.

For mathematical simplicity, we assume the heating rate $Q$ in Equation(\ref{Q}) has the separable form $p(z)S(y)e^{\rmi kx}$ like the heating function used in \cite{Wu2000},
and then we rewrite
\begin{equation}
F_{m} = F(m)S(y).
\end{equation}
Hence
\begin{equation}
\sum_{m} F(m)S(y)Z_{m}(z) = p(z)S(y),
\label{p(z) eq}
\end{equation}
where
\begin{equation}
F(m)= \sqrt{\frac{2}{\pi}} \int_0^\infty p(z) Z_{m}(z) dz
\label{fm eq}
\end{equation}

By calculating $F(m)$, the projection of $p(z)$ onto the vertical eigenfunction $Z_{m}(z)$, we obtain the contribution from each vertical $m$ mode. As shown in the preceding subsection, each vertical mode has a horizontal structure defined by a set of shallow-water equations with equivalent depth $h_{m}$ and corresponding length scale and eigenfrequencies.
We sum them up to construct the complete solutions.

\subsection{Heating function}
The heating function $J$ in Equation(\ref{Q}) describes the strength and distribution of the thermal forcing due to the stellar irradiation.
In our linear model, the zonally-uniform heating (i.e., $s=0$) provides the equilibrium temperature, which is used as the background temperature $T_0$ of the isothermal atmosphere. As has been stated in the paper, we focus ourselves on the $s=1$ thermal forcing in the present work.

\begin{figure}[ht]
\centering
\begin{tabular}{cc}
\subfigure{\scalebox{0.4}{\includegraphics{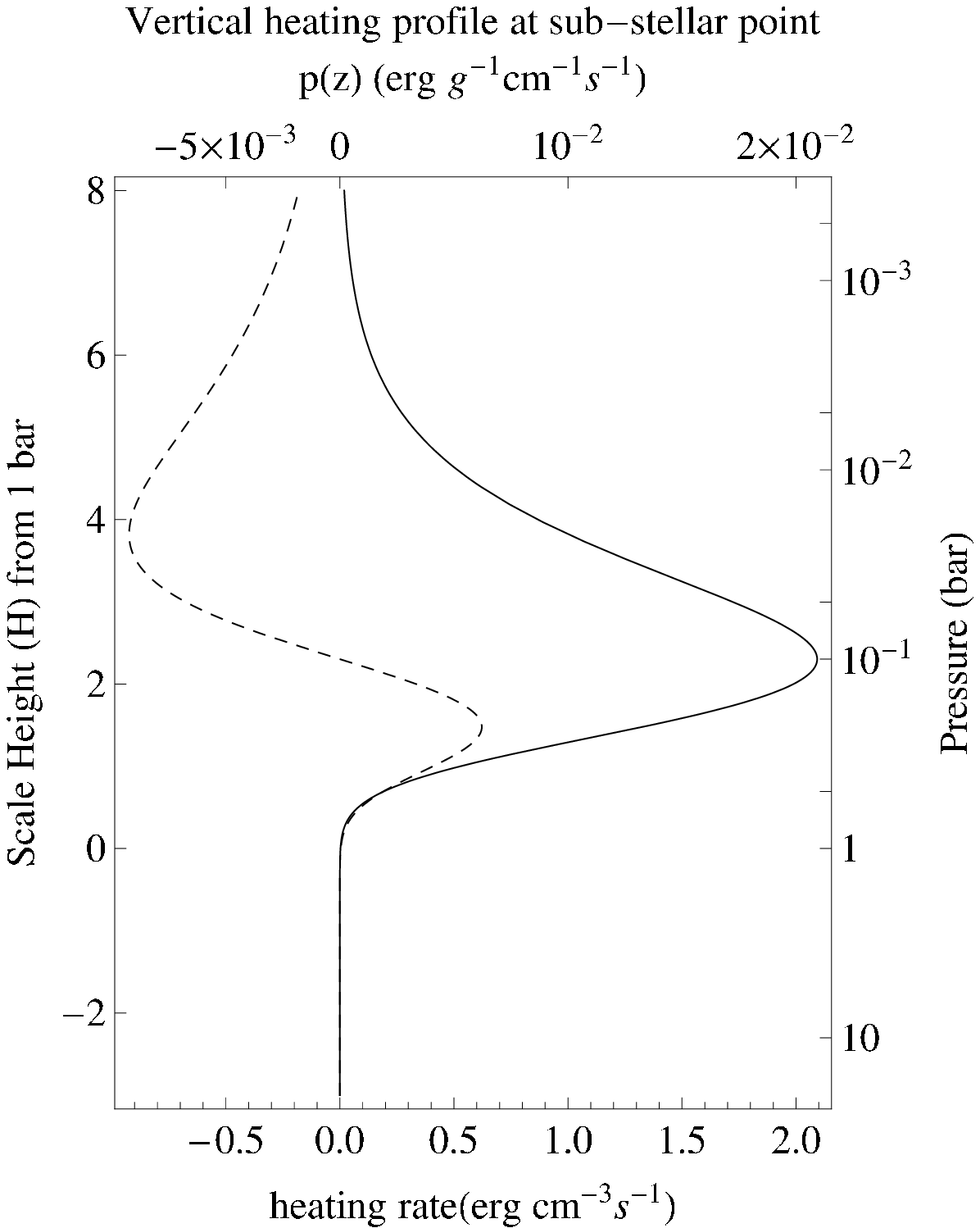}}}
&
\includegraphics[height= 5.6cm, width= 5.4cm]{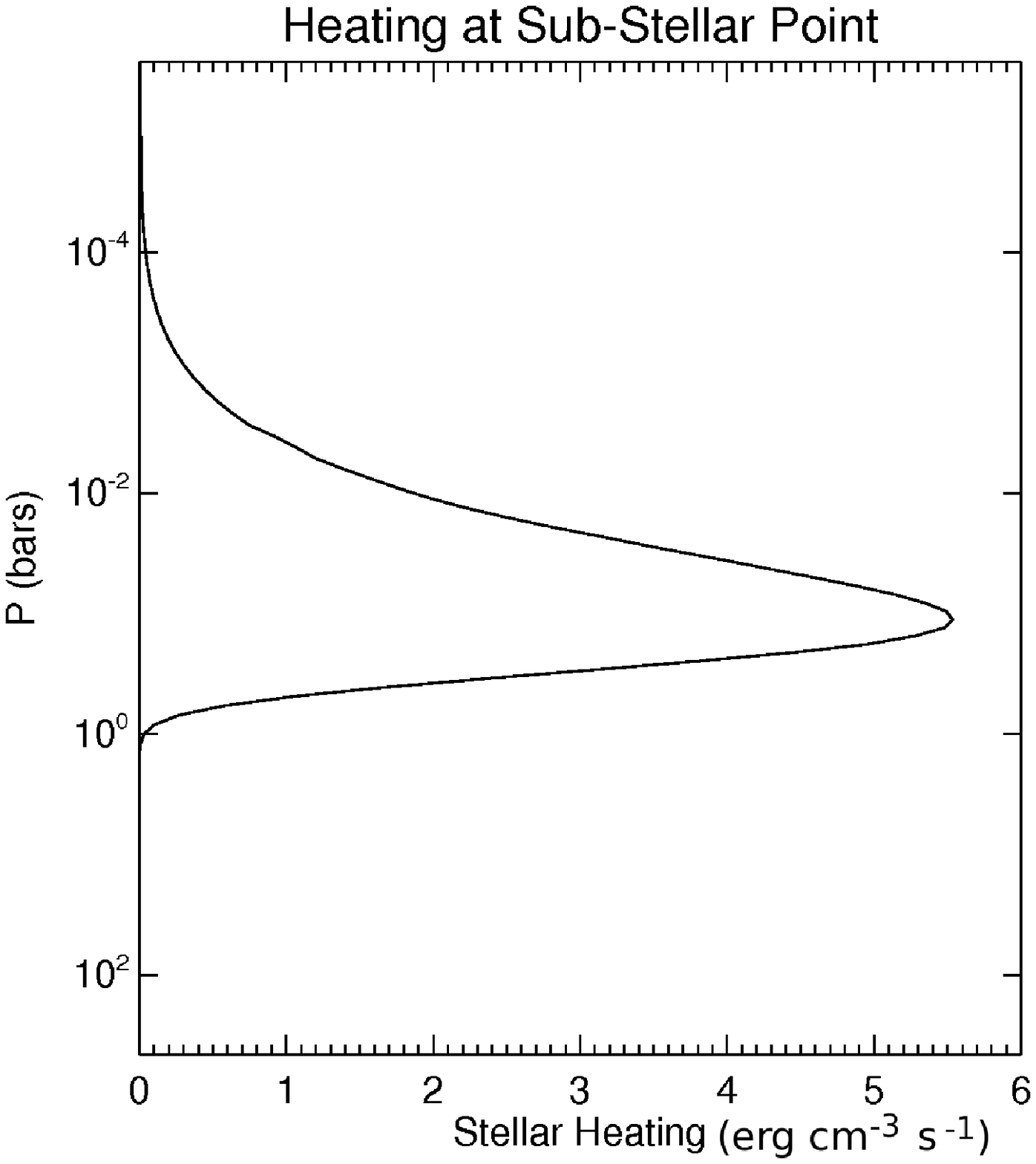}\\
\subfigure{\scalebox{0.54}{\includegraphics{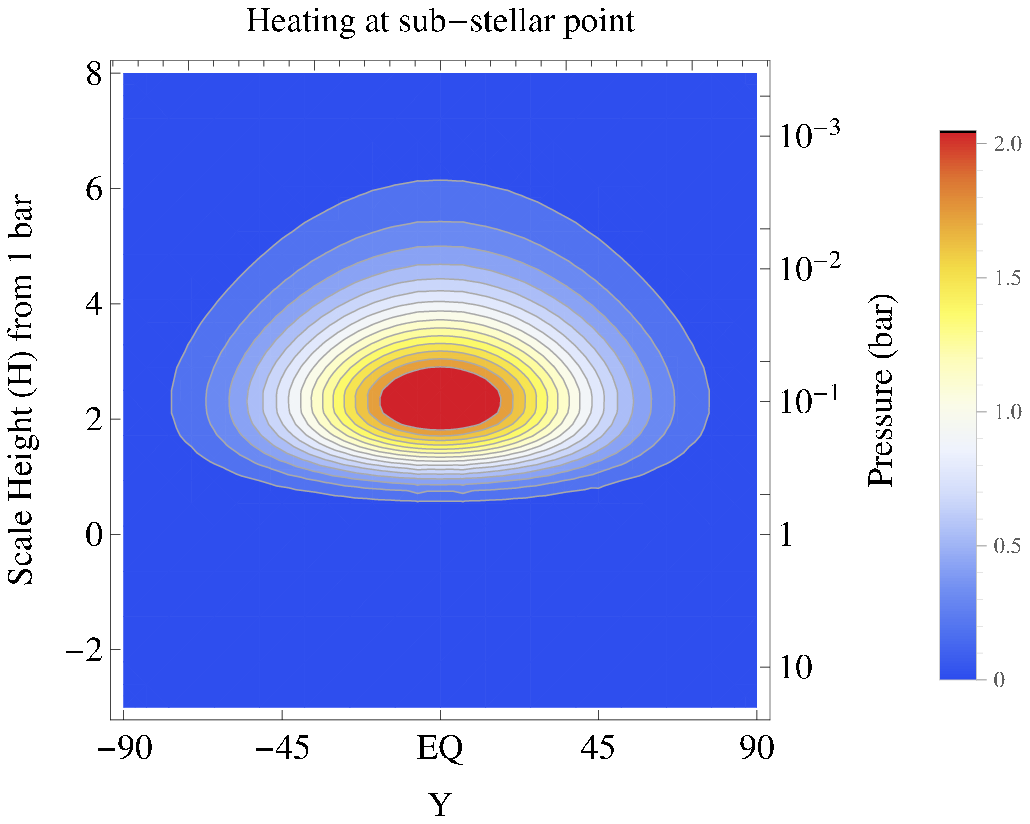}}}
&
\subfigure{\scalebox{0.27}{\includegraphics{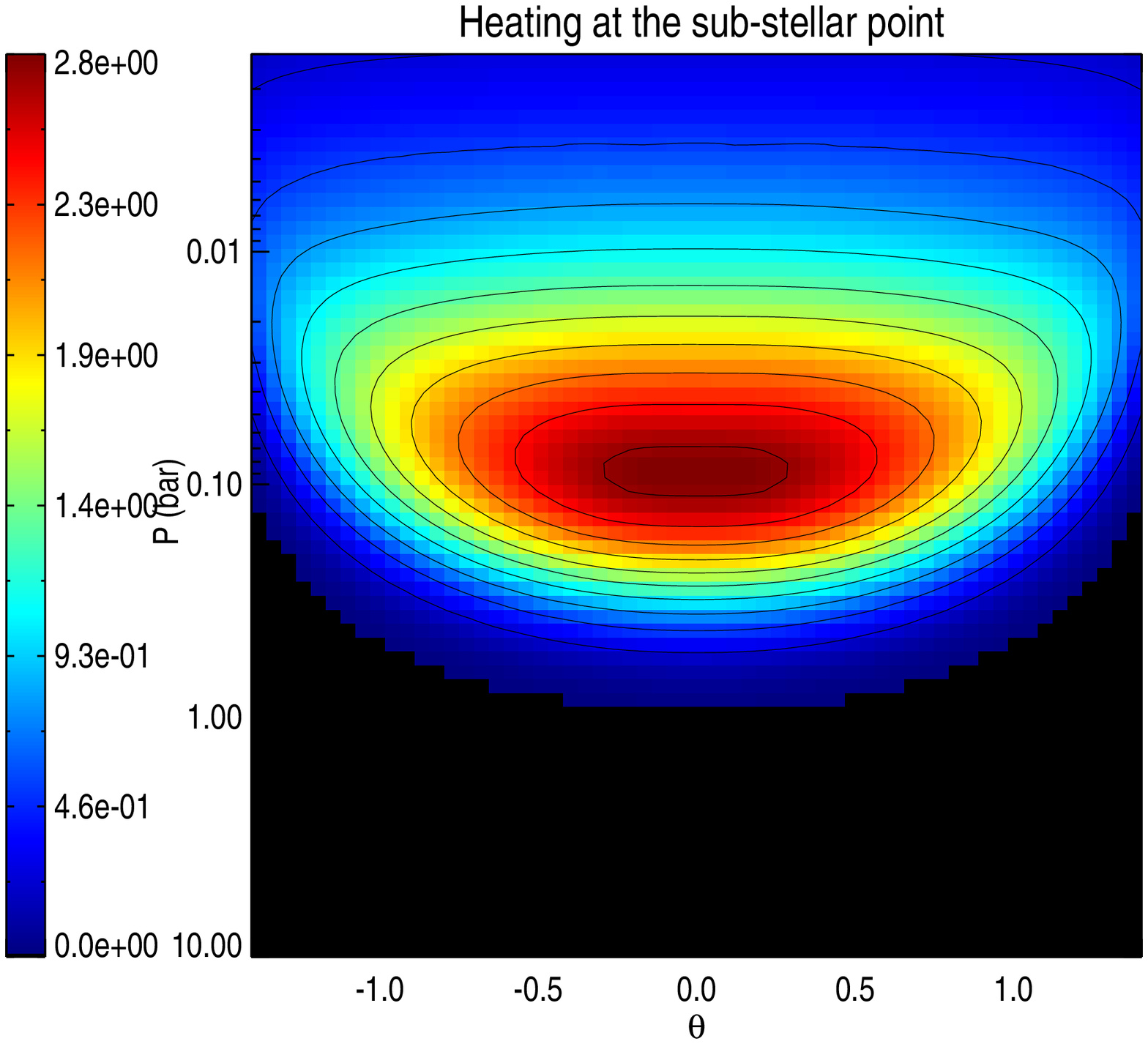}}}\\
\subfigure{\scalebox{0.54}{\includegraphics{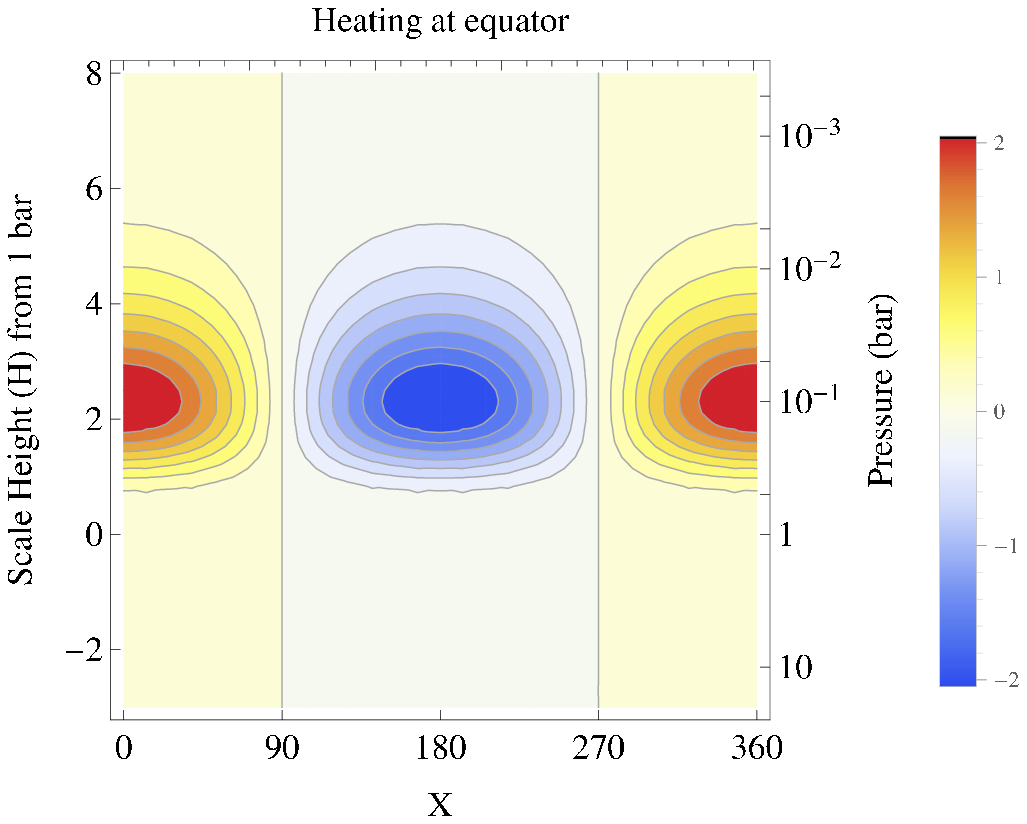}}}
&
\subfigure{\scalebox{0.27}{\includegraphics{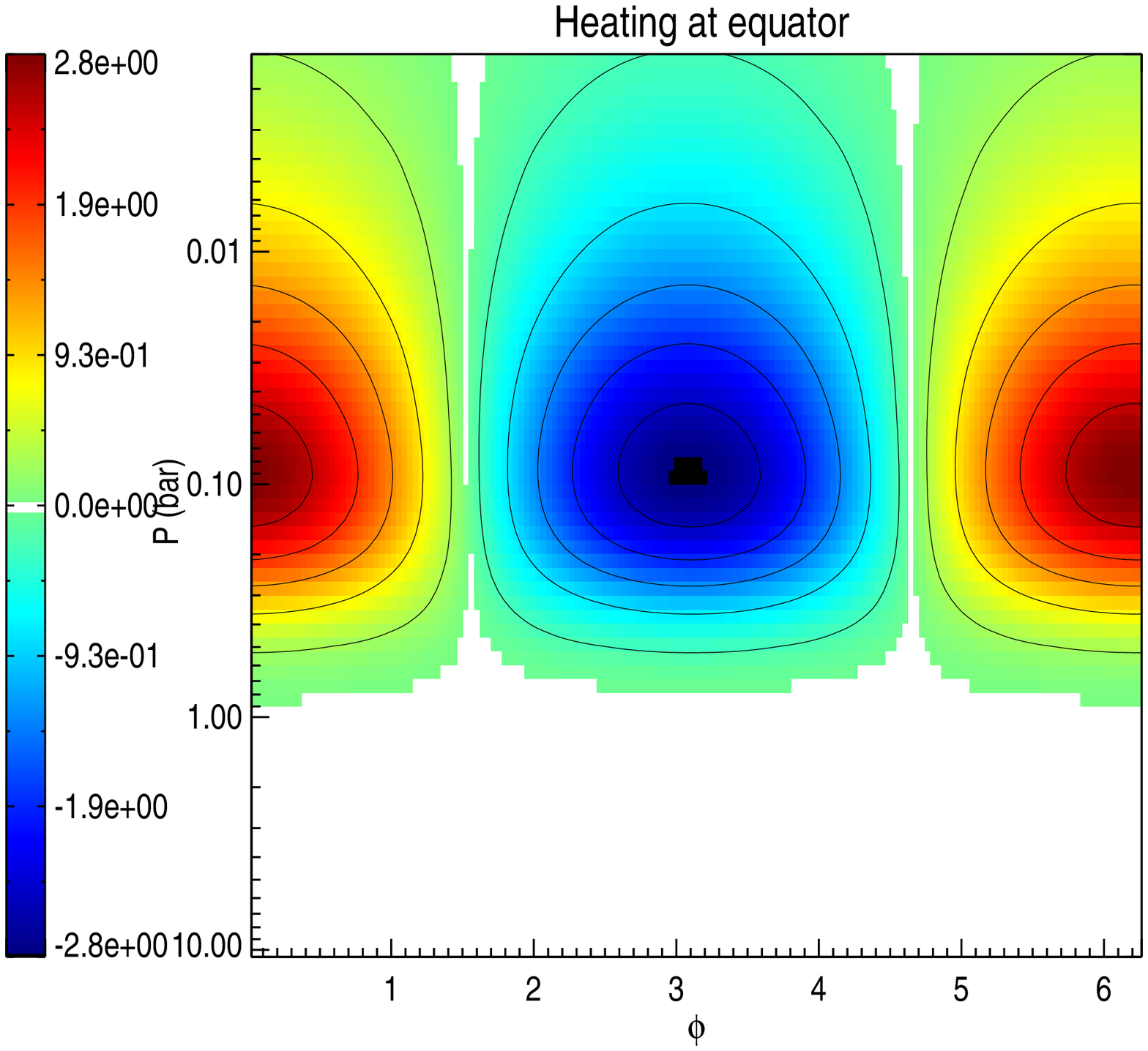}}}
\end{tabular}
  \caption{Structure of the 3D heating profile for the linear analysis (left column) in comparison with that from the radiative hydrodynamical simulation (right column). The top left panel gives the vertical profile of the $s=1$ heating $\rho J$ in Equation(\ref{eq heat}) (solid) and $p(z)$ (dashed) in Equation(\ref{p(z) eq}), while the top right panel gives the vertical profile of the maximum (at sub-stellar point) of the stellar heating term used in the simulation. The middle panels depict the $s=1$ heating profile on the $y$-$z$ cross-sectional plane at sub-stellar point. The bottom panels show the $s=1$ heating profile on the $x$-$z$ plane at the equator. The color bars denote the heating rate in erg cm$^{-3}$ s$^{-1}$.}
  \label{heat plot}
\end{figure}

Figure~\ref{heat plot} shows
the $s=1$ distribution of the stellar heating rate per unit volume absorbed in the atmosphere, i.e. $\rho_{0} J$, adopted in the linear analysis (left column) in comparison with that in the radiative hydrodynamic simulation (right column). We apply a simple analytical function to mimic the vertical profile of $s=1$ thermal forcing in \citet{Dobb} as shown in the top row of Figure~\ref{heat plot}; namely,
\begin{equation}
\rho_{0}(z) J(z) =\rho_0(z) F_{*,s=1} \kappa_e \exp{[\kappa_{e} H(\rho_{0}(z_{t})-\rho_{0}(z))]}
\label{eq heat}
\end{equation}
where $F_{*,s=1}$ is the $s=1$ component of the stellar flux, $\kappa_{e}$ is a parameter presenting the opacity effect, and $z_{t}$ is the vertical location of the top boundary. The vertical coordinate is set to zero at 1 bar, while the bottom boundary where the free-slip boundary condition applies lies at the pressure level of 20 bars.\footnote{The free-slip boundary condition is placed nearly at 500 bars in DA and the impermeable boundary condition is placed at 200 bars in SP \citep[also see][]{SF09}.} Hereafter, whenever a vertical structure of the atmosphere is presented, we show the results in the vertical domain from the bottom boundary to the altitude at $3\times 10^{-4}$ bars. $\kappa_{e}$ is set as $0.0175$ cm$^2$/g for a close fit to the vertical heating profile in the simulation, as shown in the top panels of Figure~\ref{heat plot}. The heating-center level of the stellar heating rate per unit volume lies at $p \approx 0.1$ bar. Most of the stellar irradiation is absorbed between 0.1 and 1 bar.

The meridional heating profiles in the linear theory are assumed to be Gaussian functions (see the middle left panel for $\rho J$ at the substellar point where $x=0$), with location and width matching the stellar heating term used in the simulation (see the middle right panel for $\rho J$ at $\phi=0$)\footnote{$\phi$ is the azimuthal angle of the spherical coordinate in the simulation, and should not be confused with the geopotential perturbation $\phi'$.}.  The slight difference between the latitudinal heating profile in the linear theory and in the simulation is due primarily to the fact that the atmosphere in the simulation is slightly colder and thus vertically thinner at higher latitudes, whereas the background atmosphere in the linear theory is assumed independent of latitudes.  As illustrated in the figure, the width of the intense stellar heating in the $y$ direction is large-scale, comparable to the equatorial Rossby deformation radius, as previously noted by SP. A sinusoidal variation along the longitude, i.e. $\exp (\rmi k x)$, is applied in the linear theory to represent the day-night heating contrast; $\rho J$ at the equator shown in the bottom left panel is similar to the $s=1$ thermal forcing in the simulation plotted in the bottom right panel.

Since the characteristic length scale $L_m$ of each vertical $m$ mode now differs, it is not possible to have one set of free-wave  eigenfunctions ($n=\pm 1$ modes only) with the meridional scale (equatorial Rossby deformation radius) matching the heating profile, such that the solutions can be expanded with only a few terms like those in \cite{matsuno}, in \cite{Gill80}, or in SP. Theoretically, infinite terms are required in order to expand the meridional Gaussian heating. In practice, higher $n$ modes contribute much less at the equator so the solutions can be truncated. Because the eigenfunctions with even $n$ are anti-symmetric to the equator, $n$ should be odd numbers to provide the eigenfunctions for constructing a stellar heating symmetric to the equator. Modes up to $n=7$ are used to get a fairly good approximation in our analysis. The presence of higher $n$ modes because of the vertical structure decomposition does not alter the results significantly in most cases, compared with the shallow-water solutions, but differences arise when the damping term is small, which will be further discussed in Section \ref{no u0}.

\begin{figure}[ht]
\begin{center}
\includegraphics[height= 5 cm]{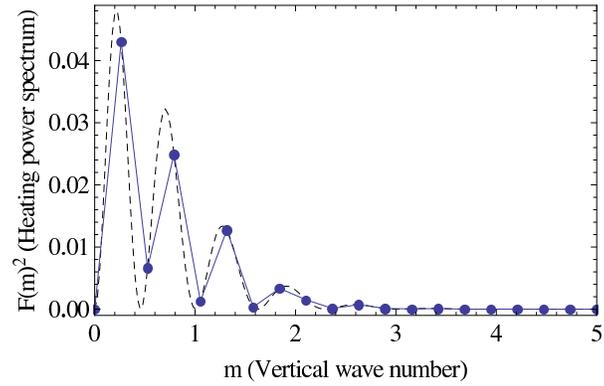}
\end{center}
\caption{Power spectrum of the vertical heating profile $F(m)^{2}$, which presents the vertical heating projected on each vertical baroclinic mode denoted by $m$ (in units of $H^{-1}$).
The power spectrum composed of 20 vertical modes (solid curve) is compared with that composed of 200 vertical modes (dashed curve).
The energy density projected on the barotropic mode (not shown here) is negligible.
}
\label{fm2 plot}
\end{figure}

The power spectrum $F^2(m)$ corresponding to the vertical heating in the linear theory is presented in Figure~\ref{fm2 plot}. As described previously, the energy spectrum of the baroclinic mode should be continuous.  Nevertheless, the 20 discrete baroclinic modes shown in Figure~\ref{fm2 plot} are able to construct a reasonably good profile of the vertical stellar heating. To further confirm that we do not terribly miss any dominant modes, the power spectrum composed of 200 vertical modes are also calculated and illustrated in Figure~\ref{fm2 plot}, showing the spectral peaks agree well with that constructed only by 20 modes. In what follows, we shall use the 20 vertical modes for the linear calculation. A baroclinic mode with a smaller $m$ has a longer vertical wavelength, and therefore has a larger equivalent depth $h_m$ and a larger meridional scale $\sim L_m$. The projection on the barotropic mode is not shown but it is negligible. Generally, only a very deep heating profile that extends to the bottom boundary would give rise to noticeable projection on the barotropic mode.

To roughly examine the sensitivity of the wave solutions to the vertical heating structure, we have further tried various values for the opacity parameter $\kappa_e$ in Equation(\ref{eq heat}). The results for different $\kappa_e$ are generally similar to each other except that the vertical location of wave patterns shifts. It is because a larger (smaller) $\kappa_e$ presents an atmosphere absorbing the stellar heating at a higher (lower) level. We have also tested the linear theory with general heating profiles using Gaussian functions with different vertical widths, and found that a narrower heating profile projects more onto shorter vertical modes as expected.
Having performed these tests, it should be kept in mind that the vertical heating profile, the efficiency of re-radiated radiation energy, and the background states are determined by the opacity of the gas. Modifying one without the other is inconsistent.
Therefore, for the purpose of the consistency that has been taken care in our radiative hydrodynamical simulations, we shall focus on the particular case of HD 189733b as the fiducial model in this study.

\subsection{Wave Resonance}
Equation(\ref{eq amp}) describes the response amplitude of each mode, which reaches a maximum when $\omega = \omega_{m,n,l}$. In the limit of no dissipation ($\alpha$ = 0), the response grows without bound in the linear theory, which implies a resonance analogous to the singularity of the topographic Rossby waves in the presence of a westerly wind \citep{Holton 2004}. The resonances result from horizontal discrete modes; as $\omega$ $\rightarrow$ $\omega_{m,n,l}$, the free mode becomes stationary relative to the heating in the reference frame of the  zonal-mean flow and will be amplified until nonlinearity becomes important. A more detailed analysis on resonances in both linear theory and multi-level GCMs, as well as the positive feedback leading to transition to superrotation can be found in \citet{Arnold} in the study of the Earth's atmosphere.

The response amplitude of each mode, mainly equatorial Rossby and Kelvin waves in the range of the mean-flow speed that we investigate, is represented in terms of $\sum_m |\sum_n a_{m,n,l}|$ and is shown in Figure~\ref{amp}. Given a $u_0$, the results are calculated by summing over all $m$ and $n$ for each type of equatorial wave. The figure is plotted using the atmospheric parameters of the hot Jupiter HD 189733b as an example (see Section 3 for the detailed information of parameters and damping rates $\alpha$).
In the plot for $\alpha=0.1$, we can see that the resonances of Rossby and Kelvin waves occur when the velocity of the zonal-mean flow is close to the horizontal phase velocity ($\omega_{m,n,l}/k$) of each free wave ($u_{0}\approx -1000$ to $-2000$ m/s for the Kelvin wave and $u_{0}\approx$600 m/s for the Rossby wave). When $\alpha$ is small ($= 0.01$), the resonance is the strongest since the frequency term outweighs the damping term in Equation(\ref{eq amp}). Thus the resonance occurs at numerous vertical modes associated with various horizontal phase velocities as explained in Section~\ref{sec:basic}. In general, smaller-scale modes have smaller equivalent depths $h_m$, and thus smaller phase speeds. Therefore, multiple resonance peaks are present in the top left panel. In contrast, stronger damping rates suppress the resonant effects more significantly, i.e., a large $\alpha$ diminishes the frequency term in Equation(\ref{eq amp}). This is illustrated in the bottom left panel for $\alpha=0.5$.

Generally, owing to the Rossby wave resonance, the response amplitude of the Rossby wave is greater than that of the Kelvin wave when $u_{0}>0$ and when the damping is weak or modest ($\alpha =$ 0.01 or 0.1). This is illustrated in the top two panels in the right column of the figure. In contrast, wave resonances are substantially suppressed by strong damping, leading to the response amplitude of each mode simply proportional to the thermal forcing amplitude of the corresponding mode. The Kelvin mode couples better with the assumed meridional heating distribution because similar to the heating profile, the Kelvin mode also varies in latitude as a Gaussian function centered at the equator. It follows that the response amplitude of the Kelvin wave becomes stronger than that of the Rossby wave when the damping is strong ($\alpha = 0.5$). This is demonstrated in the bottom panel of the right column in the figure.

\subsection{Wave momentum transport and equilibrium states of the atmosphere}
Several early theoretical studies have investigated interactions between a mean flow and waves in the Earth's atmosphere. One of the well-known examples is the quasi-biennial oscillation (QBO), the alternating zonal wind in a period about 28 months in the tropical stratosphere, driven by the propagation and dissipation of equatorial Kelvin and Rossby-gravity waves. The interaction of the waves and the zonal-mean flow is deemed to be responsible for the QBO \citep[e.g.,][]{Lindzen Holton 1972,Andrews mid atm}. In the case of hot Jupiters, the formation of atmospheric superrotation can be dominantly attributed to the Reynolds stress associated with planetary-scale equatorial waves driven by stellar irradiation as explained by SP. Omitting vertical and meridional mean flows and ignoring the friction acting on the mean flow, the acceleration of a zonal-mean zonal flow is governed by \citep[e.g.][]{andrews}
\begin{equation}
\frac{ \partial \overline{u} }{\partial t}  = - \partial_{y} \overline{u' v'} - \frac{1}{\rho_{0}} \partial_{z} \rho_{0} \overline{u' w'}.
\label{zonal mean}
\end{equation}
The first and second terms on the right-hand side are the negative divergence of the horizontal and vertical components of the wave momentum flux, respectively. By definition, $u_0=\overline{u}$. Because the Reynolds stress arises from the linear theory to second order, Equation(\ref{zonal mean}) describes a slow evolution of $u_0$ on a timescale much longer than the relaxation time of the atmosphere in response to the change of $u_0$ in a linear wave problem. The different timescales in the linear theory enables us to estimate the acceleration of the zonal flow based on the steady-state wave solutions for a given $u_0$. The relaxation time of the atmosphere will be defined and discussed in Section 3.3.2 where the implication to the evolution of $u_0$ is presented.

The divergence of the wave momentum fluxes in Equation(\ref{zonal mean}) can be related to thermal forcing and dissipation. More specifically, combing the momentum, continuity and energy equations in Equation(\ref{tot eq}), we first obtain the zonal-mean relation between the energy fluxes, thermal forcing, and dissipation for stationary forced waves as follows
\begin{equation}
\frac{\partial}{\partial y}\overline{\phi ' v'} + \frac{\partial}{\partial z}\overline{\phi ' w'} - \frac{\overline{\phi ' w'}}{H} =
-{\alpha \over 2}(u'^2+v'^2)
+ \frac{1}{N^{2}}\left[ \overline{Q' \frac{\partial \phi '}{\partial z}} - {\alpha \over 2} \left( \frac{\partial \phi '}{\partial z} \right)^{2} \right],
\label{E flux eq}
\end{equation}
where $Q' \equiv (\kappa/H)J$ and the square of a complex number denotes the square of the amplitude of the number. The wave energy fluxes on the left-hand side of the above equation can then be replaced by the wave momentum fluxes through the procedure similar to the Eliassen-Palm first theorem \citep{Lindzen90}, which can be derived from
the linearized momentum equation. Namely,

\begin{equation}
\overline {\phi' v'} = - u_{0}( \overline{u' v'}) - \overline{(\epsilon u') v'},
\label{EP1b}
\end{equation}
and
\begin{equation}
\overline {\phi' w'} = - u_{0}( \overline{u' w'}) - \overline{(\epsilon u') w'} + \overline{(\beta_{i} v') w'} ,
\label{EP1a}
\end{equation}
where $\epsilon \equiv \alpha/(ik)$ and $\beta_i \equiv \beta y/(ik)$.
Substituting Equations(\ref{EP1b}), (\ref{EP1a}), and (\ref{E flux eq}) into Equation(\ref{zonal mean}) for the divergence of wave momentum fluxes\footnote{In the absence of the thermal forcing and dissipation, Equations(\ref{E flux eq}), (\ref{EP1b}), (\ref{EP1a}) give rise to the conservation of the horizontal and vertical momentum fluxes, i.e. $\overline{\rho_0 u'v'}$ and $\overline{\rho_0 (u'-\beta y \eta)w'}$, where $\eta= v'/iku_0$ is the latitudinal displacement of an air parcel due to the Coriolis effect in the steady state. It should be noted that the vertical momentum flux with the extra Coriolis term can be interpreted as the vertical momentum flux measured by following the fluid element (i.e. a Lagrangian mean). It is the Lagrangian mean of the momentum flux that is relevant to the conservation law \citep[e.g.,][]{Bretherton}. In contrast, the zonal mean denoted by an overbar is the Eulerian mean, the average taken at a fixed point \citep{andrews78}. It is also worth noting that the equations with zero $u_0$ present special solutions in which the thermal forcing and/or dissipation must exist to maintain a steady state, and thus it is meaningless to describe the conservation of wave momentum fluxes when $u_0=0$.},
we obtain
\begin{equation}
{\partial u_0 \over \partial t}={1\over u_0}\left[ -{\alpha \over 2} (u'^2+v'^2)+{\overline{Q'\partial_z \phi'}-(\alpha/ 2) (\partial_z \phi')^2 \over N^2}
+\partial_y \overline{(\epsilon u')v'}+\partial_z \overline{(\epsilon u')w'}-\partial_z\overline{(\beta_iv')w'}-{\overline{(\epsilon u')w'} \over H} \right].
\end{equation}
The right-hand side of the above equation contains the terms involving the thermal forcing ($\overline{Q'\partial_z \phi'}$), dissipation (terms associated with $\alpha$), and Coriolis effect (the term with $\beta_i$). Once these second-order terms are balanced out, the net acceleration of the zonal-mean zonal flow is zero and thus there exist equilibrium solutions in Equation(\ref{zonal mean}) for the atmosphere. The stability of the equilibrium solutions is relevant to the evolution of $u_0$, which will be explored in Section 3.3.


\section{Linear Analysis Results}
We apply our linear analysis to the atmosphere of the hot Jupiter HD 189733b as our fiducial model. The atmospheric parameters are listed in Table~1. With the basic state, $H$ is about $2 \times 10^{5}$ m and $N$ is about $5\times 10^{-3}$ sec$^{-1}$. These values are comparable to the values between 10 to $10^{-3}$ bars in the atmosphere of the hot Jupiter in our 3D radiative hydrodynamical simulation. As has been stated in Section 2.3, the vertical domain for our linear analysis ranges from 20 to $3\times 10^{-3}$ bars. In the region below the bottom boundary, the simple setup using the constant temperature and buoyancy frequency for the background state in the linear theory becomes unrealistic.

In the following, we first show our 3D linear results at 0.03 bar pressure level (i.e. slightly above the heating-center level) in comparison with the 2-D linear results in SP in the absence of a zonal-mean zonal flow, recreated here by neglecting the vertical component of our solution. The 3D linear results with a zonal flow are then compared with the $s=1$ wave structures (the predominant mode) extracted by the fast Fourier transform from our 3D radiative hydrodynamical simulation based on DA, which solves the Navier-Stokes equations with viscosity rather than the primitive equations with hypervisocisty used by SP.
The simulations of DA examine the atmosphere of HD 189733b utilizing the fully compressible 3D Navier-Stokes equations coupled to a two-stream, frequency dependent radiative transfer scheme. Radiative opacities are assumed to be a combination of molecular with a solar composition \citep{SB07}, a uniform grey component mimicking the effect of hazes, and a Rayleigh scattering component that scales as the 4th power of the inverse of wavelength. The planet is assumed to be tidally locked to its host star with a rotation period of 2.22 days. The equations are solved on a fixed spherical grid, extending from pressures of 500 to $10^{-4}$ bars. Spectra and phasecurves calculated using results from this model compare favorably to observations.

The magnitude of the kinematic viscosity $\nu$, included self-consistently as a source of drag in the momentum equation and a heat source in the energy equation, in DA was $10^7$ cm$^2$/s; here we have run a series of additional models with viscosities ranging from $10^8$ to $10^{10}$ to explore a wide range of dissipative scenarios.
Likewise, we solve the linear problem in terms of three 
normalized damping rates: $\alpha=$ 0.01, 0.1, and 0.5, hereafter referred to as the weak, modest, and strong damping scenarios.
The dimensional unit to normalize $\alpha$ is $5 \times 10^{-5}$ sec$^{-1}$, which is about the inverse of the characteristic wave travel time (see Equation~\ref{eq:T} later in the paper). Therefore, $\alpha=$ 0.01, 0.1, 0.5 correspond to the dissipation time constants of 20 Earth days, 2 Earth days, and half an Earth day, respectively.

\subsection{No zonal-mean zonal flow: horizontal wave patterns and zonal phase shifts}
\label{no u0}
We begin by investigating the case with no zonal-mean zonal flow.
We show our solutions at 0.03 bar pressure level to compare with the linear shallow-water
results in SP. The three panels in the left column of Figure~\ref{u0 plot} 
display the solutions on the $x$-$y$ plane at the pressure level of 0.03 bar
from the 3D linear model in the three damping scenarios. For comparison,
the three panels in the right column of the figure are solutions obtained from linear shallow-water equations with no vertical structure, similar to the results with $\tau$ = 100, $\tau$ = 10 and $\tau$ = 1 in Figure~3 of SP (corresponding to the dissipation time constants of 1 month, 3 Earth days, and 6 hours), where $\tau=1/\alpha$. We merely elaborate
the change of the zonal phase of wave responses relative to the thermal forcing as $\alpha$ changes, which is not mathematically analyzed by SP.

\begin{figure}
\begin{tabular}{cc}
\scalebox{0.5}{\includegraphics{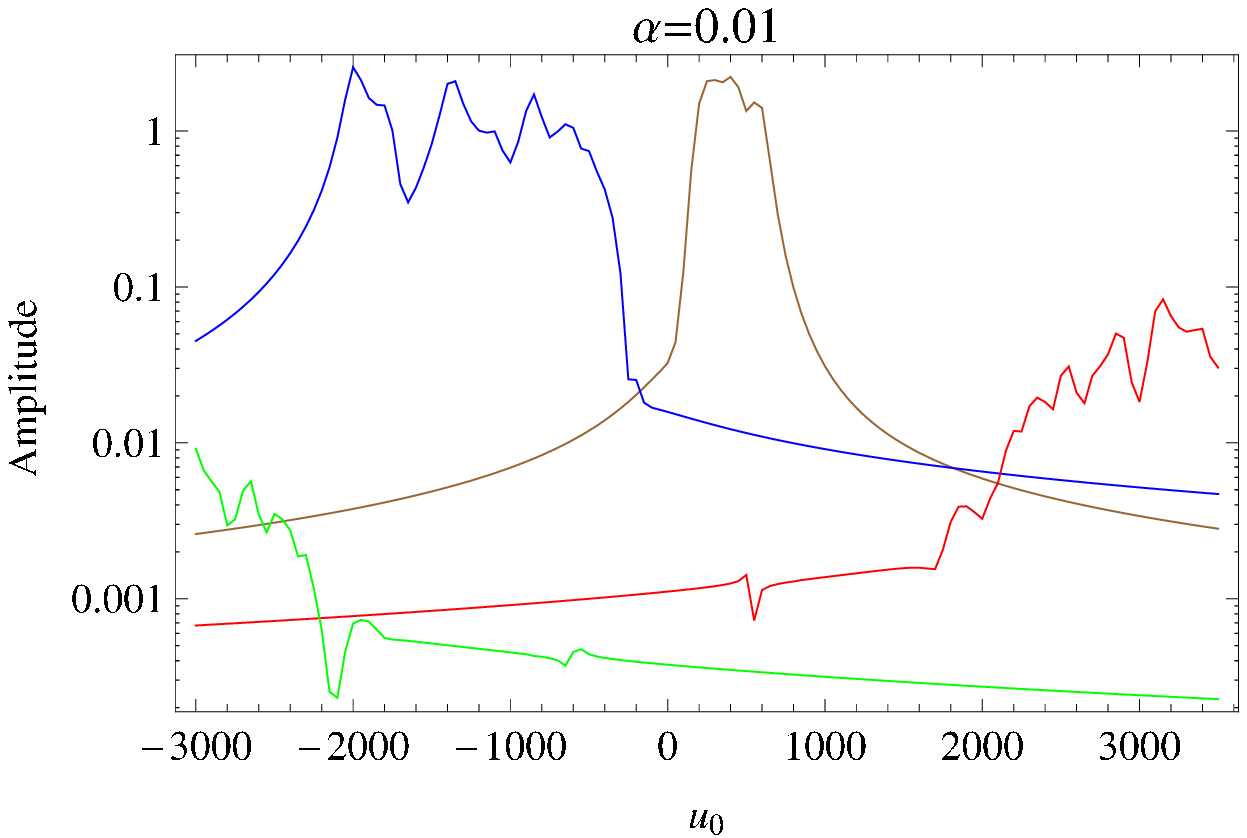}}
&
\scalebox{0.5}{\includegraphics{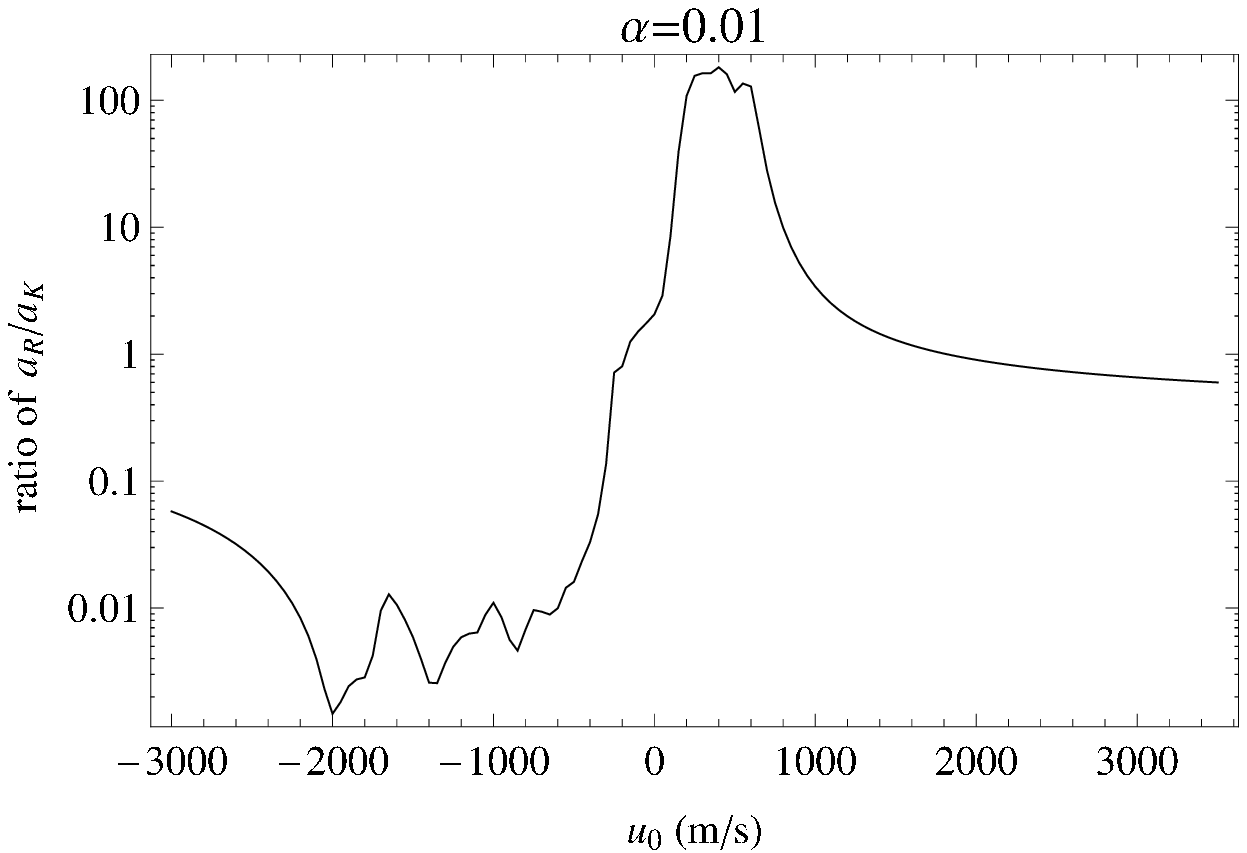}}\\
\scalebox{0.5}{\includegraphics{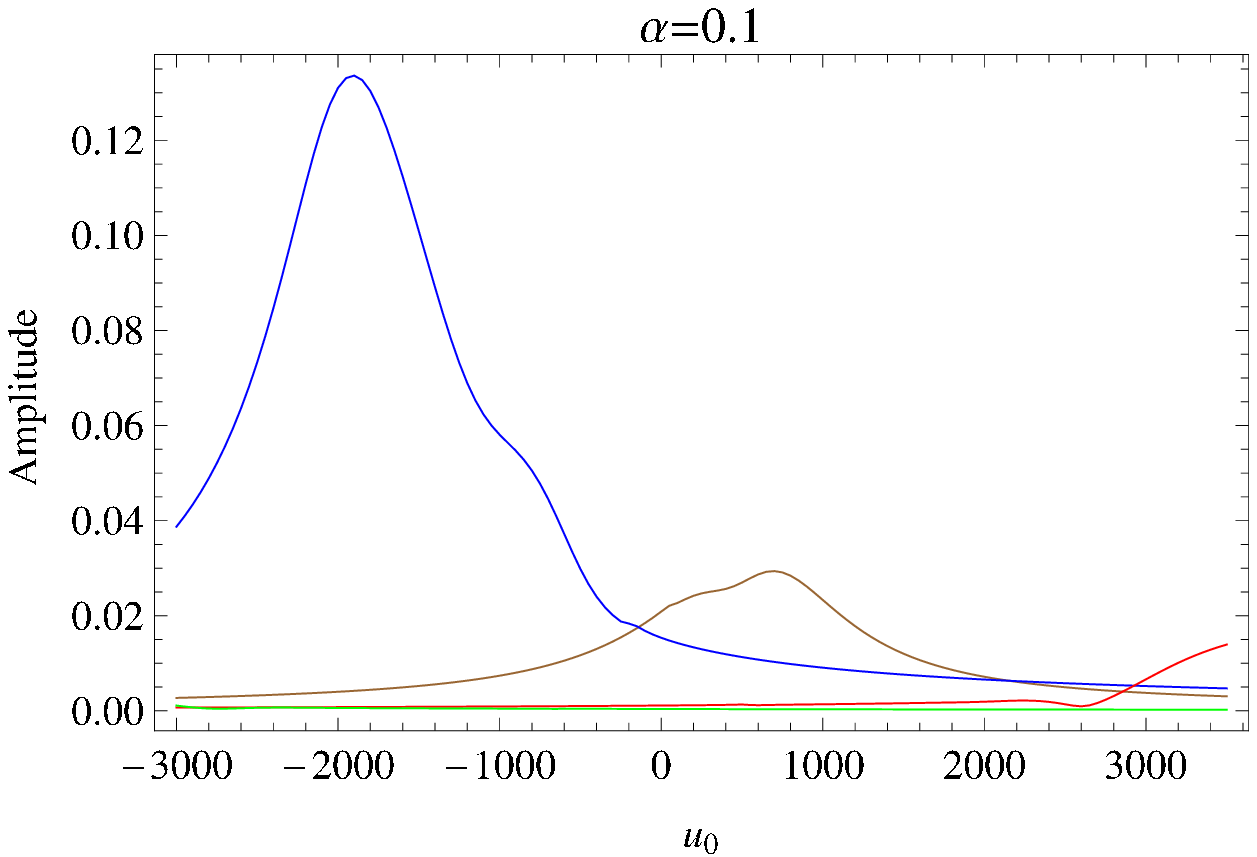}}
&
\scalebox{0.5}{\includegraphics{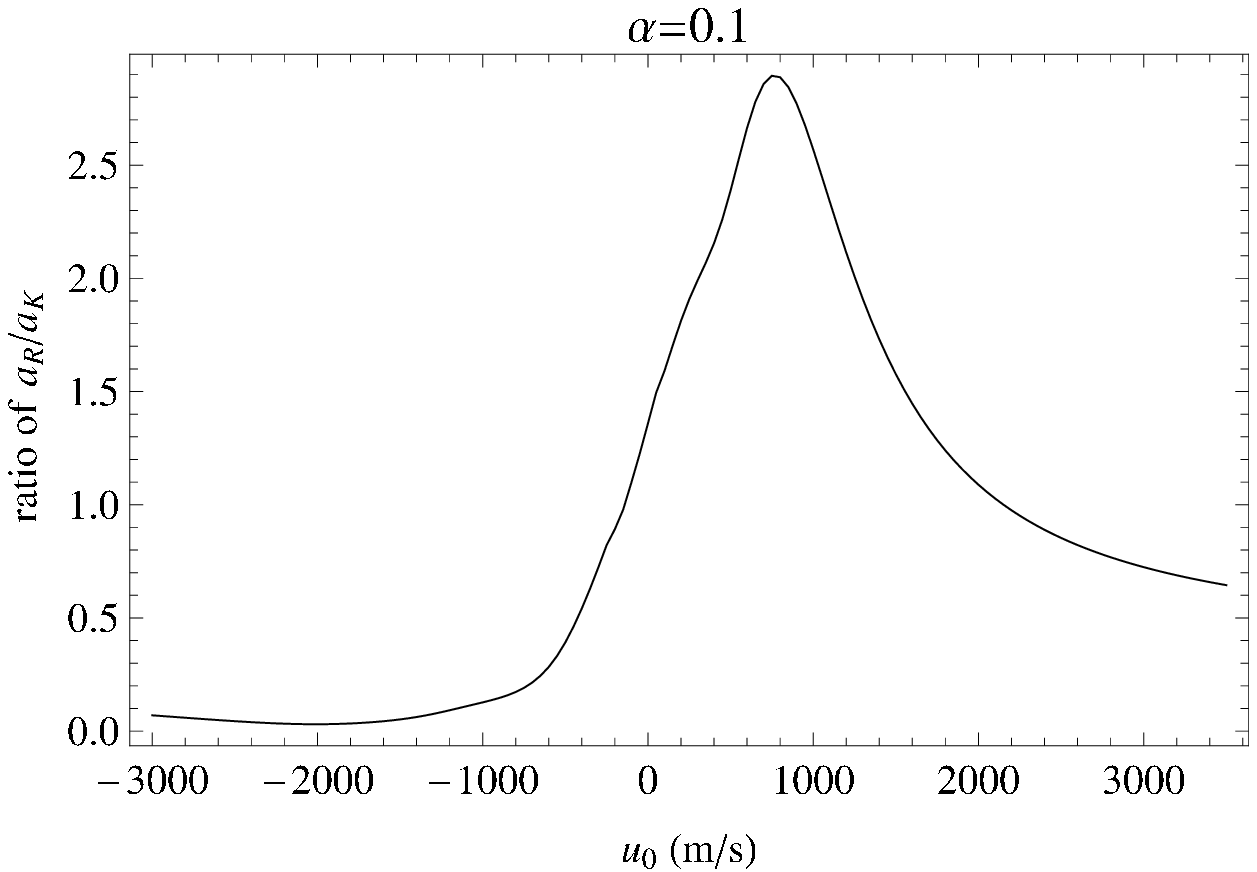}}\\
\scalebox{0.5}{\includegraphics{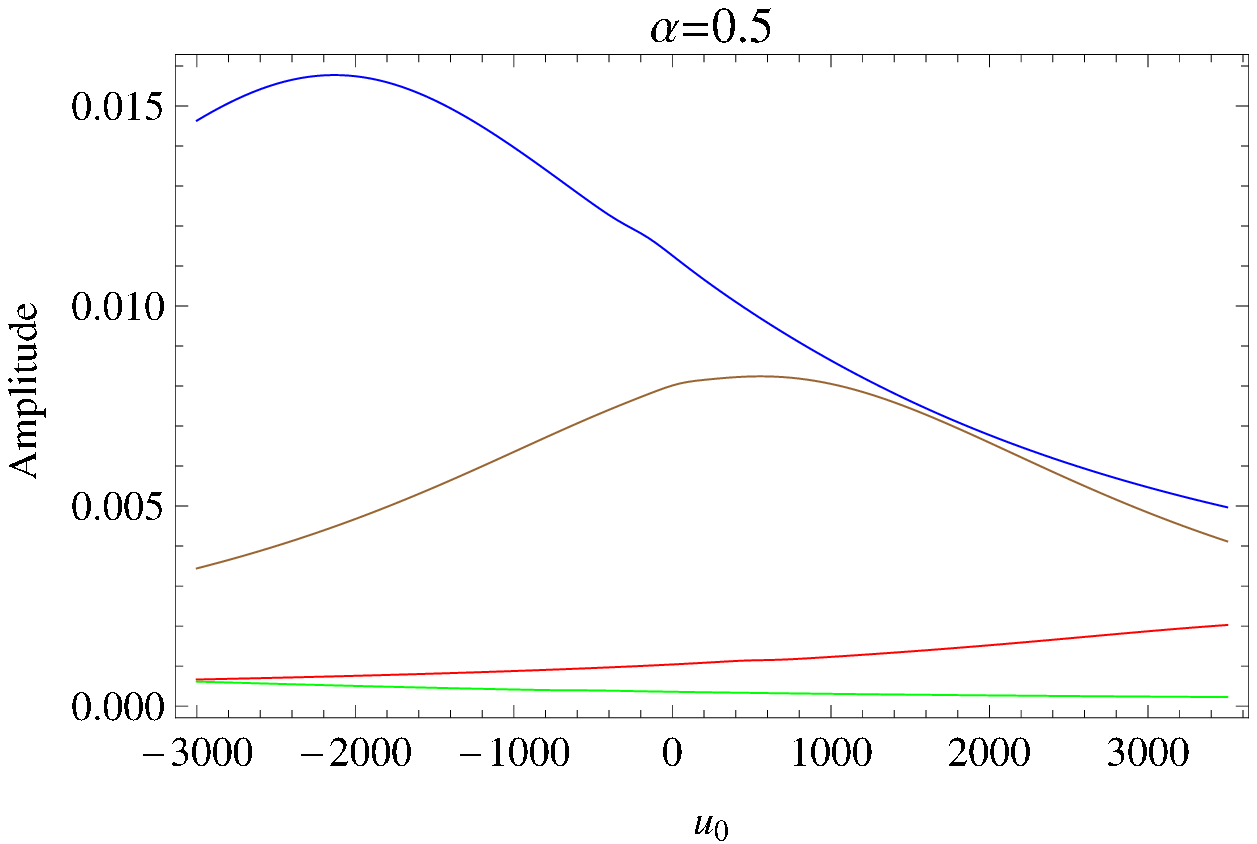}}
&
\scalebox{0.5}{\includegraphics{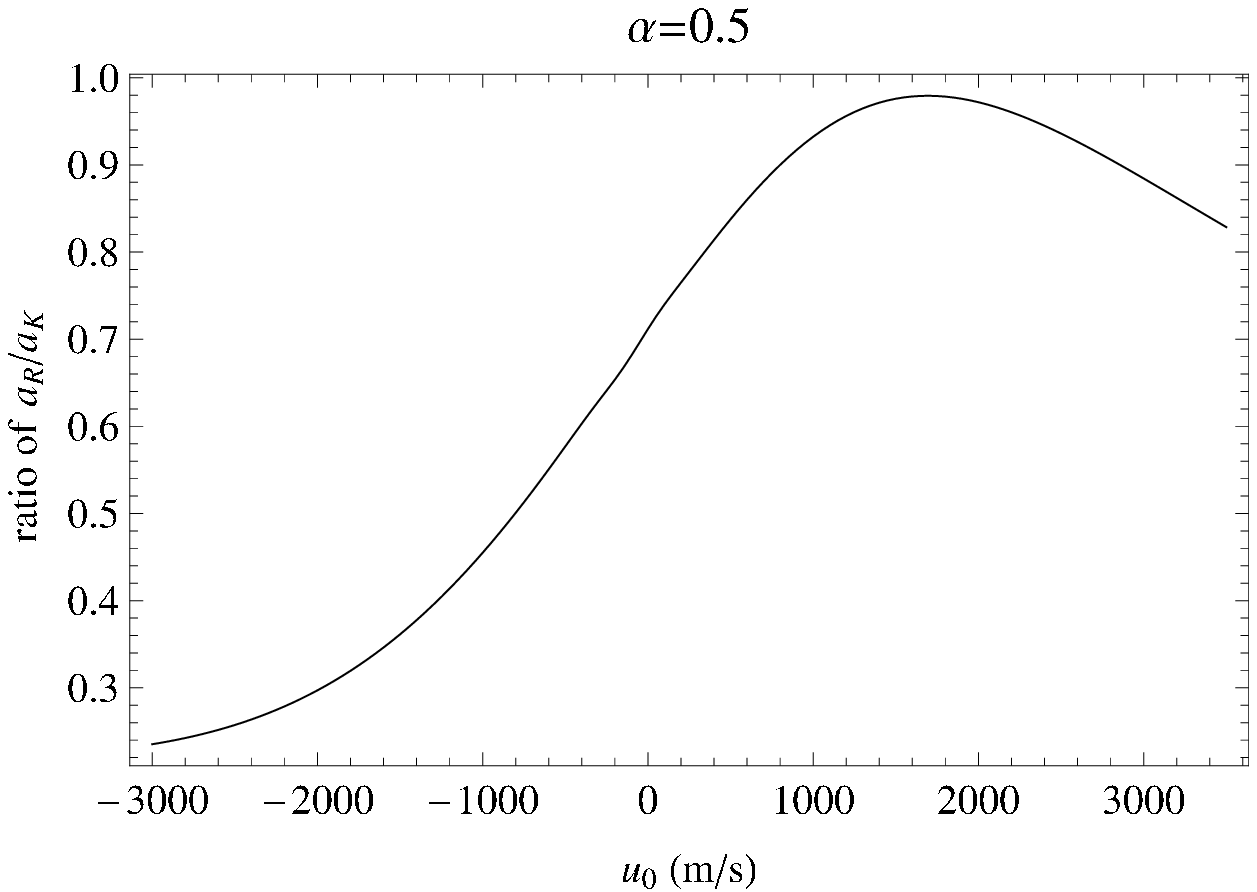}}
\end{tabular}

\caption{Response amplitude of each mode for a range of zonal-mean zonal flow speeds are shown in the left column for different damping rates.
Brown presents the Rossby wave, blue represents the Kelvin wave, red and green denote westward and eastward inertia-gravity waves. The plots in the left column show weak damping (top, shown in the logarithmic coordinate for a clearer view), modest damping (middle) and strong damping (bottom) scenarios. The resonant features manifested by the amplitude peaks are suppressed as the damping is stronger. The three panels in the right column  illustrate the ratio of the response amplitude of Rossby components to Kelvin components for the damping rates corresponding to the panels on the left.
As the resonant features are more suppressed by increasing the damping rate, the ratio becomes smaller.}
\label{amp}
\end{figure}

\begin{figure}[ht]
\begin{center}
\begin{tabular}{cc}
\includegraphics[height= 4.5cm]{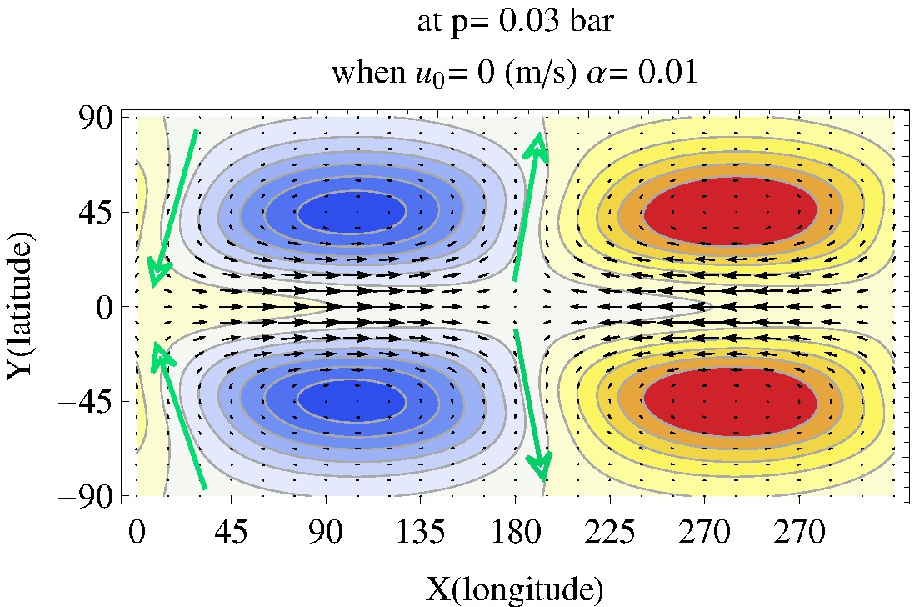}
\raisebox{0.35\height}{\includegraphics[height= 2.75cm, width= 1.05cm]{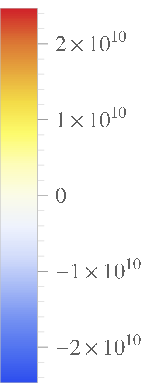}} &
\includegraphics[height= 4cm]{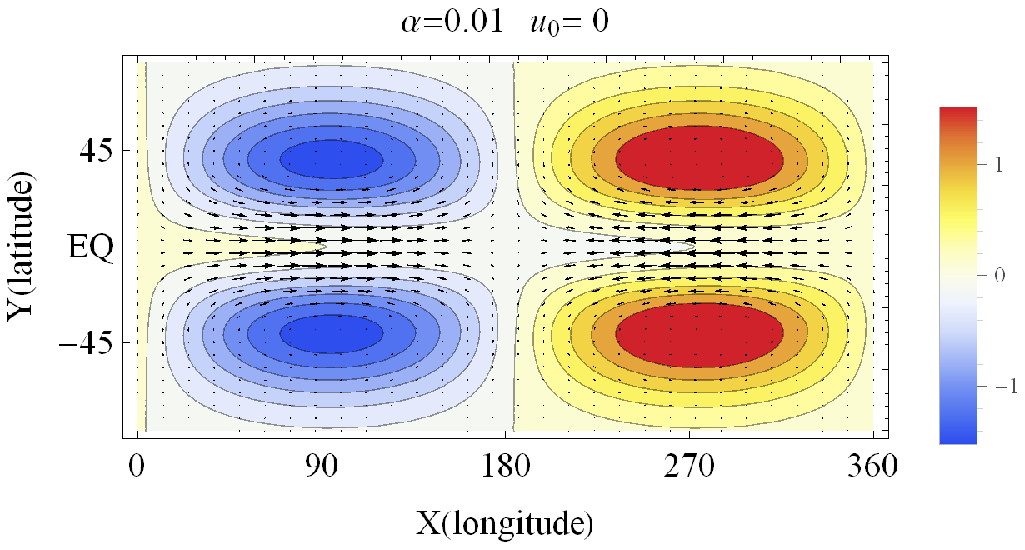} \\
\includegraphics[height= 4.5cm]{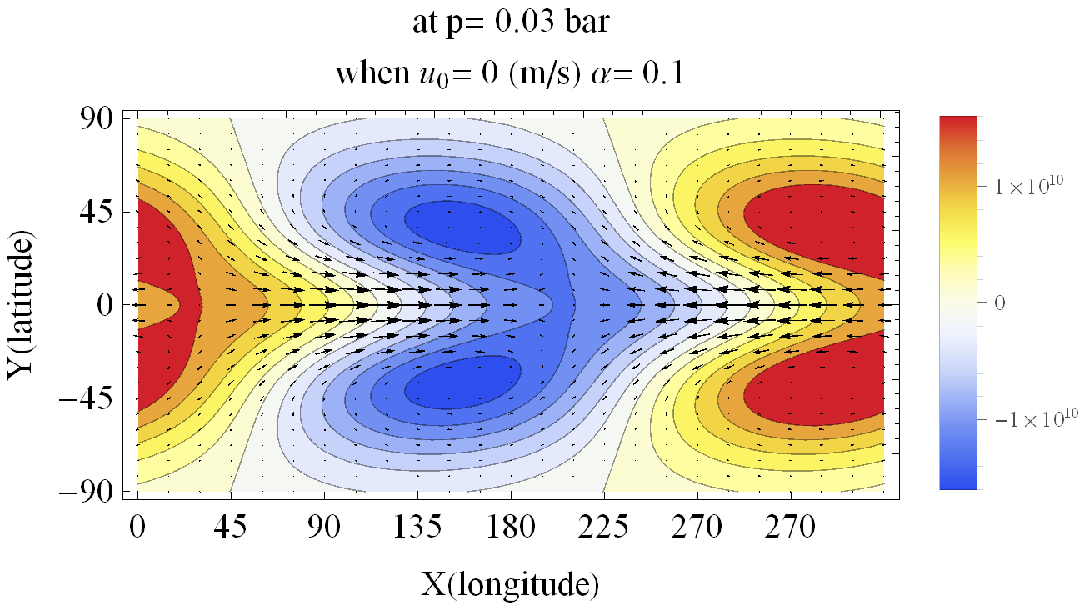} &
\includegraphics[height= 4cm]{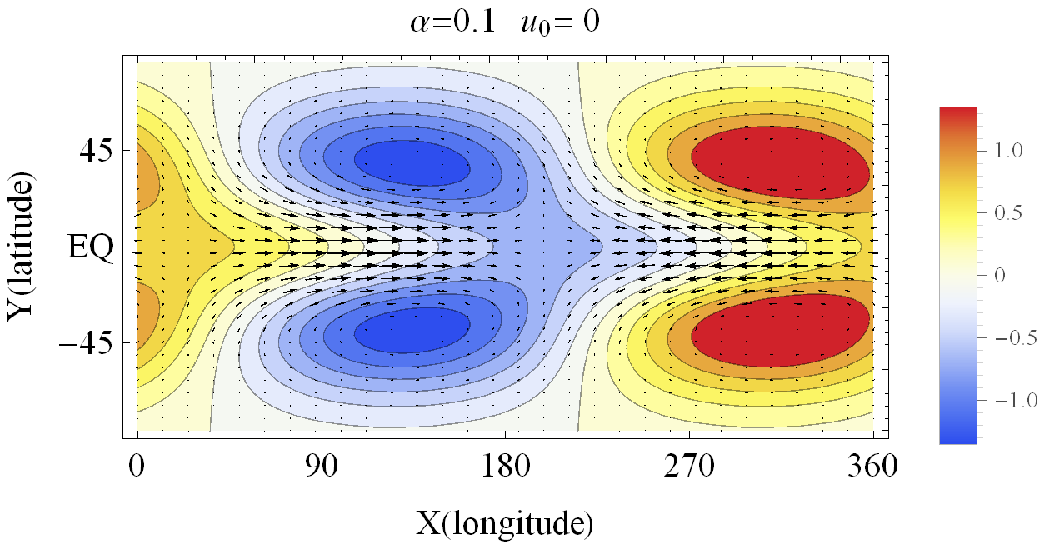} \\
\includegraphics[height= 4.5cm]{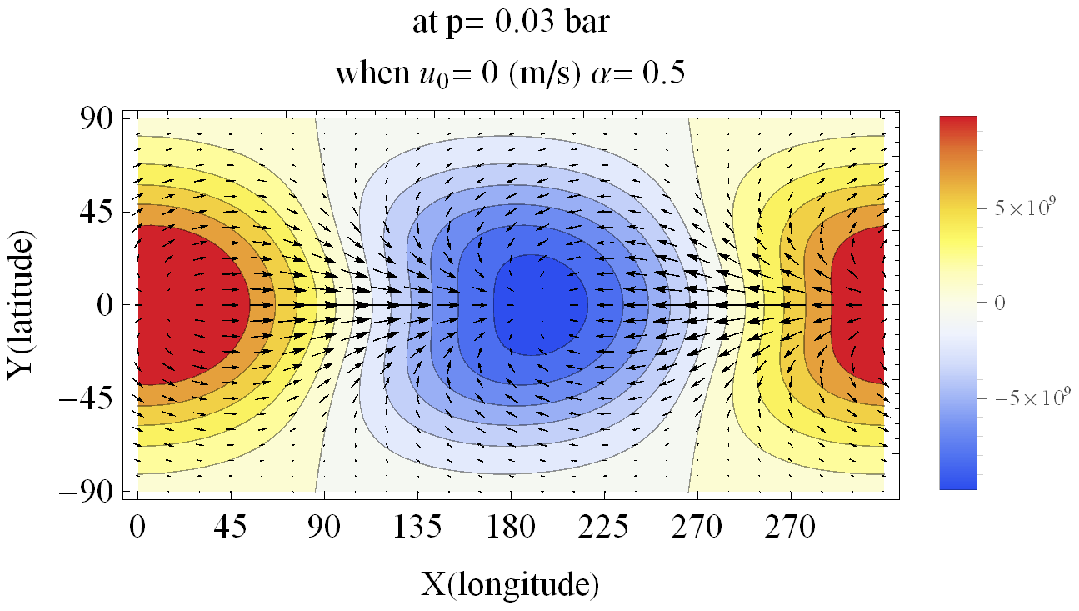} &
\includegraphics[height= 4cm]{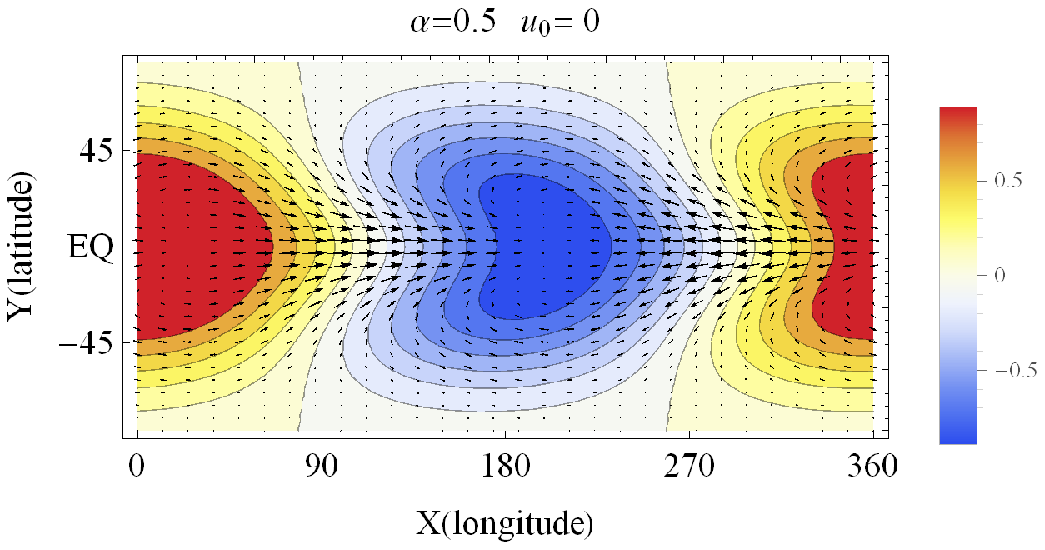}
\end{tabular}
\end{center}
\caption{Effect of the damping rate $\alpha$ on the equatorial wave patterns in the absence of a zonal-mean zonal flow. Wave velocities (arrows) and geopotentials (contours) at height about the heating center are calculated from our linear theory with vertical structures (left column) and from linear shallow-water equations (right column). Color bars give the magnitude scales of wave geopotentials in units of cm$^{2}$s$^{-2}$, which will apply to all the following figures for geopotentials in the rest of the paper. The sub-stellar point lies at $x=0$. The magnitude of velocity fields in each panel is displayed in terms of relative values and hence should not be compared among different panels.}
\label{u0 plot}
\end{figure}

We inspect the two extreme damping scenarios first. In the strong damping scenario, according to Equation(\ref{eq amp}), the response amplitude
is dictated by the factor 1/$\alpha$, so the response of each mode almost has the same phase pattern as the perturbed heating profile. In other words, the geopotential response simply resembles the horizontal heating structure in our 3D linear theory (or the radiative equilibrium height in the linear shallow-water model).
In the weak damping scenario, the response amplitude
is instead dominated by the factor $-1/\rmi \omega_{m,n,l}$ in Equation(\ref{eq amp}). In addition, the Rossby-wave response is stronger than other waves in the weak damping scenario (see Figure~\ref{amp}). Hence, the geopotential response is mainly manifested by planetary-scale Rossby gyres, which have a 90 degree phase difference westward to the horizontal heating due to the imaginary term $-1/\rmi \omega_{m,n,l}$.

In the modest damping scenario, $\alpha$ and $\omega_{m,n,l}$ have the same order of magnitude. The responses from Kelvin and Rossby waves are comparable in magnitude (see Figure~\ref{amp}). Since $\omega_{Kelvin}$ and $\omega_{Rossby}$ have opposite signs (eastward for Kelvin wave and westward for Rossby waves), the two types of waves with opposite propagating directions have competing magnitude, thus leading to the formation of the Matsuno-Gill pattern, described as a ``chevron" pattern of velocity fields (northwest-southeast tilts in the northern hemisphere and southwest-northeast tilts in the southern hemisphere) in SP. This particular pattern, producing $\overline{u'v'} < 0$ in the north of the equator and $\overline{u'v'} > 0$ in the south of the equator, is suggested by SP (also referred to Equation~\ref{eq:uv_flux}) to be crucial for equatorward momentum transport that induces superrotation.

The above results based on the 3D linear model with $u_0=0$ agree fairly well with the solutions from the linear shallow-water model as shown in the panels in the right column of Figure~\ref{u0 plot},
and are therefore in reasonable agreement with the analytical solutions when the drag and radiative time constants are equal as shown in Figure~3 of SP.

\begin{figure}[ht]
\begin{center}
\subfigure[]{\includegraphics[height= 4cm]{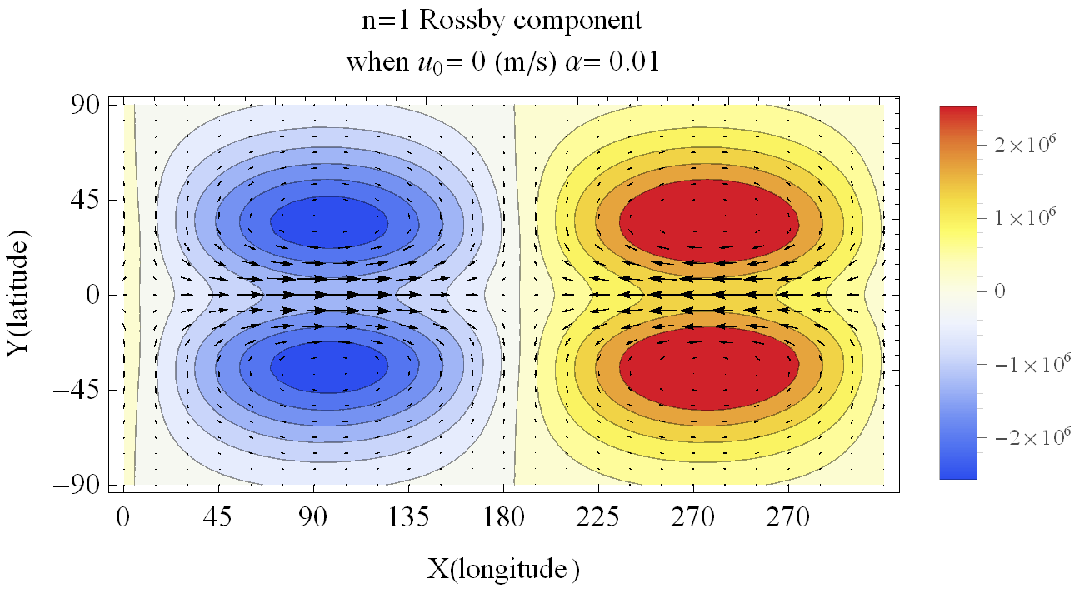}}
\subfigure[]{\includegraphics[height= 4cm]{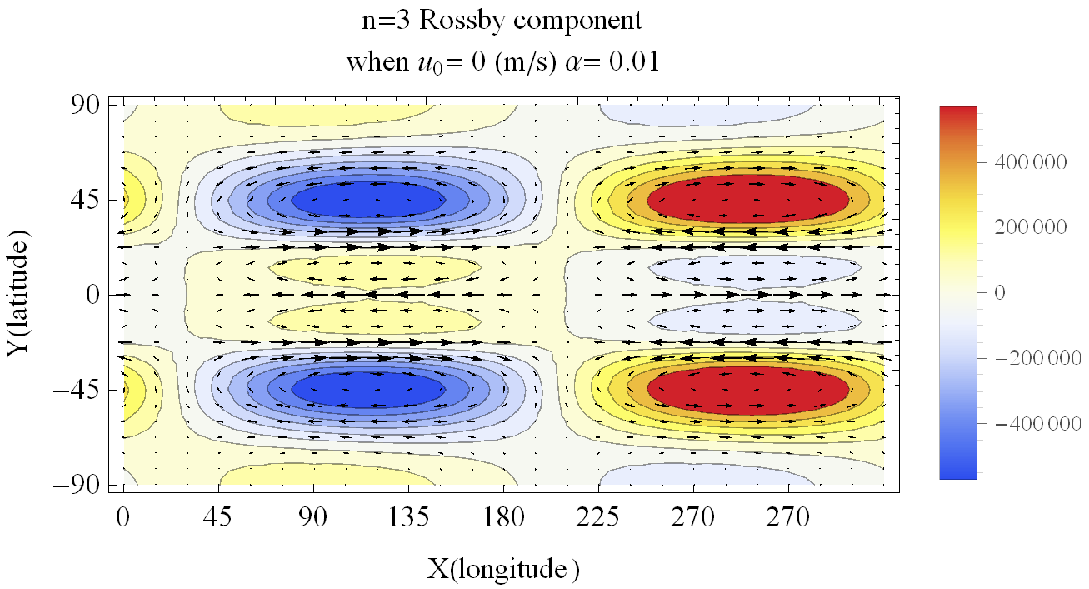}}
\subfigure[]{\includegraphics[height= 4cm]{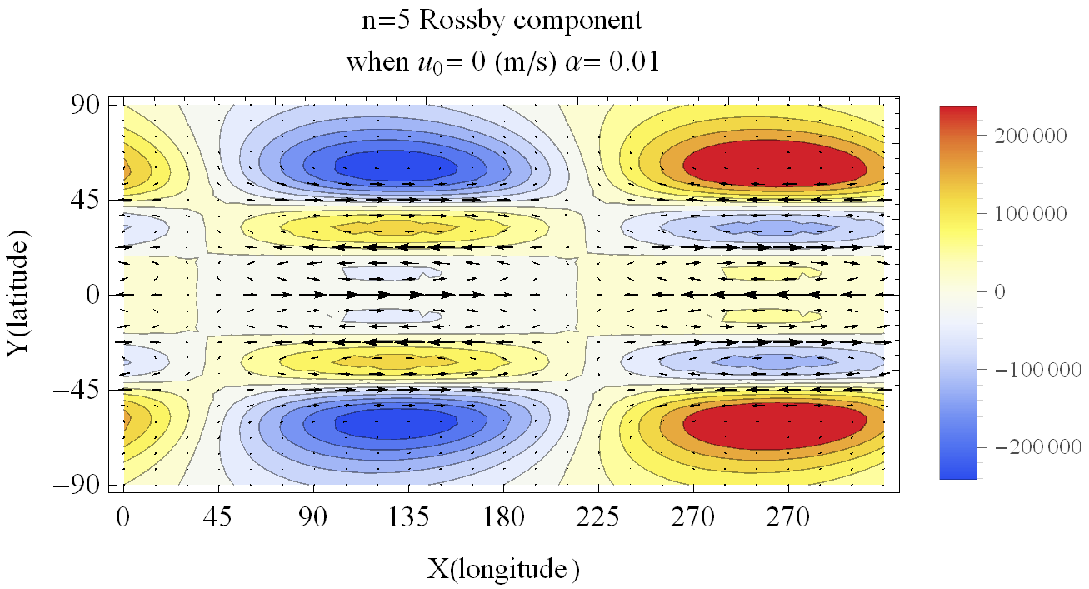}}
\subfigure[]{\includegraphics[height= 4cm]{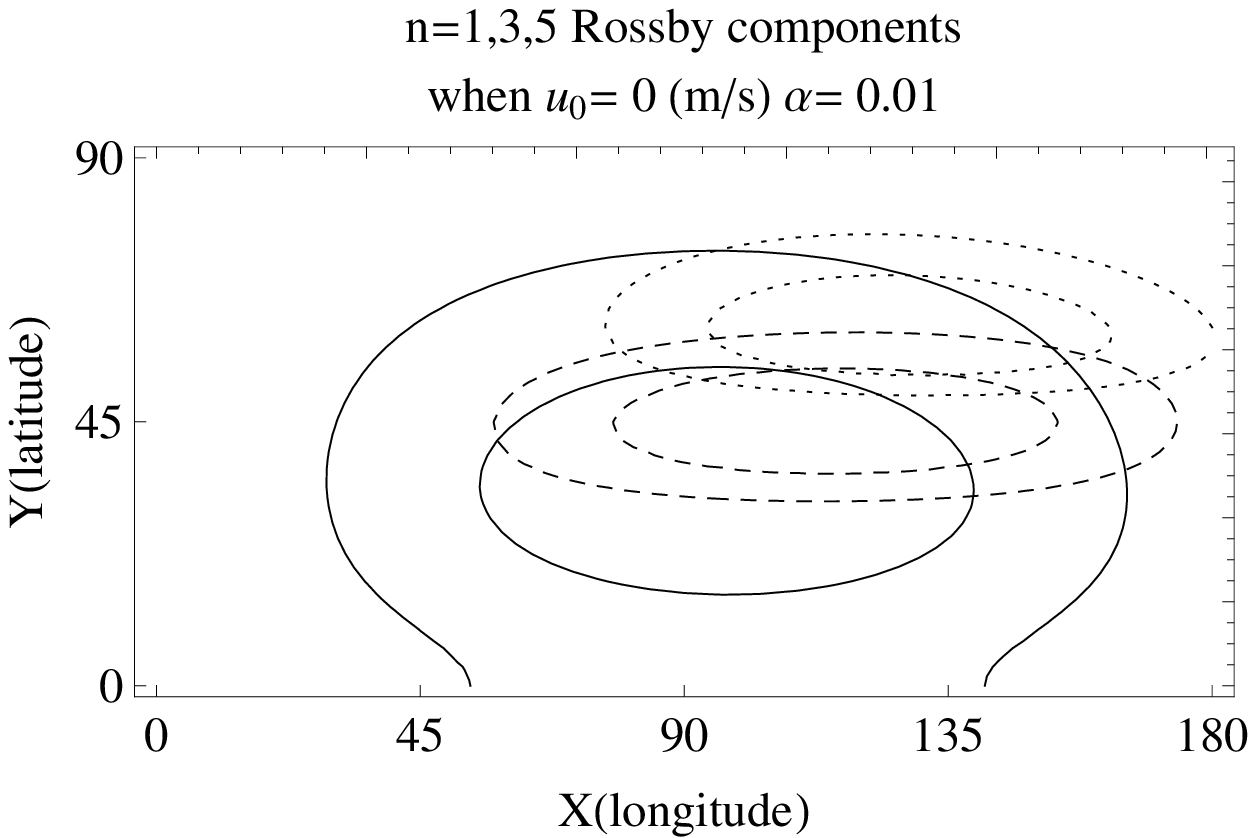}}
\end{center}
\caption{Rossby components at $p=0.03$ bar for $n=1$, 3, 5 in panels (a),(b),(c), respectively. The Rossby gyres for $n=1$, 3, 5 (solid, dashed, dotted contours) are illustrated in panel (d) to schematically explain the eastward tilt in the modest damping case (i.e. $\alpha$ = 0.01.)}
\label{n135 Rplot}
\end{figure}

Having shown that the solutions around the heating-center level from the three-dimensional linear equations are generally consistent with the shallow-water solutions, there actually exists a slight difference in the weak damping scenario. Unlike the gyres shown in the top right panel of Figure~\ref{u0 plot}, the gyres from three-dimensional solutions exhibit a minor eastward tilt in
the longitudinal direction, as can be seen from the shape of the gyres in the top left panel, guided by the green arrows which roughly point to the local flow directions.
In fact, the eastward tilt in the presence of vertical stratification is a result of the combination of Rossby waves summed over all vertical modes, which gives rise to higher orders of $n$, as discussed in Sections 2.1 \& 2.3. The phenomenon is illustrated in Figure~\ref{n135 Rplot}. Rossby waves of higher $n$ have wider latitudinal extension and slightly slower phase velocity \citep{matsuno}, such that they
have
smaller westward
shifts than the Rossby waves of lower $n$. Referring to Equation(\ref{eq amp}) again, with a small damping rate and no mean flow, we can see that Rossby modes with higher $n$ have
smaller westward
shifts.
The eastward tilt is not seen in the linear shallow-water model (e.g., top right panel of Figure~\ref{u0 plot} or Figure 3 in SP) since there are no $n>1$ waves in the shallow-water model forced by the Gaussian heating function with the same latitudinal extent as the equatorial Rossby deformation radius. For the same horizontal heating profile, vertical stratification introduces the eastward tilt at high latitudes.

\subsection{Including a zonal-mean zonal flow}
We have demonstrated that the atmospheric responses based on our 3D linear theory around the heating-center level in the absence of a zonal flow qualitatively agree with the 2D linear analysis in SP. Now we extend our linear analysis to study one of the key issues of our paper, namely, to examine the atmospheric responses in the presence of a uniform eastward mean zonal flow.

\subsubsection{Horizontal structures around the heating-center level: eastward shift of Rossby modes as the eastward zonal flow speed increases}
\label{sec: 2D z3}

\begin{figure}[ht]
\begin{center}
\begin{tabular}{cc}
\includegraphics[height= 4.5cm]{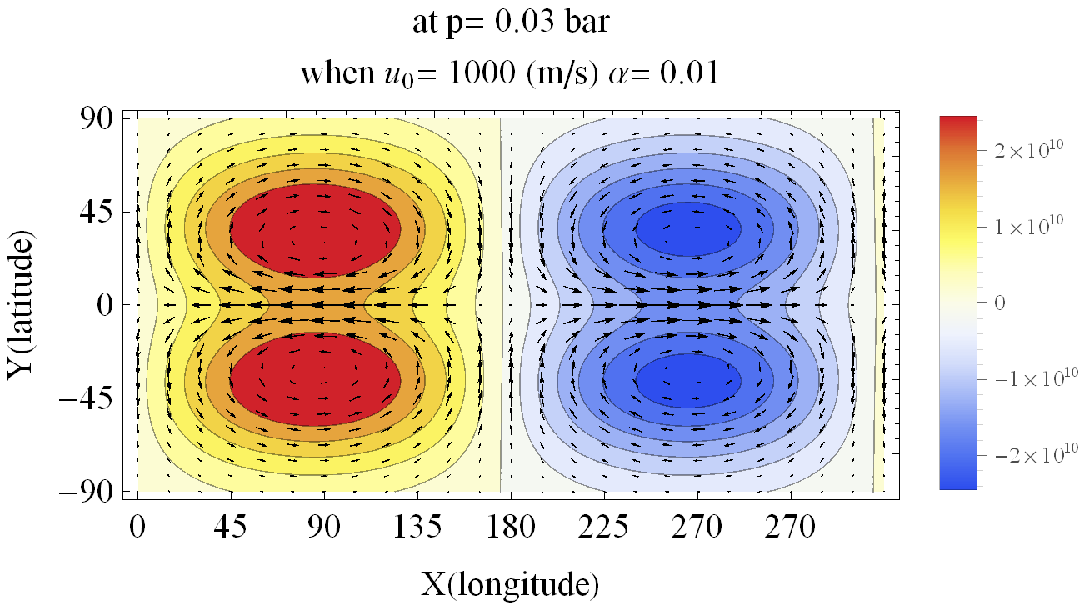}&
\includegraphics[height= 4cm, width= 8cm]{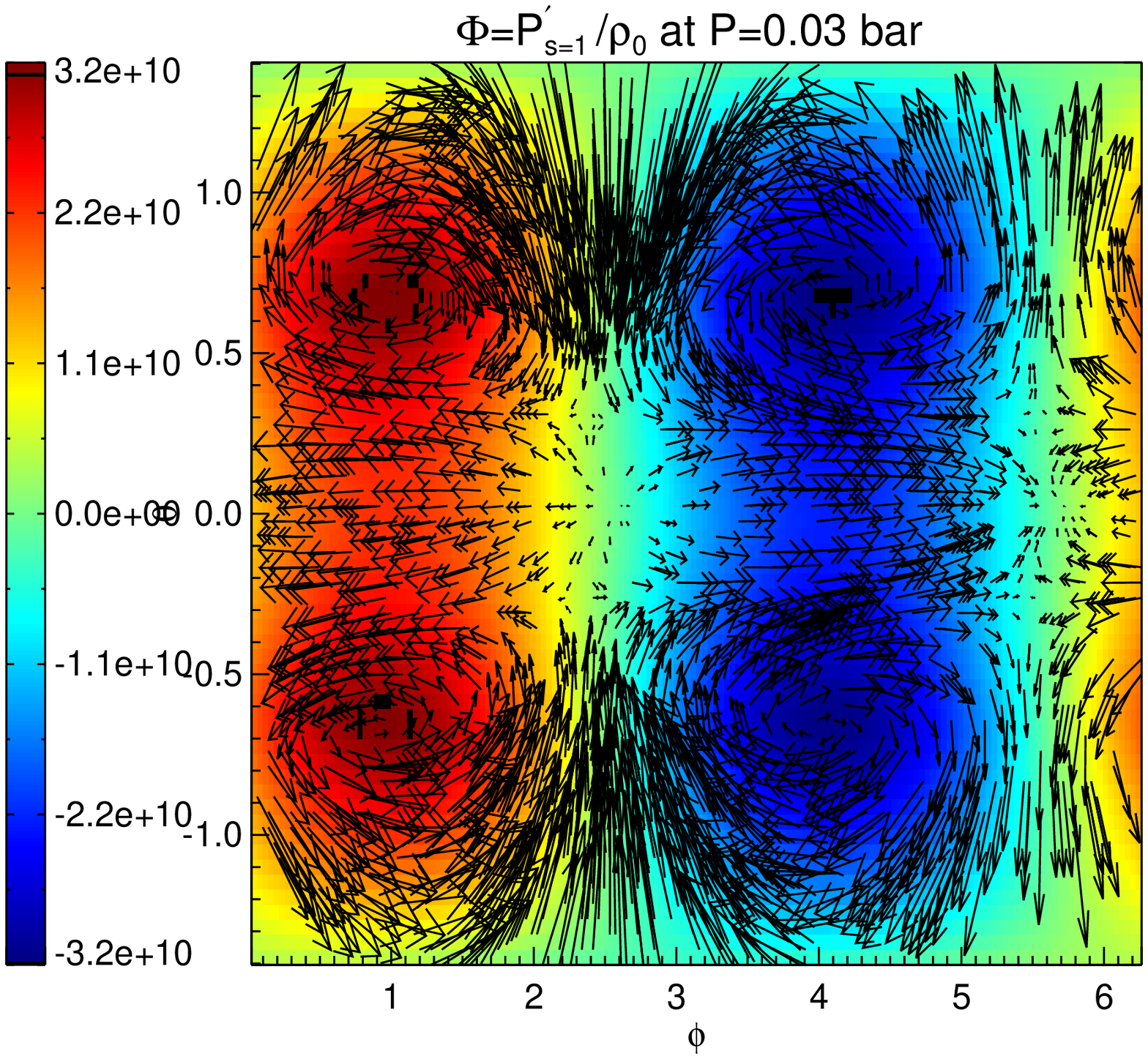} \\
\includegraphics[height= 4.5cm]{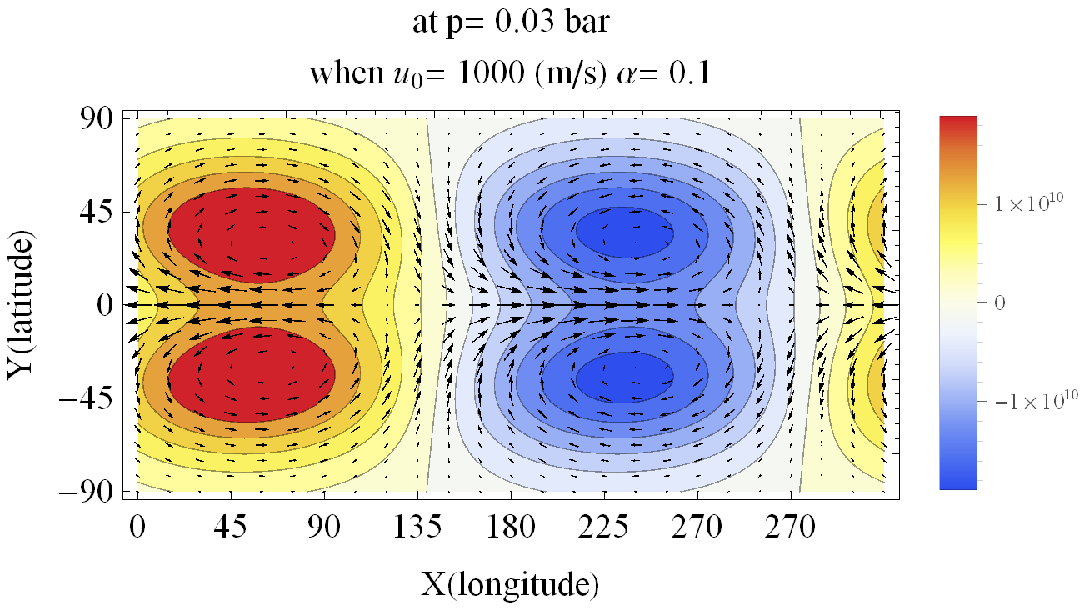}&
\includegraphics[height= 4cm, width= 8cm]{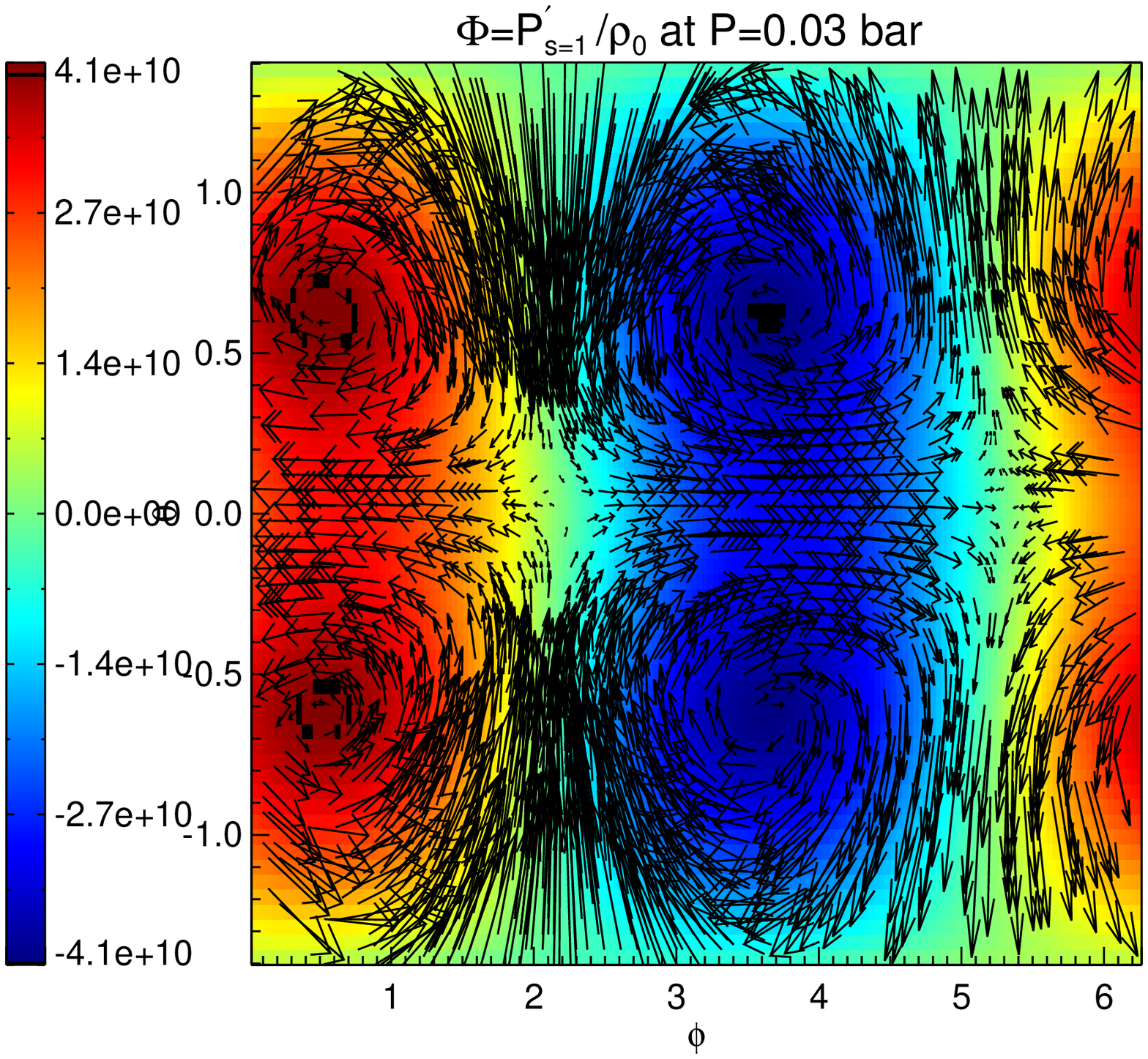}\\
\includegraphics[height= 4.5cm]{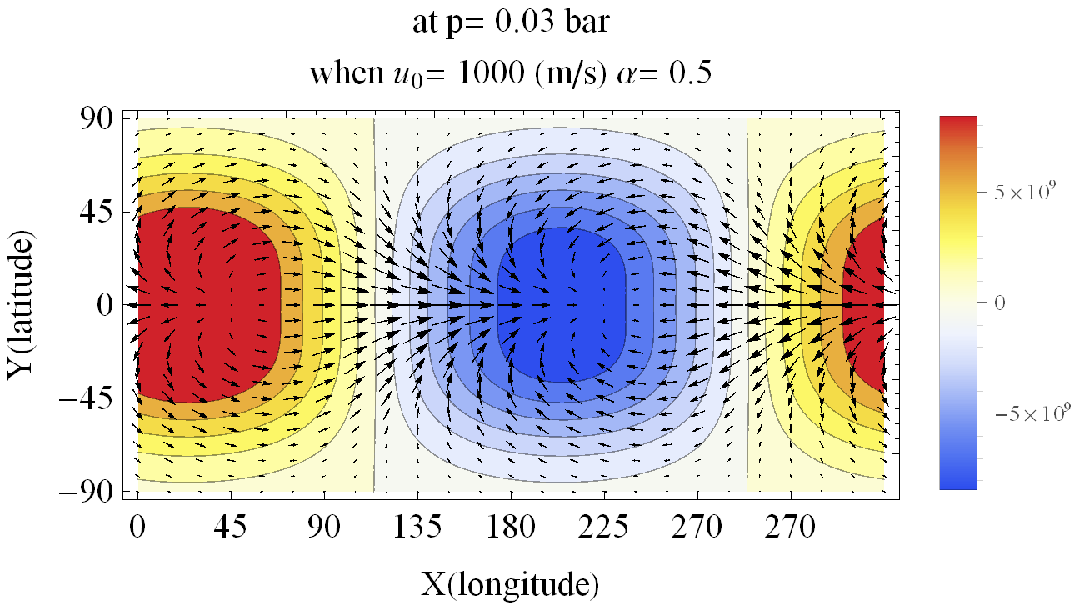}&
\includegraphics[height= 4cm, width= 8cm]{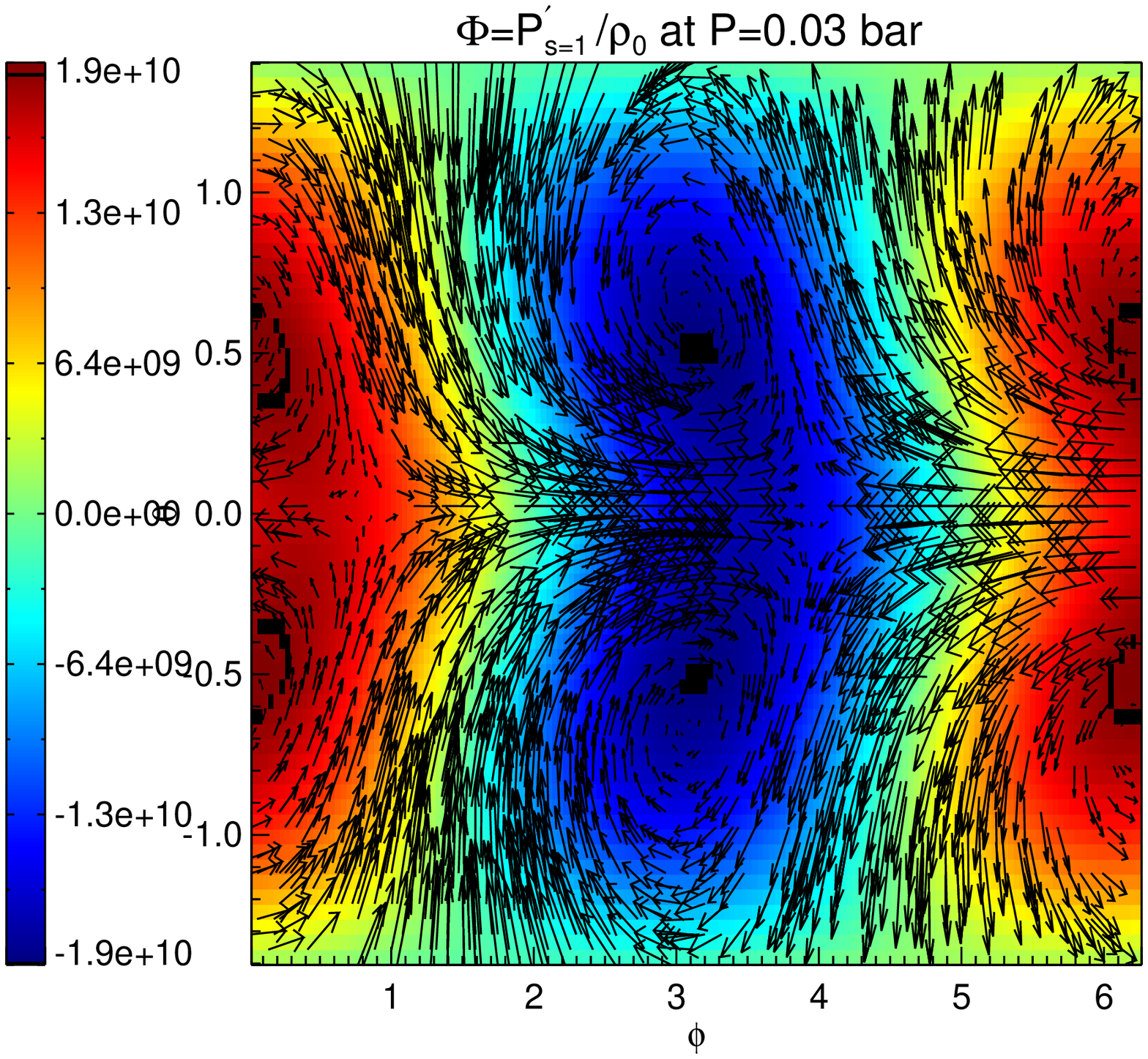}
\end{tabular}
\end{center}
\caption{Three panels on the left are the same as the three panels on the left of Fig.\ref{u0 plot} except for the presence of a uniform zonal-mean flow at the speed of $u_{0}$ = 1000 m/s. The three panels on the right are the corresponding results extracted from the simulation with the viscosity $\nu$ equal to $10^8$ (top), $10^9$ (middle), and $10^{10}$ (bottom) cm$^2$/s.}
\label{u1000 plot}
\end{figure}

Figure~\ref{u1000 plot} compares the linear solutions with the steady-state simulation results around the heating-center level ($p=0.03$ bar for both of the linear model and simulation). The figure shows the wave velocity fields superimposed on the color-coded geopotential contours in the cases of different damping rates. We adopt the typical zonal-mean flow speed 1000 m/s for $u_{0}$ in the linear model. The simulation results shown in the figure are Fourier transformed for the $s=1$ component.

Despite the fact that the nature of viscous dissipation is different from linear damping and that the equatorial super-rotating velocities are not the same in the numerical simulations with different viscosities, we present the simulation results for three different viscosities progressively differing also by a factor of 10 to qualitatively compare with the linear results for the three damping scenarios.
Figure~\ref{u1000 plot} shows that the wave velocity at higher latitudes are weaker in the linear theory than those in the numerical simulations. It is because the linear solutions with an {\it a prior} global uniform zonal flow on the equatorial $\beta$-plane decline with latitude due to the zero-velocity boundary condition \citep{matsuno,Holton 2004}, whereas in the atmospheric simulations in the spherical geometry, there actually exist flows across the polar regions and hence no global zonal flows are present at high latitudes.
With all the uncertainties and discrepancies, Figure~\ref{u1000 plot} nonetheless shows that the linear solutions resemble the overall structure around the heating-center level in the numerical simulation.

We further compare Figures~\ref{u0 plot} and \ref{u1000 plot} to investigate the changes of atmospheric responses after a eastward zonal flow is developed.
As shown in the figures for weak and modest damping scenarios, the solutions with a zonal flow exhibit cyclones and anticyclones like the case for $u_0=0$ but with almost opposite phases. In the weak damping scenario shown in the upper panels of Figure~\ref{u1000 plot}, since now $a_{m,n,l}$ is predominately determined by the factor $-1/\rmi( \omega_{m,n,l} - \omega)$ in response to the thermal forcing, a sufficiently large $\omega=u_0k$ (i.e., $\omega > \omega_{m,n,l}$ due to the large $u_0$) leads to a sign change and thus the opposite phase relative to the no-zonal-mean-flow case. In addition, the pressure variations increase at the equator compared to the no-zonal-mean-flow case in the weak damping scenario. The reason is that a weak frictional force only requires a small pressure gradient to balance out in the absence of Coriolis forces and a zonal flow at the equator.
When $u_{0}$ is nonzero, the pressure gradient consequently increases to maintain the force balance against advection flows.
In the modest damping scenario, the middle panels of Figure~\ref{u1000 plot} show that the structure not only exhibits a phase shift eastward to the structure for the non-zero-mean flow case (refer to Equation \ref{eq amp}), the Rossby component also exceeds the Kelvin component due to the Rossby resonance (see Figure~\ref{amp}). Therefore the chevron pattern is no longer present once the zonal flow is established. In the strong damping scenario, $\alpha$ is significantly large, and the presence of $u_0$ only causes a slightly eastward phase shift as shown in the bottom panels of Figure~\ref{u1000 plot} in comparison with the bottom panels of Figure~\ref{u0 plot}. In this strong damping regime, the wave velocity deviates significantly from the geostrophic configuration and generally flows from the day to night side, driven primarily by the pressure gradient with a slight influence of the Coriolis force.

Figure~\ref{2D K R zf} further presents the solutions with increasing speed of the zonal-mean flow from 0, 500, to
1000 m/s in the modest damping rate scenario.
We decompose the solutions (left column) in terms of Kelvin (middle column) and Rossby (right column) components
As indicated by the strength of wave geopotentials, waves in the case of $u_0=500$ m/s displayed in the middle row
have the strongest response due to the resonant effect.
When $u_{0}=0$, the geopotential maximums of Kelvin component and Rossby component have a phase difference by about 90 degrees, forming the chevron pattern (i.e. $\sin(\theta_{m,n,l}-\theta_{m,n,l'})\approx 1$ in Equation~\ref{eq:uv_flux}). However, the phase of Rossby component progressively shifts eastward with increasing $u_{0}$, while that of Kelvin component remains almost stationary. This is because the pattern speeds of the Kelvin component are generally too large to be affected by the change of $u_0$ smaller than 1000 km/s.
As a result, when the mean flow reaches a certain speed, say $u_{0}$=1000 m/s in this example, the maxima of Kelvin component and Rossby component have almost the same zonal phase. Thus, the chevron pattern progressively weakens as the speed of the zonal-mean flow increases, superseded by more symmetric Rossby gyres. This evolution can be found in Figure 3 of \citet{Arnold} as well. Therefore the structure provides weaker horizontally equatorward momentum transport as a result of the smaller zonally averaged flux $\overline{u'v'}$, in agreement with Equation(\ref{eq:uv_flux}) with $\sin(\theta_{m,n,l}-\theta_{m,n,l'})\approx 0$. The weakening of the zonal acceleration due to a change of the equatorward momentum transport after the superrotating flow is developed
was first mentioned in SP. The authors reported that the Kelvin-wave structure is shifted eastward and the midlatitude velocity structure differs significantly in the final equilibrated state by observing the morphology of the temperature and velocity fields in their simulations. Although it seems to appear in that way, our linear-mode analysis suggests that the same change of the atmospheric structure near the equator is in fact attributed to the eastward shift of the Rossby wave with almost no zonal phase shift of the Kelvin wave. The topic of momentum fluxes will be discussed in more detail in Section~\ref{sec: momentum flux}.

\subsubsection{Vertical structures on the $x$-$z$ plane: baroclinic equatorial waves and wavefront tilt}

In the subsection, we study the vertical structures of equatorial waves projected onto the $x$-$z$ plane along the equator, primarily focusing on the tilt of wavefronts (constant phase lines) of baroclinic equatorial waves, which do not appear in the shallow-water analysis.

Figure~\ref{XZ U1000 plots} displays the vertical wave structures on the $x$-$z$ plane corresponding to those in Figure~\ref{u1000 plot} at the equator. $\phi'$ superimposed with velocities $u'$ and $w'$ are plotted for 3 different damping scenarios in the linear model to compare to our simulation results with three viscosities. Once again,
$u_{0}=1000$ m/s is adopted in the linear model to represent the final equilibrium state in which the equatorial jet has fully developed.
Inspection of Figure~\ref{XZ U1000 plots} indicates that
the linear solutions qualitatively resemble the simulation results.
The nature of the equatorial waves may be identified from these plots. Despite the Coriolis effect, the vertical structures of baroclinic equatorial waves appear similar to those of internal gravity waves according to dispersion relations \citep[e.g.,][]{Holton92}. Since equatorial Kelvin modes have eastward horizontal phase velocities and Rossby modes have westward horizontal phase velocities, such properties along with those for internal gravity waves can determine the tilt of wavefronts in their vertical structures. As the baroclinic waves propagate vertically downwards from the level of excitation, the wave fronts, which are perpendicular to the wave vectors, exhibit an eastward or westward tilt depending on the directions of their horizontal phase velocities. As indicated by the $\phi'$ contours in Figure~\ref{XZ U1000 plots}, the wavefronts tilt westward with increasing depth from the heating level when $\alpha = 0.01$ and $0.1$ but tilt eastward with increasing depth from the heating level when $\alpha = 0.5$.

The difference in the vertical structures arises because the Rossby-wave response is stronger than the Kelvin-wave response in both the weak and modest damping scenarios but weaker in the strong damping scenario, as has been explained earlier in the paper. To assert the relation between the tilt of wavefronts and the types of waves, we decompose $\phi'$ into the Rossby and Kelvin components ($\phi'_K$ and $\phi'_R$, respectively) in the moderate and strong damping scenarios and plot them in Figure~\ref{XZ RK mode}. The same decomposition is performed for the velocity field. The figure confirms that the wave front of $\phi'_R$ below the heating level exhibits a downward-westward tilt and that of $\phi'_K$ exhibits an downward-eastward tilt. $\phi'_R$ is larger (smaller) than $\phi'_K$ in the modest (strong) damping scenario. Thus, the wave front of $\phi'$ ($=\phi'_R+\phi'_K$) below the heating level preferentially exhibits a downward-westward tilt in the modest damping scenario and a downward-eastward tilt in the strong damping case, as shown in Figure~\ref{XZ U1000 plots} for $\alpha=0.1$ and $\alpha=0.5$, respectively.

$u'$ and $w'$ also exhibit the same wavefront tilts as $\phi'$ but with different phase differences depending on the degree of dissipation, albeit difficult to recognize by arrows in Figure~\ref{XZ U1000 plots}. The horizontal velocity $u'$ around the heating-center level agrees with that shown in Figure~\ref{u1000 plot}.
The velocity field is almost
horizontal since the vertical velocity is relatively small due to strong vertical stratification and hence is hard to see in the figure. Nevertheless, upwelling (downwelling) flows primarily occur in the regions where
the horizontal westward (eastward) flows take place.
Therefore, the presence of small $w'$ leads to an overall upward-westward and downward-eastward velocity correlation, which results in downward transfer of momentum and will be discussed in Section \ref{sec:Distribution of momentum flux}.

\begin{figure}[ht]
\begin{center}
\begin{tabular}{ccc}
\includegraphics[height= 3.5cm, width=5cm]{f4d.eps} &
\includegraphics[height= 3.5cm, width= 5cm]{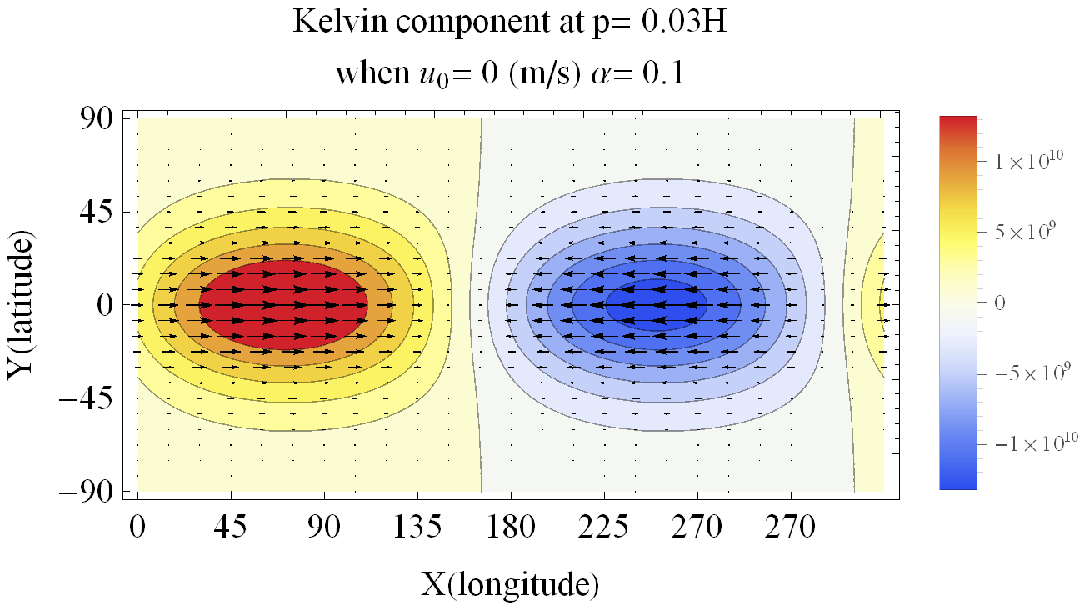} &
\includegraphics[height= 3.5cm, width= 5cm]{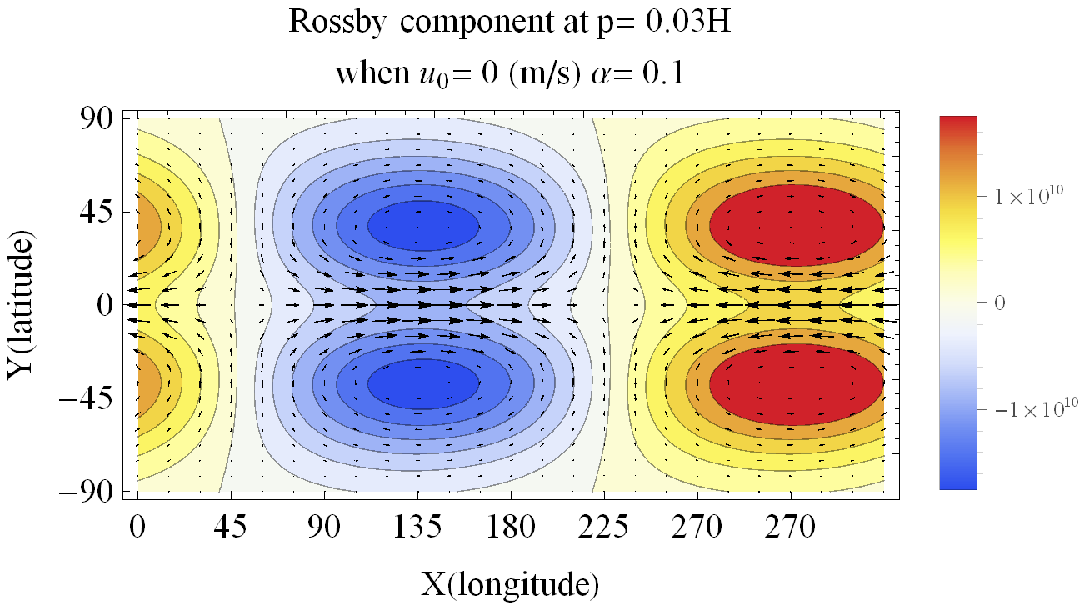} \\
\includegraphics[height= 3.5cm, width= 5cm]{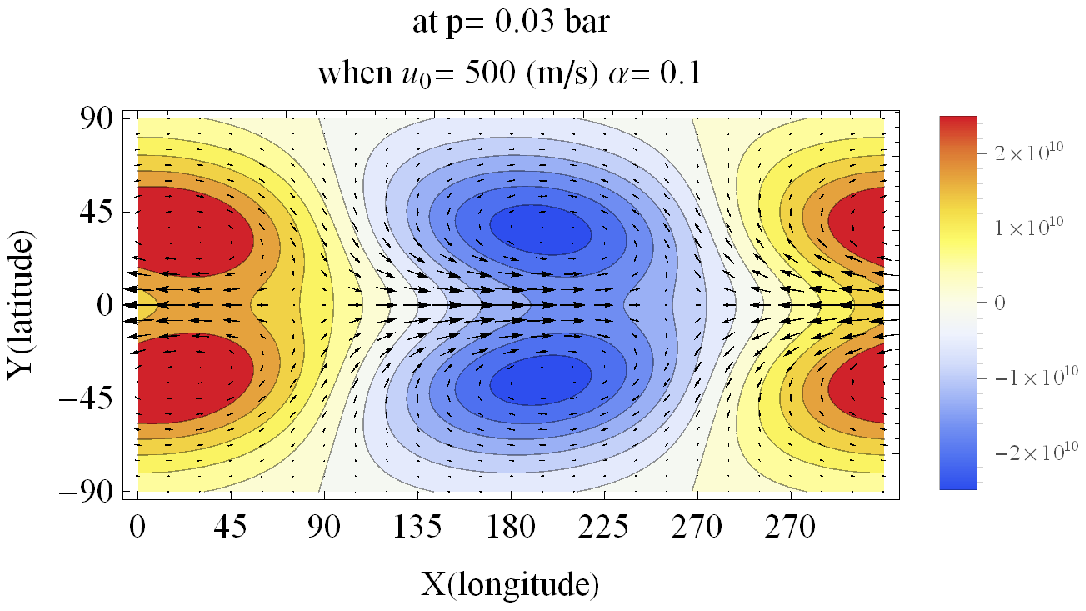} &
\includegraphics[height= 3.5cm, width= 5cm]{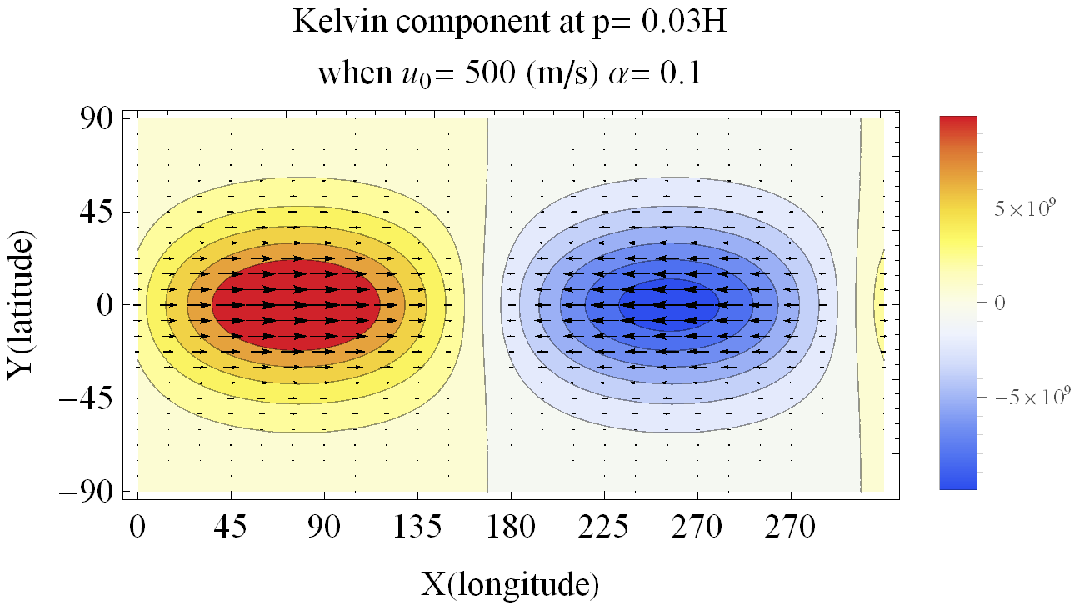} &
\includegraphics[height= 3.5cm, width= 5cm]{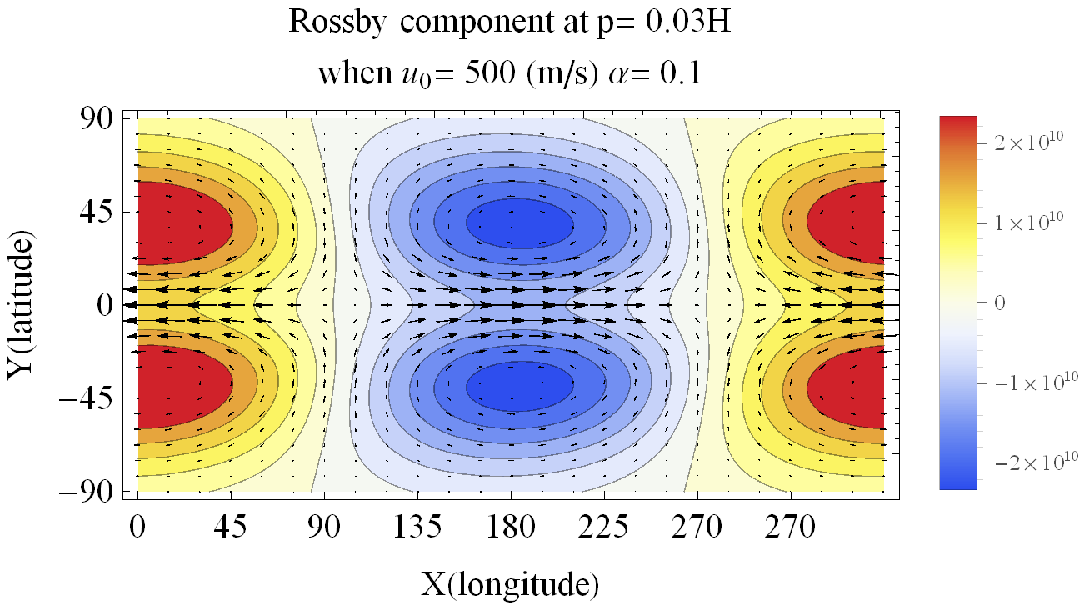} \\
\includegraphics[height= 3.5cm, width= 5cm]{f6c.eps} &
\includegraphics[height= 3.5cm, width= 5cm]{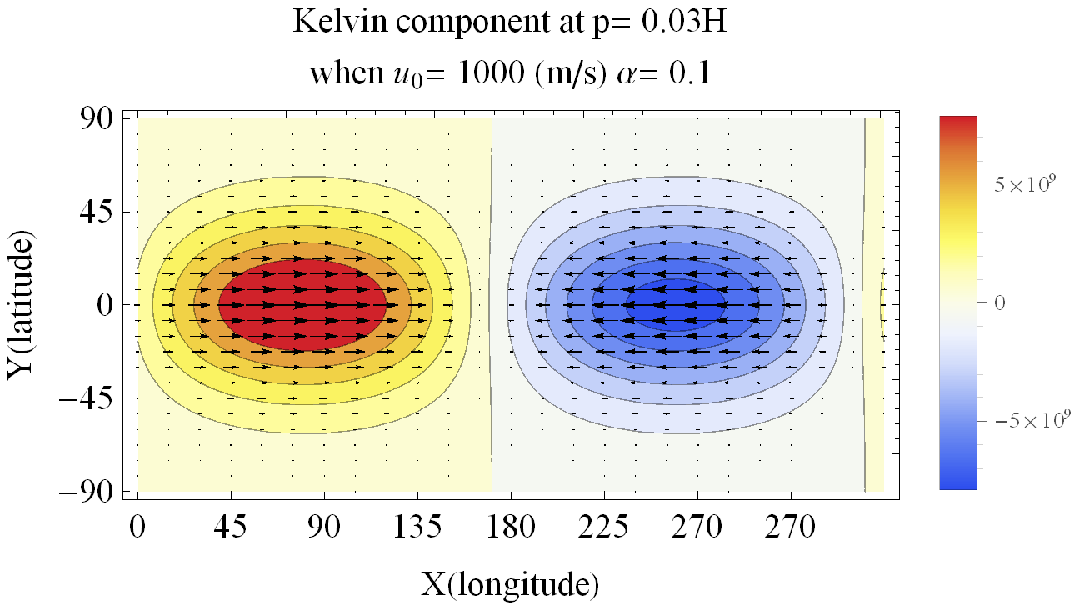} &
\includegraphics[height= 3.5cm, width= 5cm]{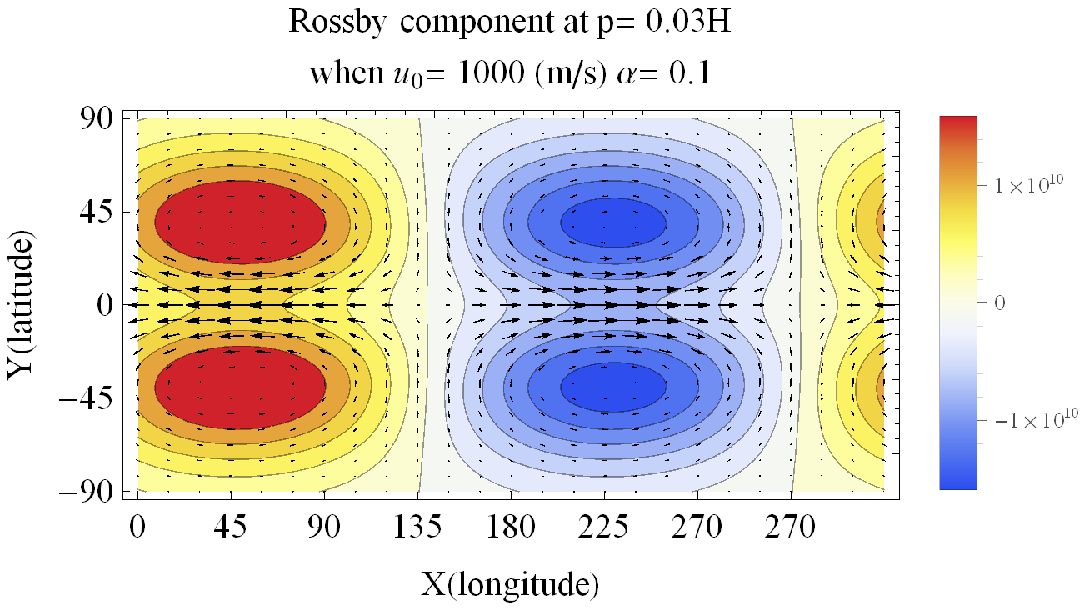}
\end{tabular}
\end{center}
\caption{Wave velocities (arrows) and wave geopotentials (contours) for three different values of $u_0$ in the modest damping scenario ($\alpha=0.1$). The wave structures are decomposed into the Kelvin and Rossby components, which are shown in the second and third columns, respectively.}
\label{2D K R zf}
\end{figure}

\begin{figure}[ht]
\begin{center}
\begin{tabular}{cc}
\includegraphics[height= 4.5cm]{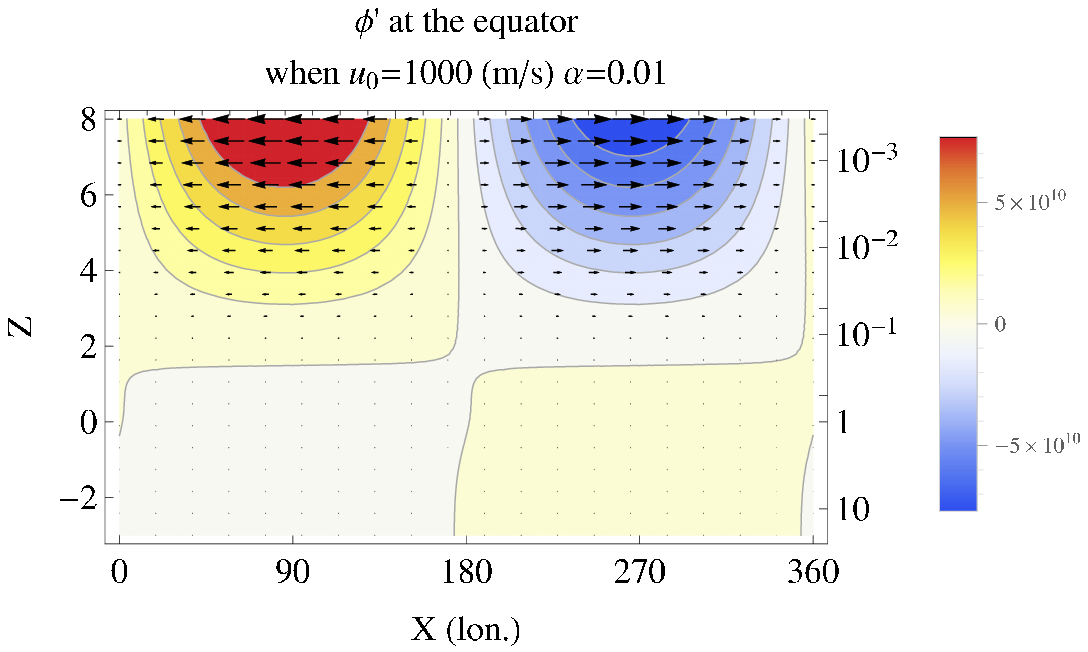} &
\includegraphics[height= 4cm, width= 8cm]{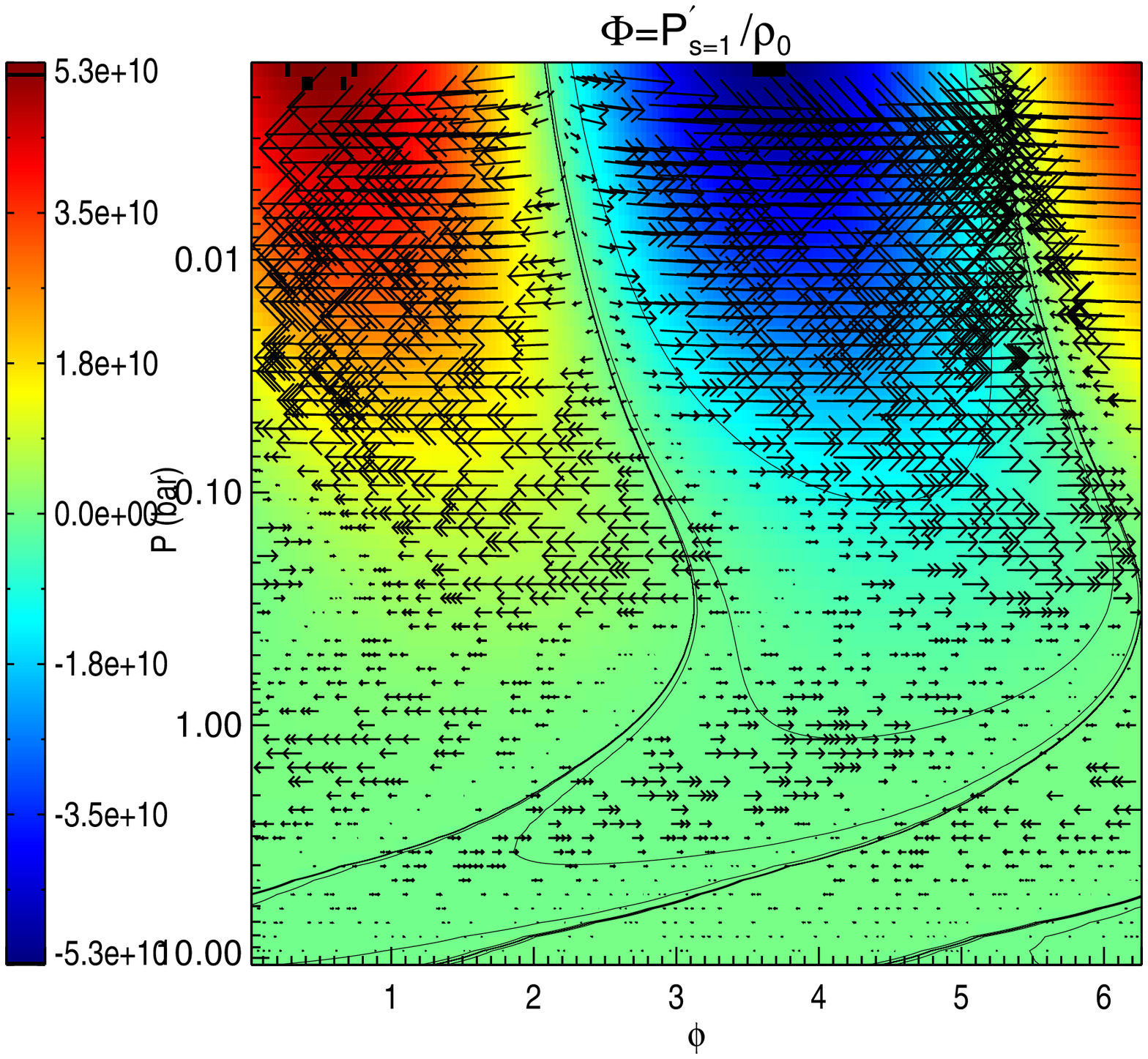} \\
\includegraphics[height= 4.5cm]{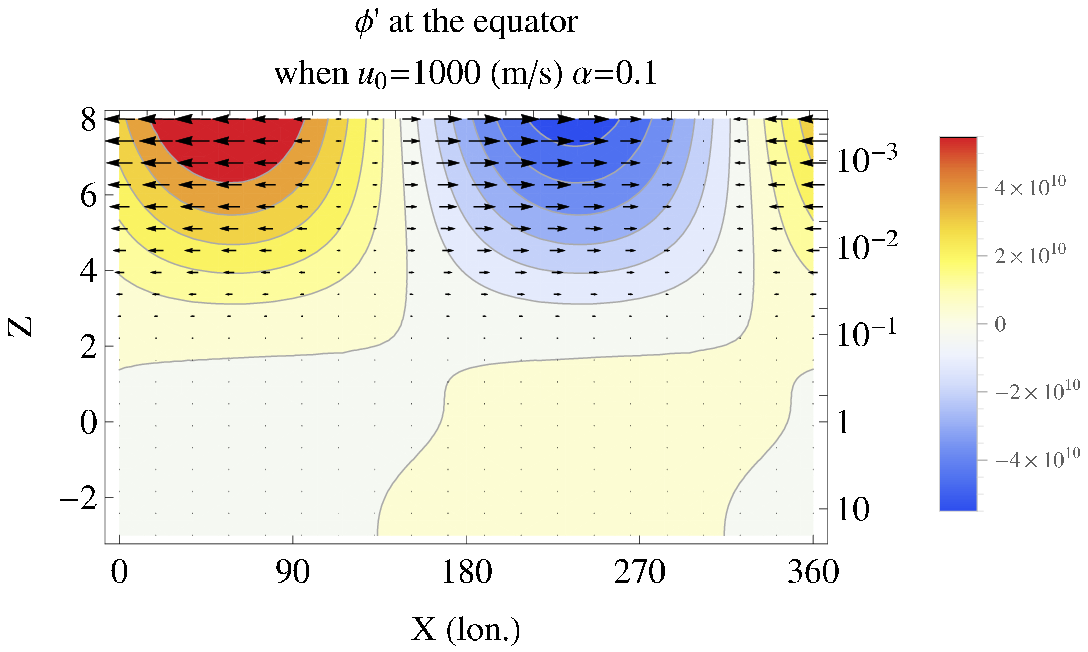} &
\includegraphics[height= 4cm, width= 8cm]{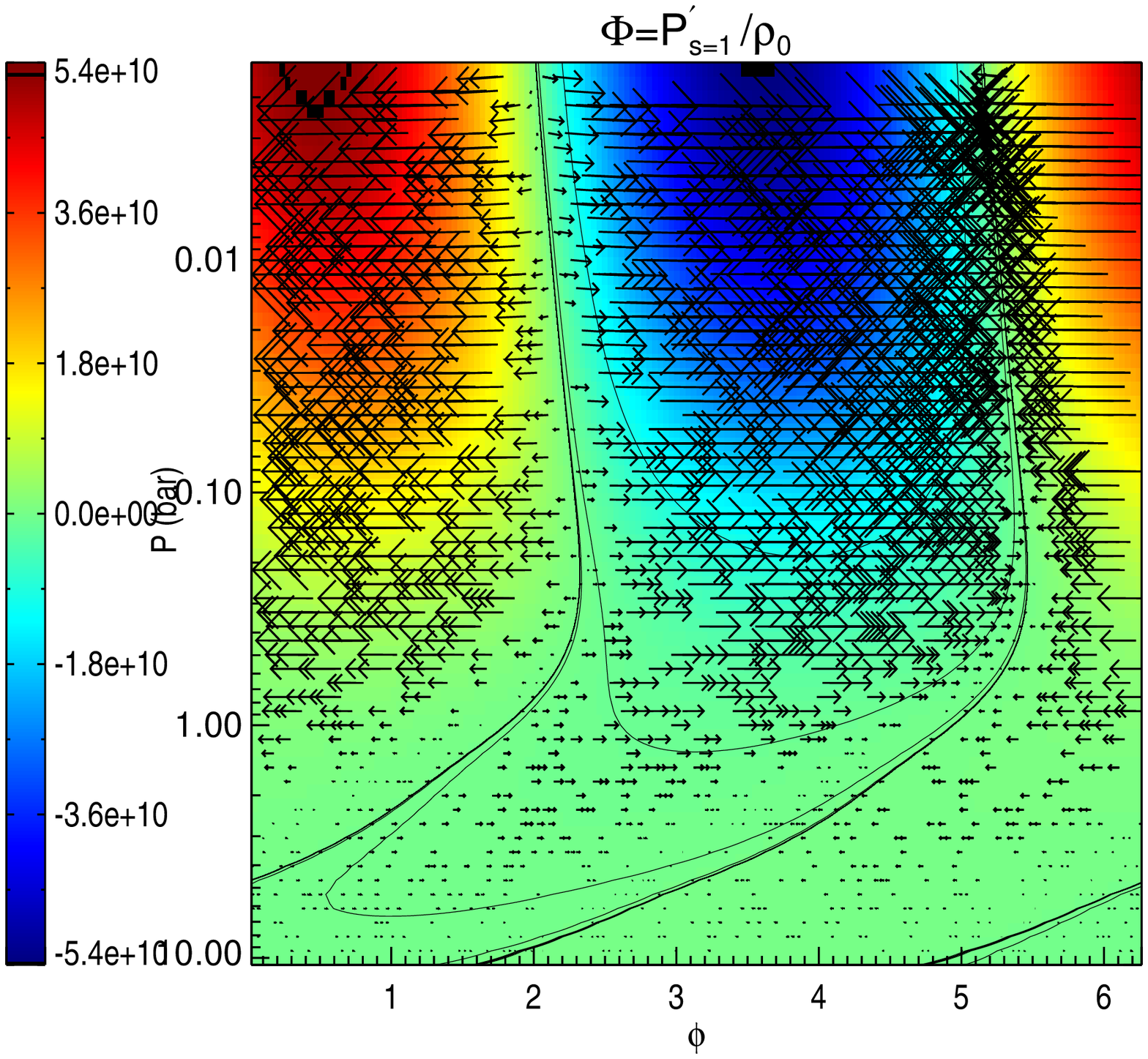} \\
\includegraphics[height= 4.5cm]{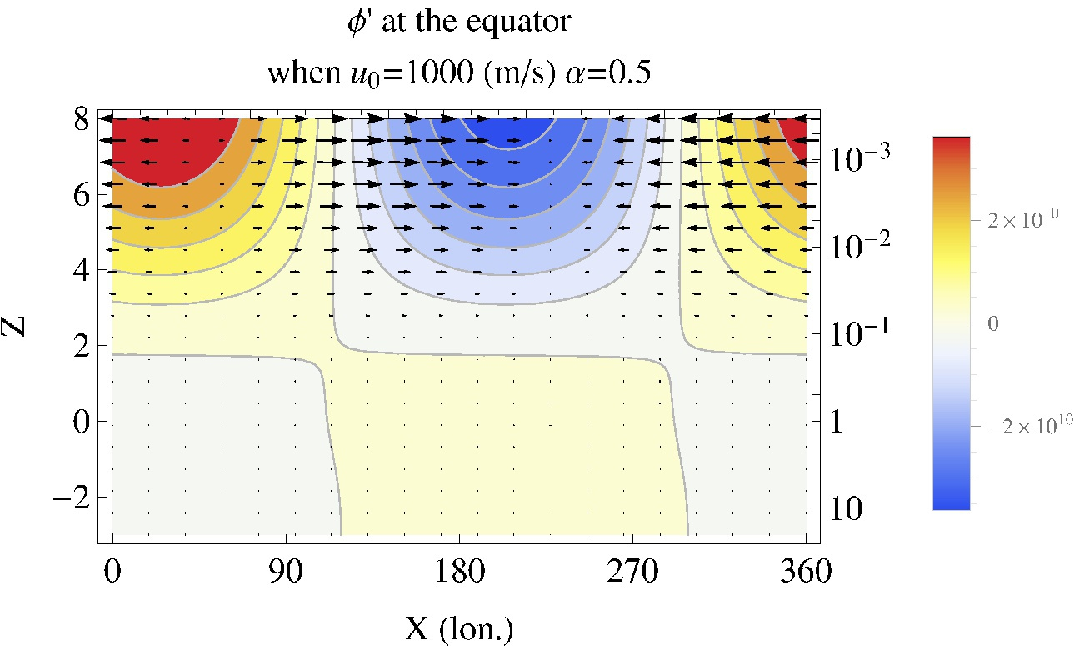} &
\includegraphics[height= 4cm, width= 8cm]{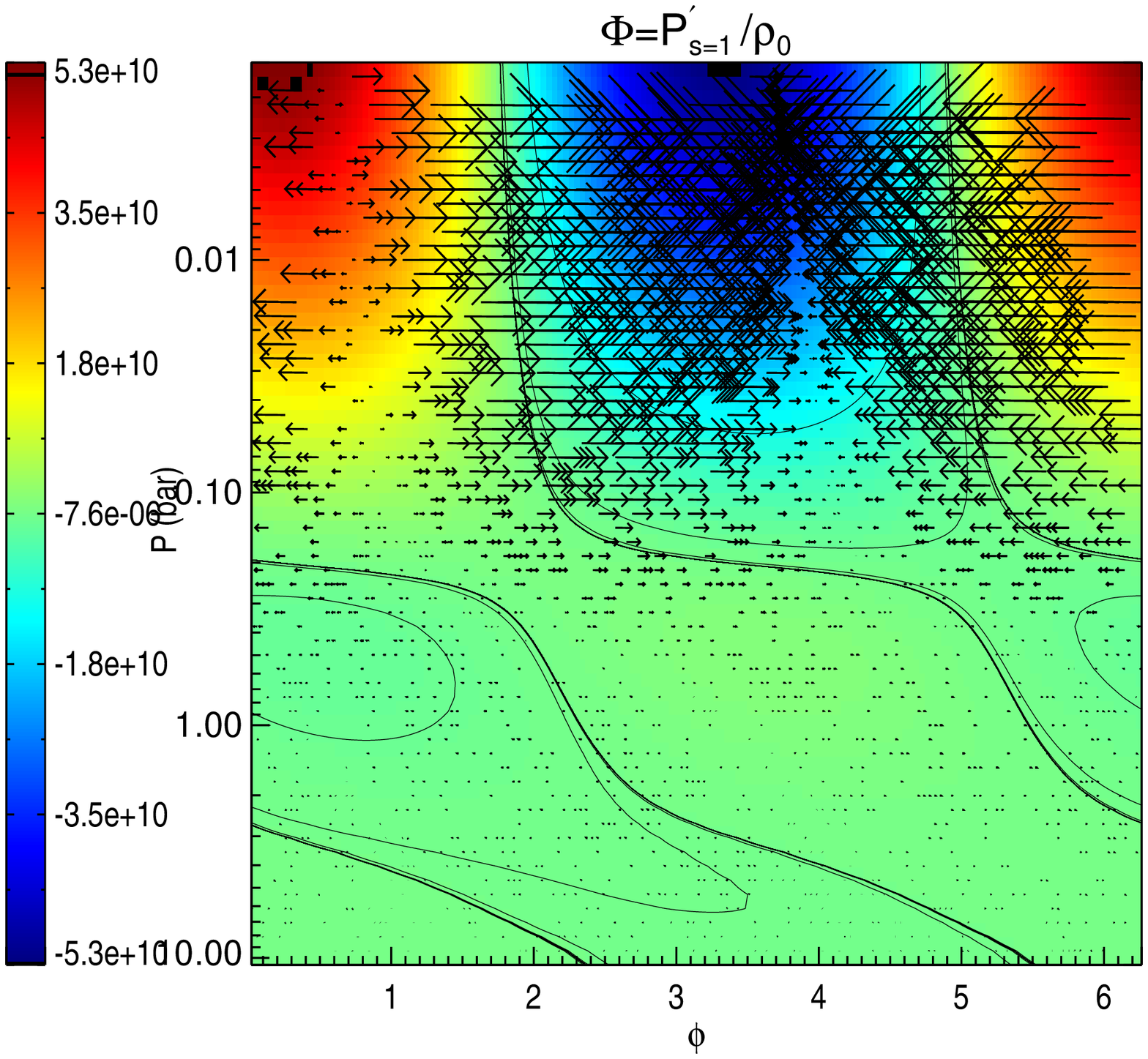}
\end{tabular}
\end{center}
\caption{Vertical structures of waves corresponding to Figure~\ref{u1000 plot} along the equator (i.e. $y=0$).
The linear solutions for $\alpha=$ 0.01, 0.1, and 0.5 are displayed on the left to compare with the simulation results for $\nu=10^8$, $10^9$, and $10^{10}$ cm$^2$/s on the right from top to bottom panels. The color contours present $\phi'$, overlaid with the velocity fields ($u'$, $w'$) shown by arrows.}
\label{XZ U1000 plots}
\end{figure}

\begin{figure}
\begin{center}
\includegraphics[height= 5cm]{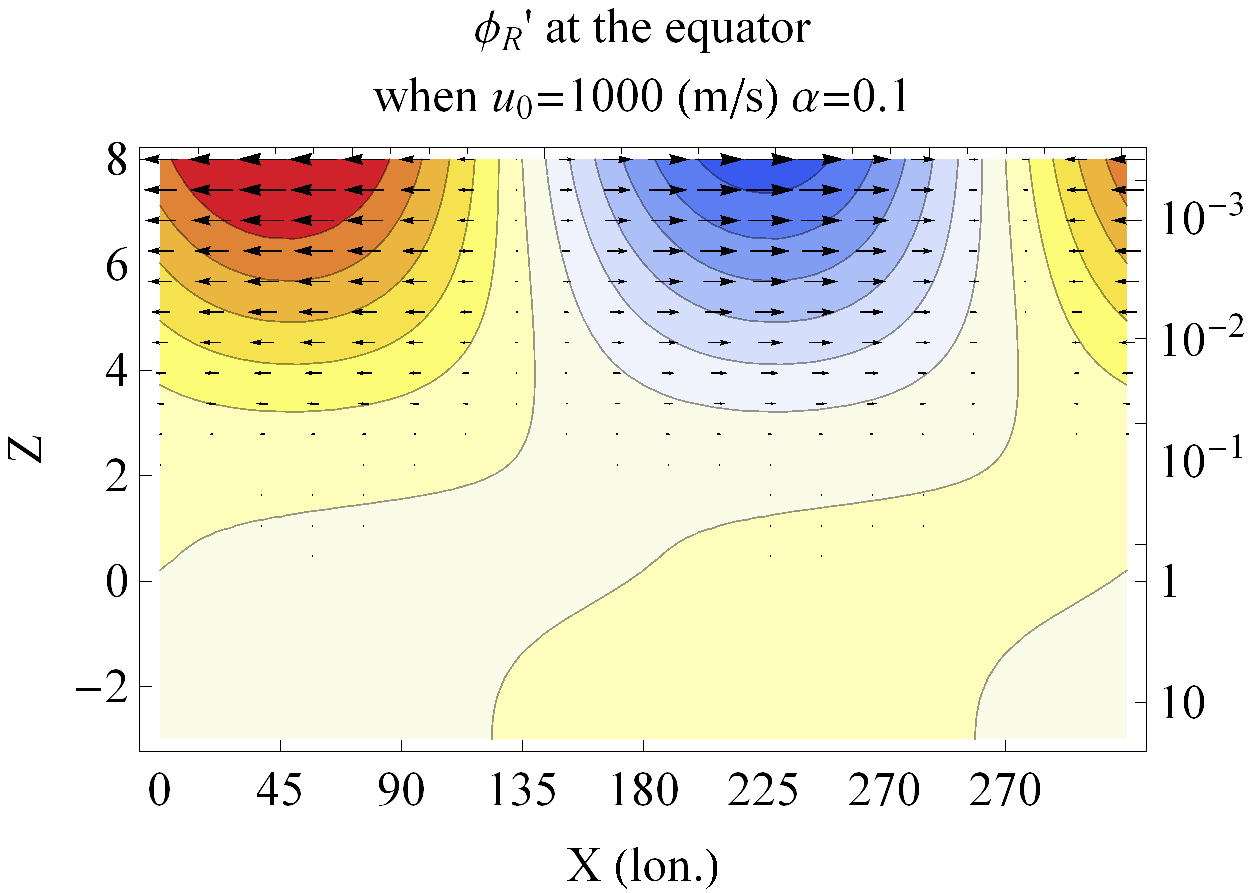}
\includegraphics[height= 5cm]{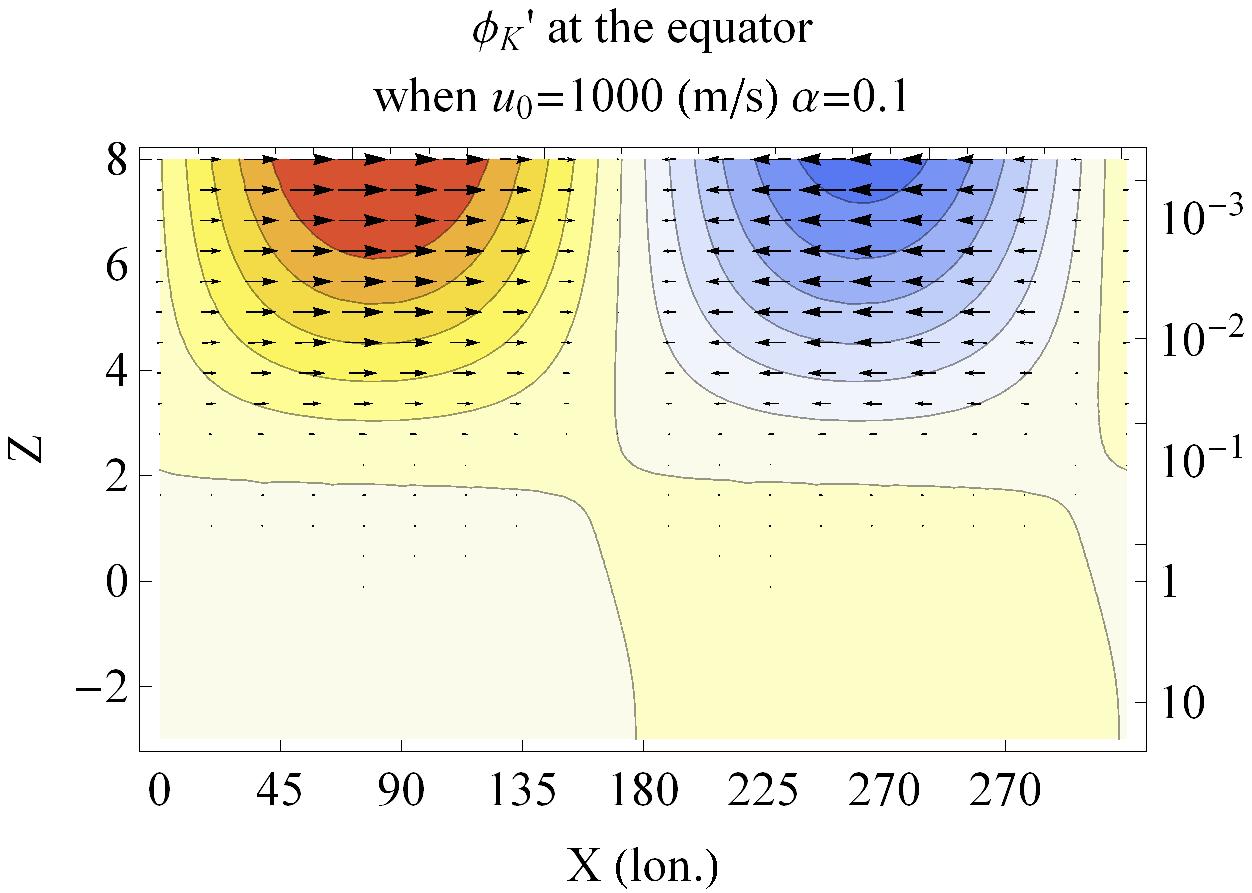}
\raisebox{0.35\height}{\includegraphics[height= 2.75cm, width=
1.05cm]{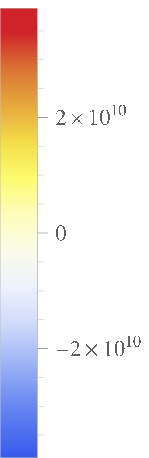}}
\includegraphics[height= 5cm]{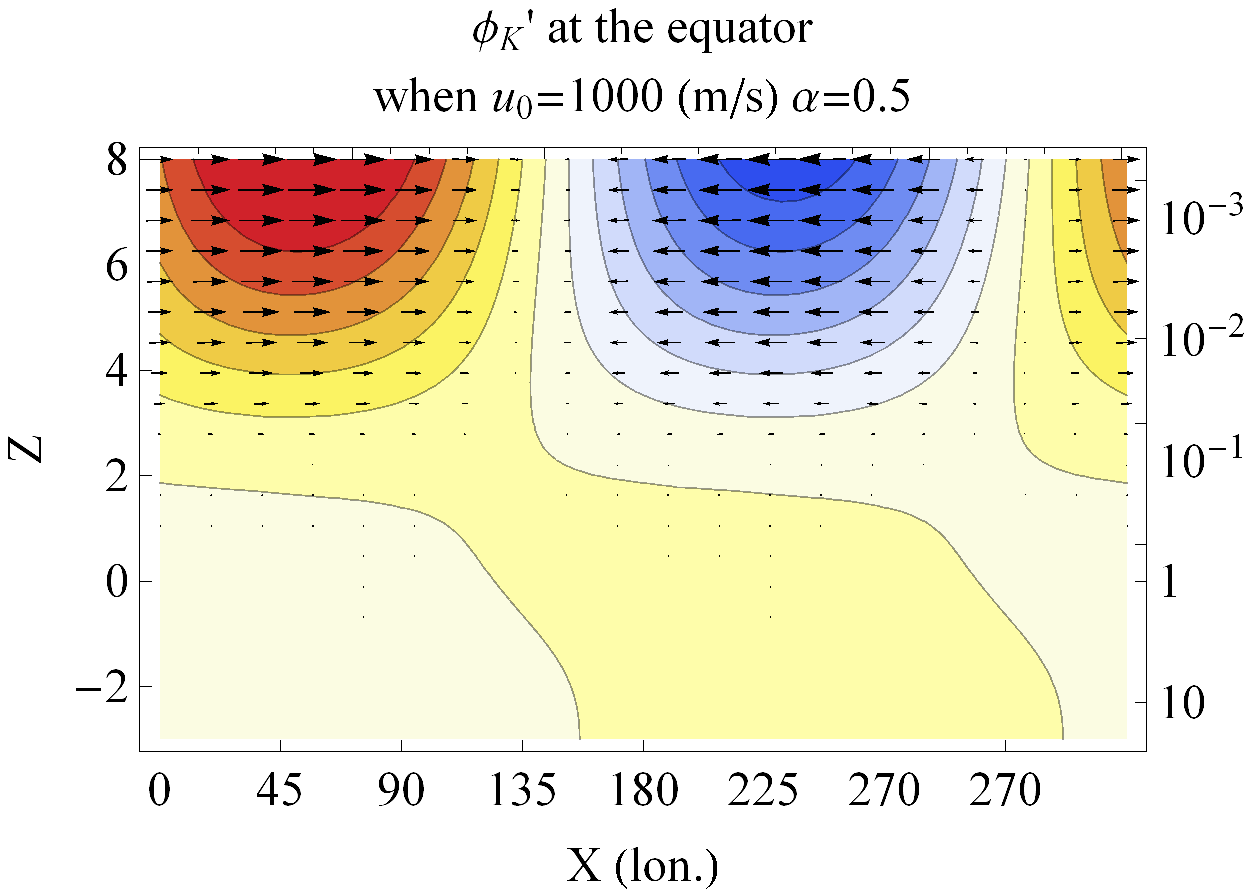}
\includegraphics[height= 5cm]{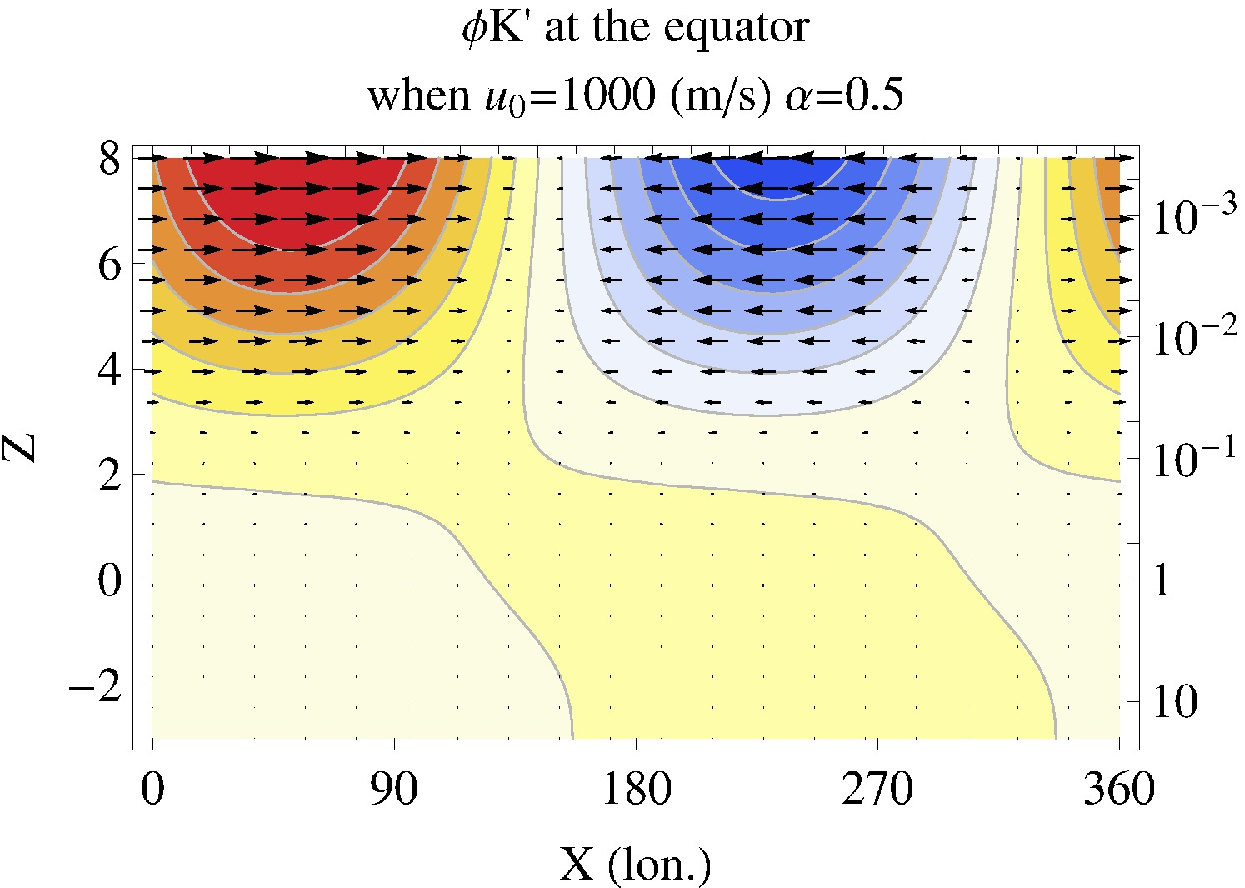}
\raisebox{0.35\height}{\includegraphics[height= 2.75cm, width=
1.05cm]{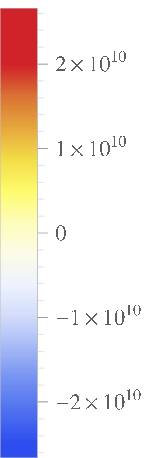}}
\end{center}
\caption{Rossby (left) and Kelvin (right) components of the wave geopotential and velocity for the modest $\alpha=0.1$ and strong $\alpha=0.5$ damping cases shown in Figure~\ref{XZ U1000 plots}. The geopotential contours are color-coded in the same scale.}
\label{XZ RK mode}
\end{figure}

\begin{figure}[ht]
\begin{center}
\begin{tabular}{cc}
\includegraphics[height= 5cm]{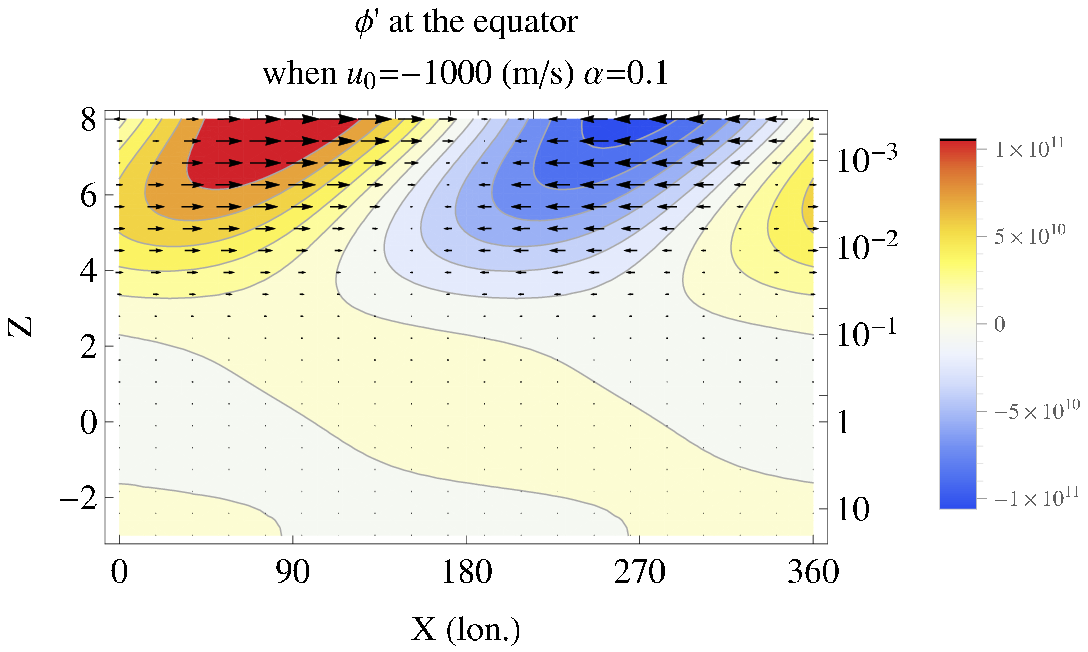} &
\includegraphics[height= 5cm]{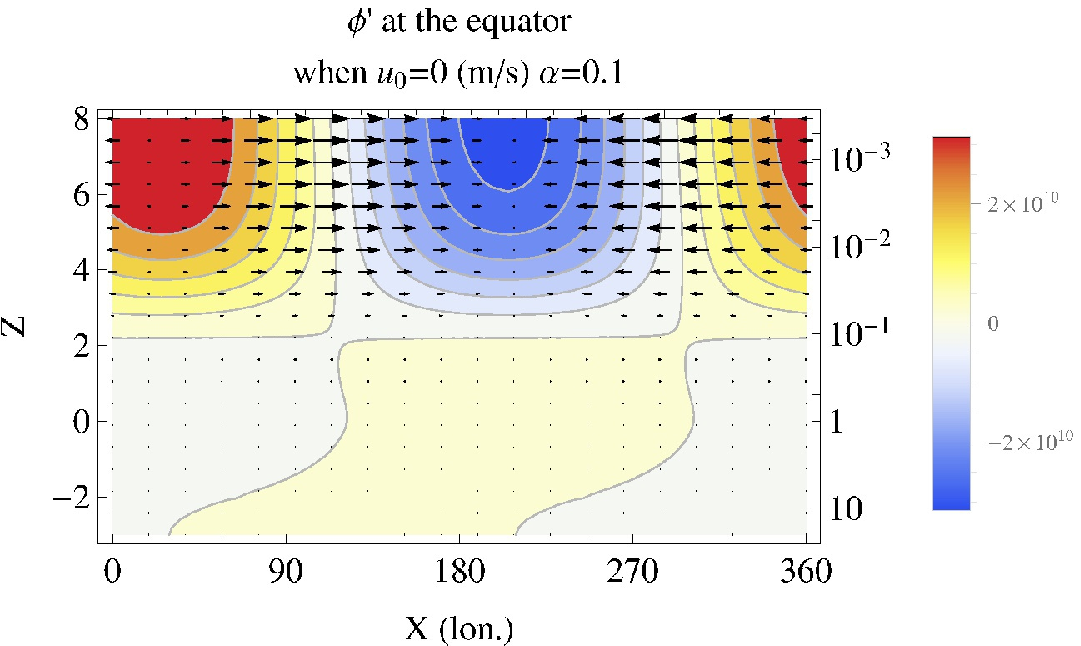} \\
\includegraphics[height= 5cm]{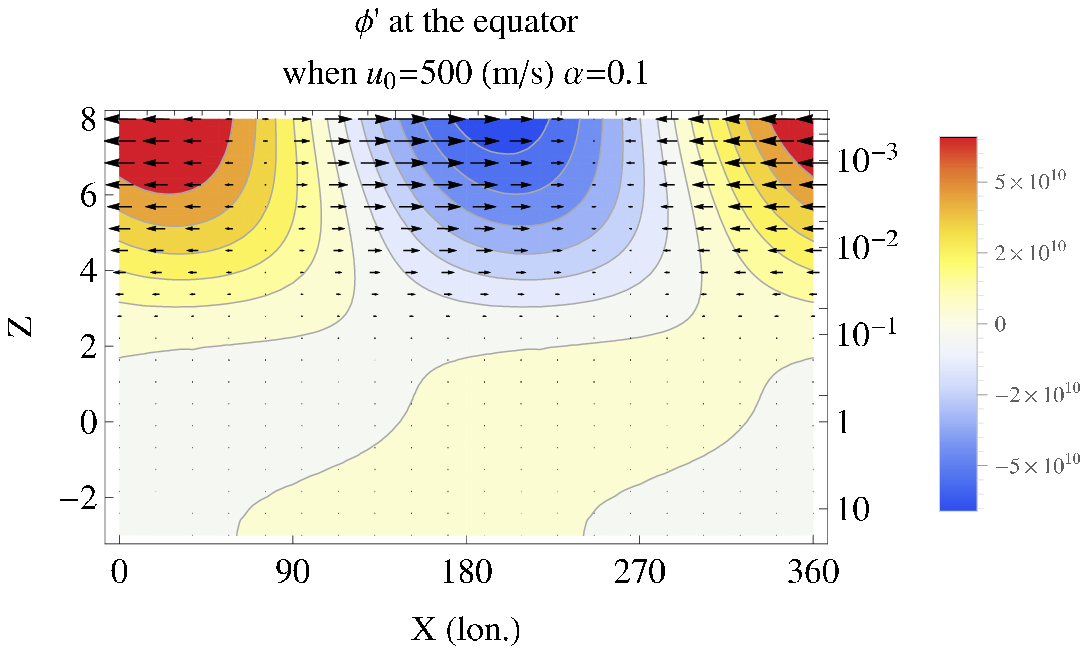} &
\includegraphics[height= 5cm]{f8c.eps}
\end{tabular}
\end{center}
\caption{Vertical structures of the wave geopotential and velocity along the equator when the zonal mean flow increases from $-1000$, 0, 500, to 1000 m/s.}
\label{XZ u u0 plots}
\end{figure}

We further present the linear results in Figure~\ref{XZ u u0 plots} for the vertical structures of waves projected on the vertical plane at equator for $u_0=-1000$, 0, 500, 1000 m/s in the modest damping scenario, as have been done for the horizontal structures of waves shown in Figure~\ref{2D K R zf}. Figure~\ref{2D K R zf} shows that when $u_0=0$, the horizontal chevron pattern leads to more symmetric horizontal flows from the day to night sides near the equator. As $u_0$ increases and therefore the Rossby-wave component shifts eastward, the day-to-night horizontal flows in the equatorial region become less symmetric. This trend of the flow changing with $u_0$ is also seen in vertical wave structures plotted in Figure~\ref{XZ u u0 plots}. As indicated from the color-coded geopotential contours, the wave response to the stellar heating for $u_0=500$ m/s is more pronounced than in other cases, and so is the magnitude of the wavefront tilt from the vertical in the heating region ($p\lesssim 0.1$ bar),
which results from the Rossby-wave resonance. The resonance amplifies the wave response so that Rossby waves can propagate farther before being dissipated \citep[cf.][]{Gill80}. This may also be roughly identified by the WKBJ dispersion relation for internal gravity waves with the vertical wavenumber $k_z$,
\begin{equation}
\frac{(\omega- u_0 k)^2}{k^2} \approx \frac{N^2}{k^2_z},
\end{equation}
where the damping rate is omitted for a simple illustration.
When a resonance occurs, $\omega- u_0 k$ is small, leading to large
$k_z$ while $k$ and $N$ are fixed. Thus the larger $k_z$ brings the
wavefront tilt more from the vertical in the heating region. We also show in the upper left panel that the largest wavefront tilt for $u_0<0$ occurs near the Kelvin-wave resonance (i.e. $u_0 \approx -1000$ km/s) in the heating region, in support of the relation between the wavefront tilt angle $\arctan (k_z/k)$ and $u_0$.

\begin{figure}
\begin{tabular}{c}
\includegraphics[height= 6cm]{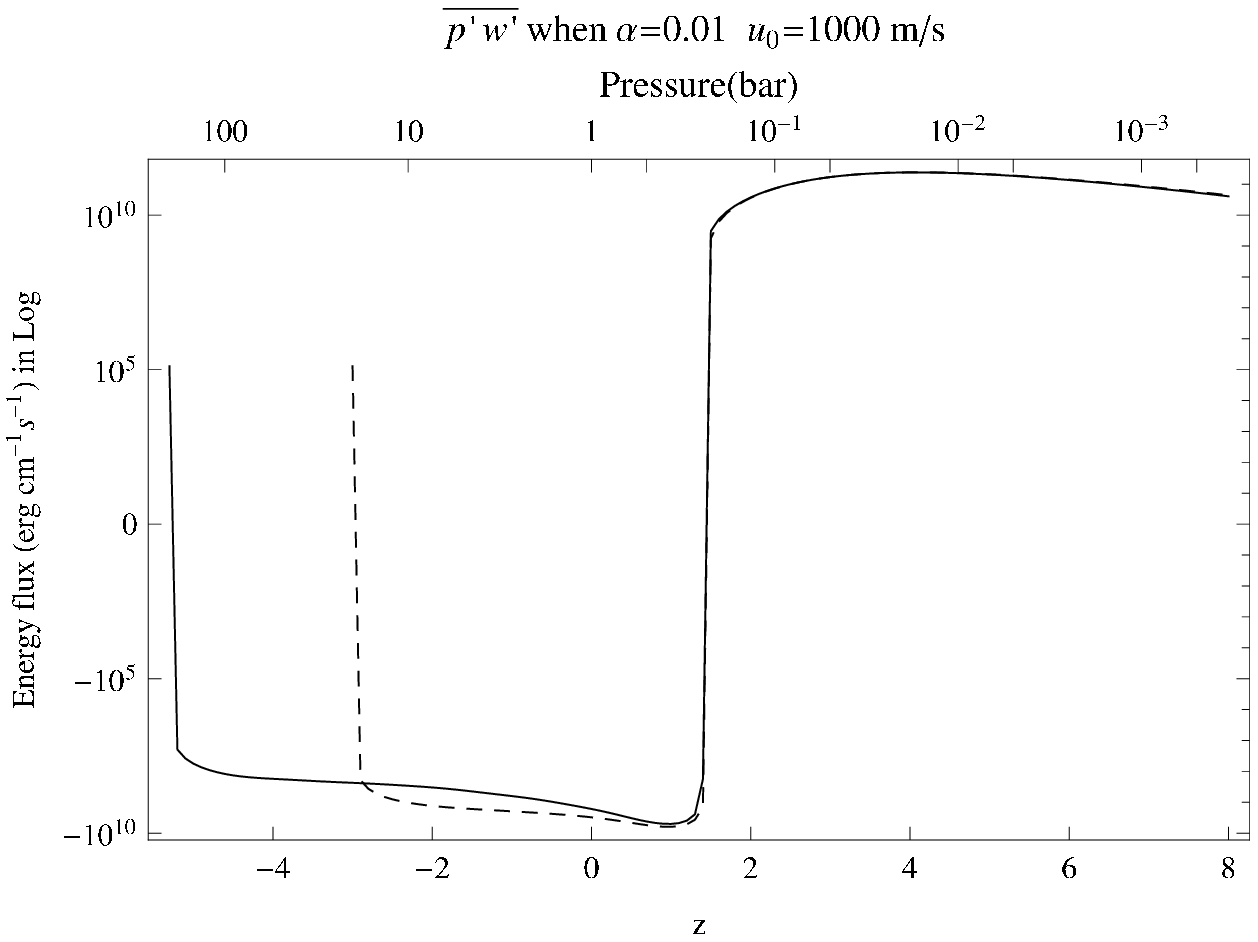} \\
\includegraphics[height= 6cm]{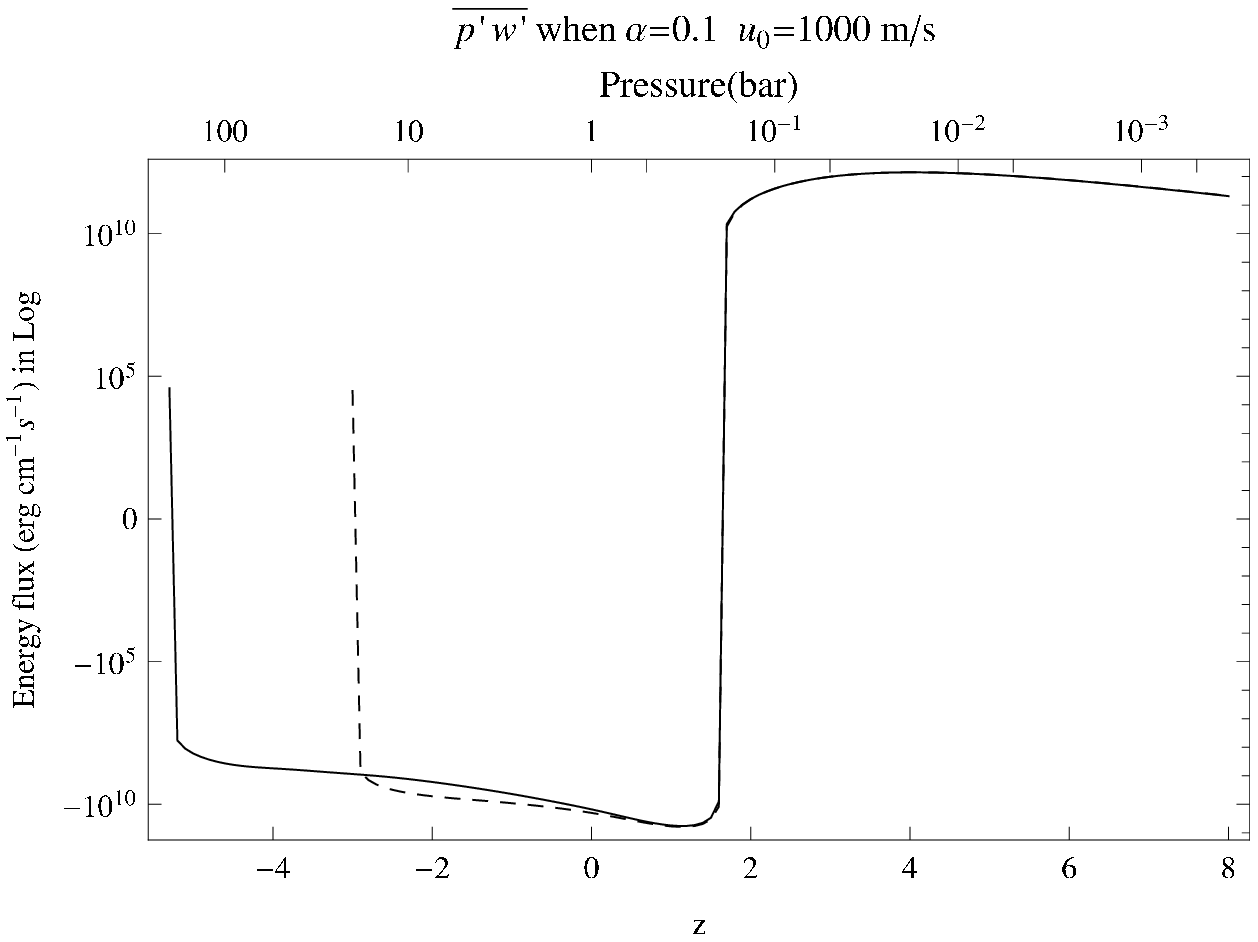} \\
\includegraphics[height= 6cm]{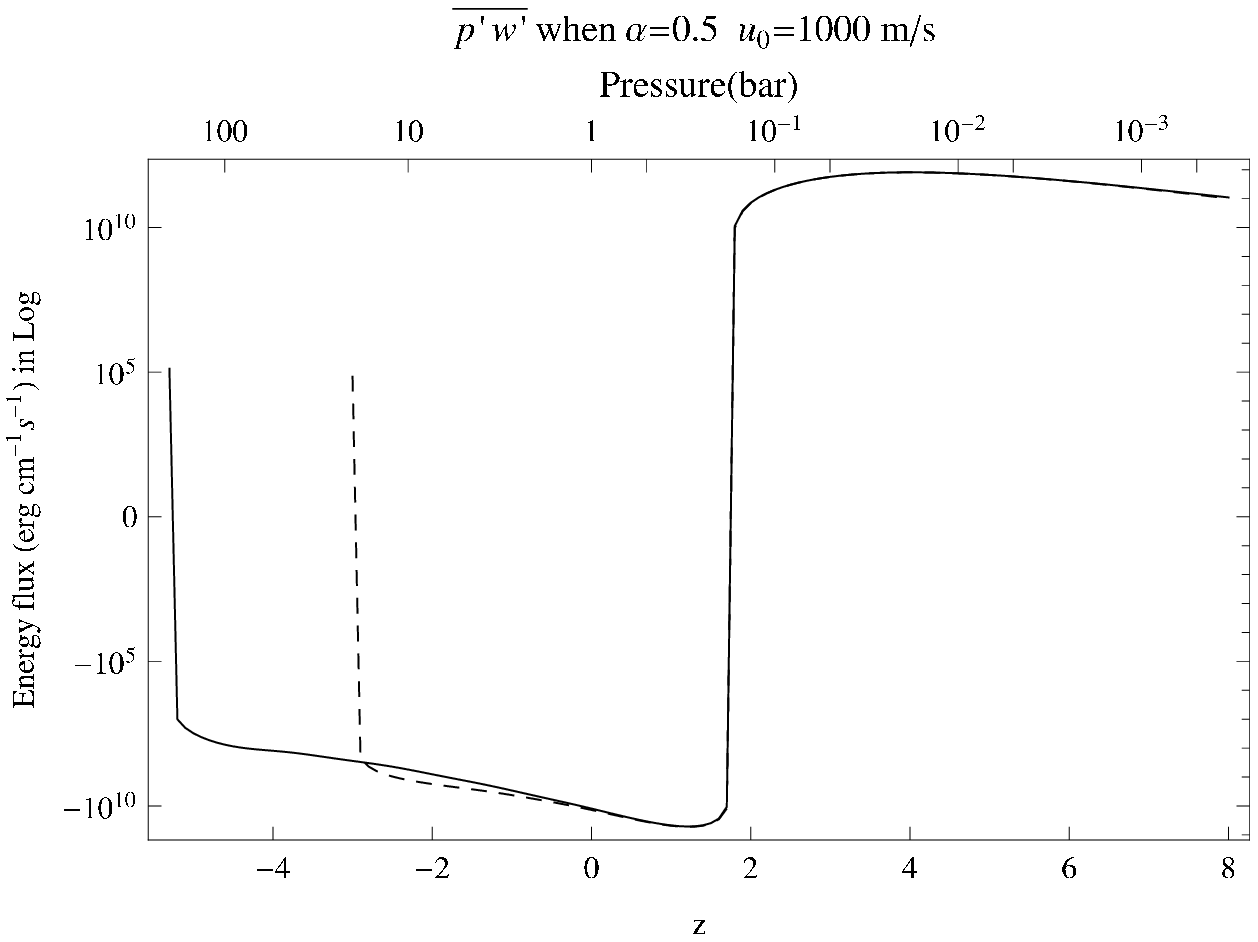}
\end{tabular}
\caption{Vertical wave energy flux $\overline{p'w'}$, integrated over all latitudes, as a function of $z$. The results for the bottom boundary placed at 20 bars (dashed line) and at 200 bars (solid line) are compared in the three damping scenarios.}
\label{fig:energy_flux}
\end{figure}

Because the bottom boundary of the linear model reflects waves and its location at $p=20$ bars eliminates the modes with longer vertical wavelengths, we examine the issue by extending the location of the bottom boundary down to 200 bars, while keeping the background states the same for a simple test. We find that the results in the original vertical domain (i.e. $p \geq 20$ bars) do not change significantly.
This arises because the waves are substantially dissipated beneath the heating region. Equation(\ref{E flux eq}) indicates that the effect of wave dissipation with the altitude in the absence of thermal forcing can be examined by inspecting the profile of the vertical wave energy flux $\overline{p'w'}$.\footnote{We investigate how the wave energy flux changes vertically, i.e $\partial_{z} \overline{p'w'}$, which is just the term $\frac{\partial}{\partial z}\overline{\phi ' w'} - \frac{\overline{\phi ' w'}}{H}$ in Equation(\ref{E flux eq}).} We integrate the wave energy fluxes over all latitudes for the three damping scenarios and plot them as a function of $z$ in Figure~\ref{fig:energy_flux}. As expected, the energy fluxes derived based on two different locations of the bottom boundary agree better for stronger damping (i.e., lager $\alpha$). Nonetheless, the flux profiles approximately coincide in each damping scenario except for the region near the bottom boundaries where the vertical flux profiles decline quickly to meet the free-slip boundary condition. Overall, the vertical wave energy flux decreases by 1-2 orders of magnitude from the altitude just below the heating region ($p\gtrsim 1$ bar) to $p\lesssim 20$ bars, inferring that with the bottom boundary at 20 bars, the amplitude of the reflected waves is small enough to not affect the wave solutions significantly. Specifically,
the vertical group velocity for the Rossby modes
is given by
\begin{equation}
\frac{\partial \omega_{Rossby}}{\partial m}
\approx - \omega_{Rossby}^{2} \frac{m(2n+1)}{ N k \sqrt{m^2+1/4H^2}}.
\end{equation}
For $n=1$ Rossby waves, the vertical group velocity is about 0.05-1 m/s. Thus, the estimated vertical wave travel time crossing
the vertical domain from the heating-center level to the bottom boundary
is roughly $10^{6}$ sec, which is comparable to the small damping time scale, and longer than the moderate and strong damping time scales.

\subsubsection{Horizontal structure below and above the heating level: dispersive phenomenon induced by baroclinic modes}
\begin{figure}[ht]
\begin{center}
\begin{tabular}{cc}
\includegraphics[height= 4cm]{f6c.eps} &
\includegraphics[height= 3.3cm, width= 7cm]{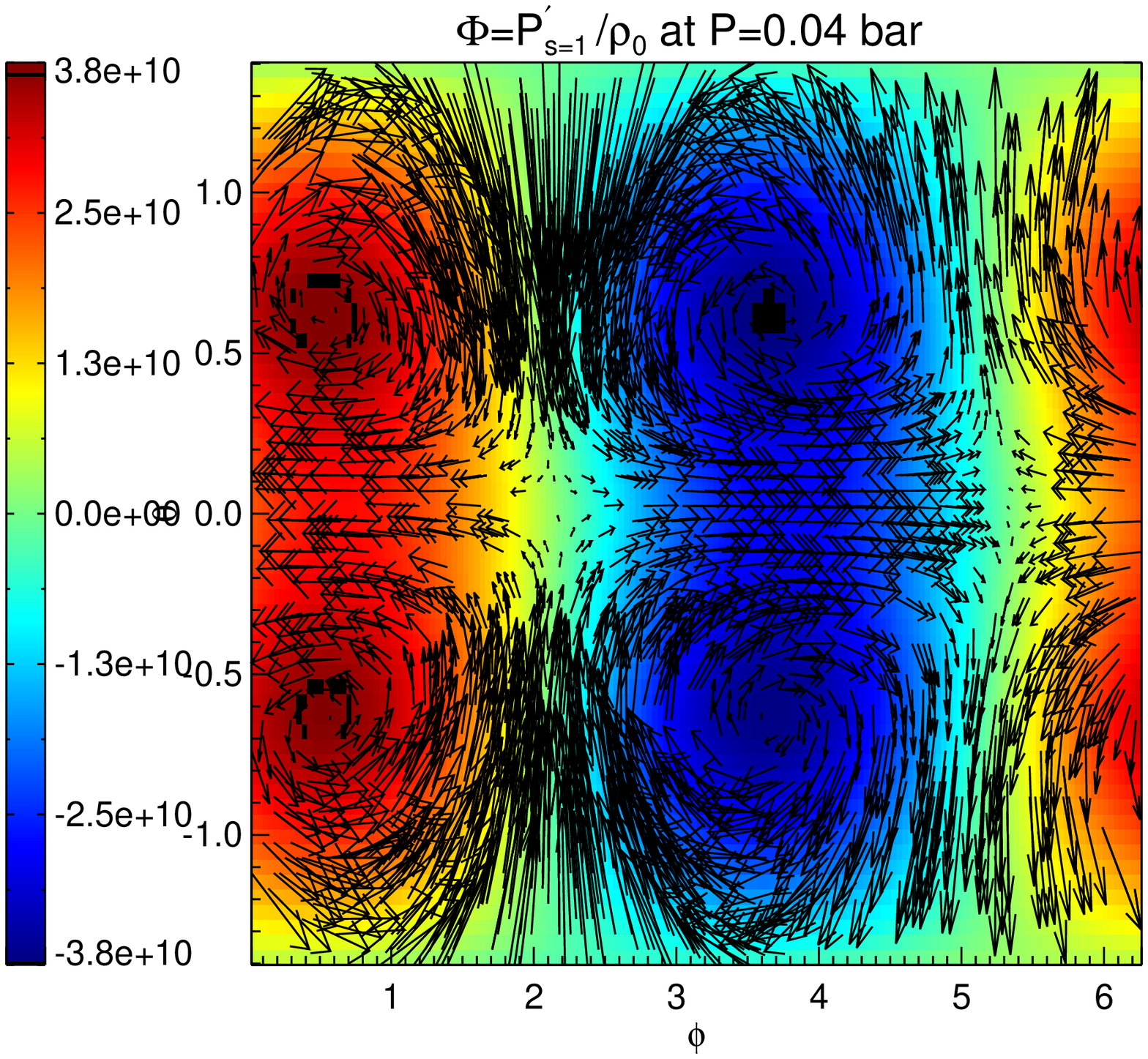}   \\
\includegraphics[height= 4cm]{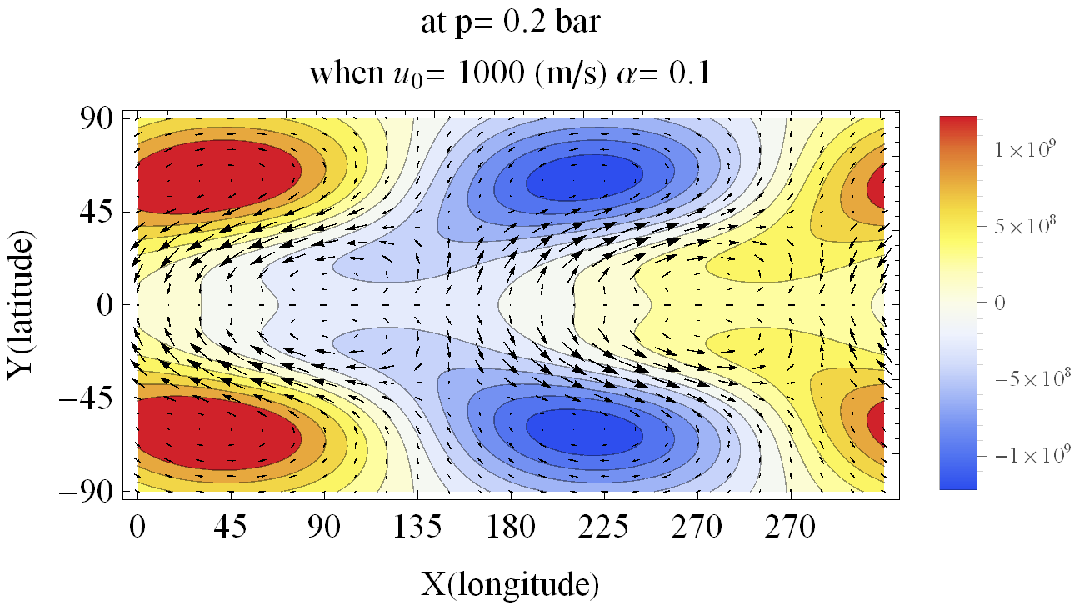}  &
\includegraphics[height= 3.3cm, width= 7cm]{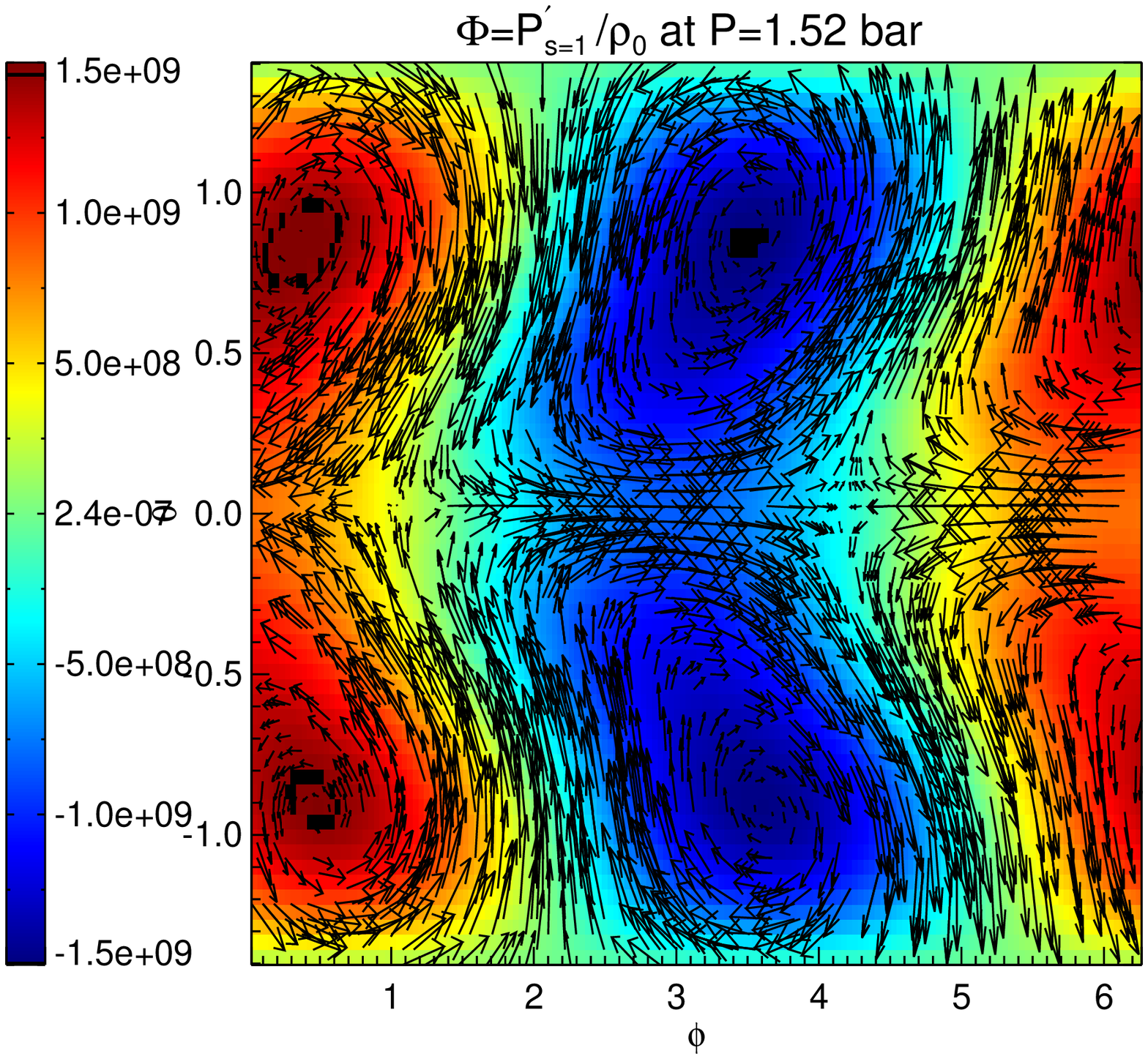}   \\
\includegraphics[height= 4cm]{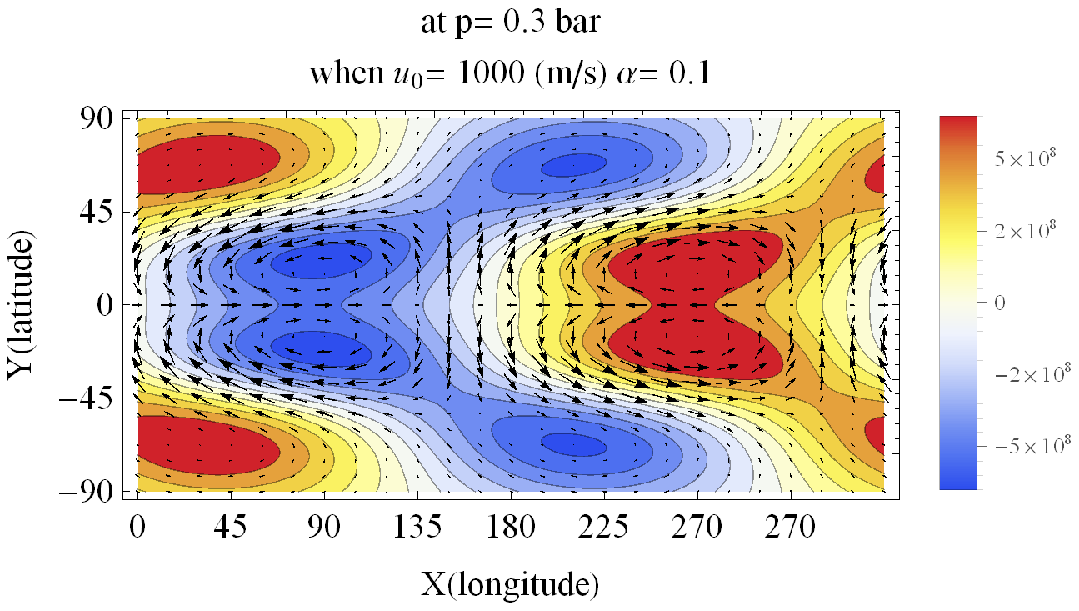}  &
\includegraphics[height= 3.3cm, width= 7cm]{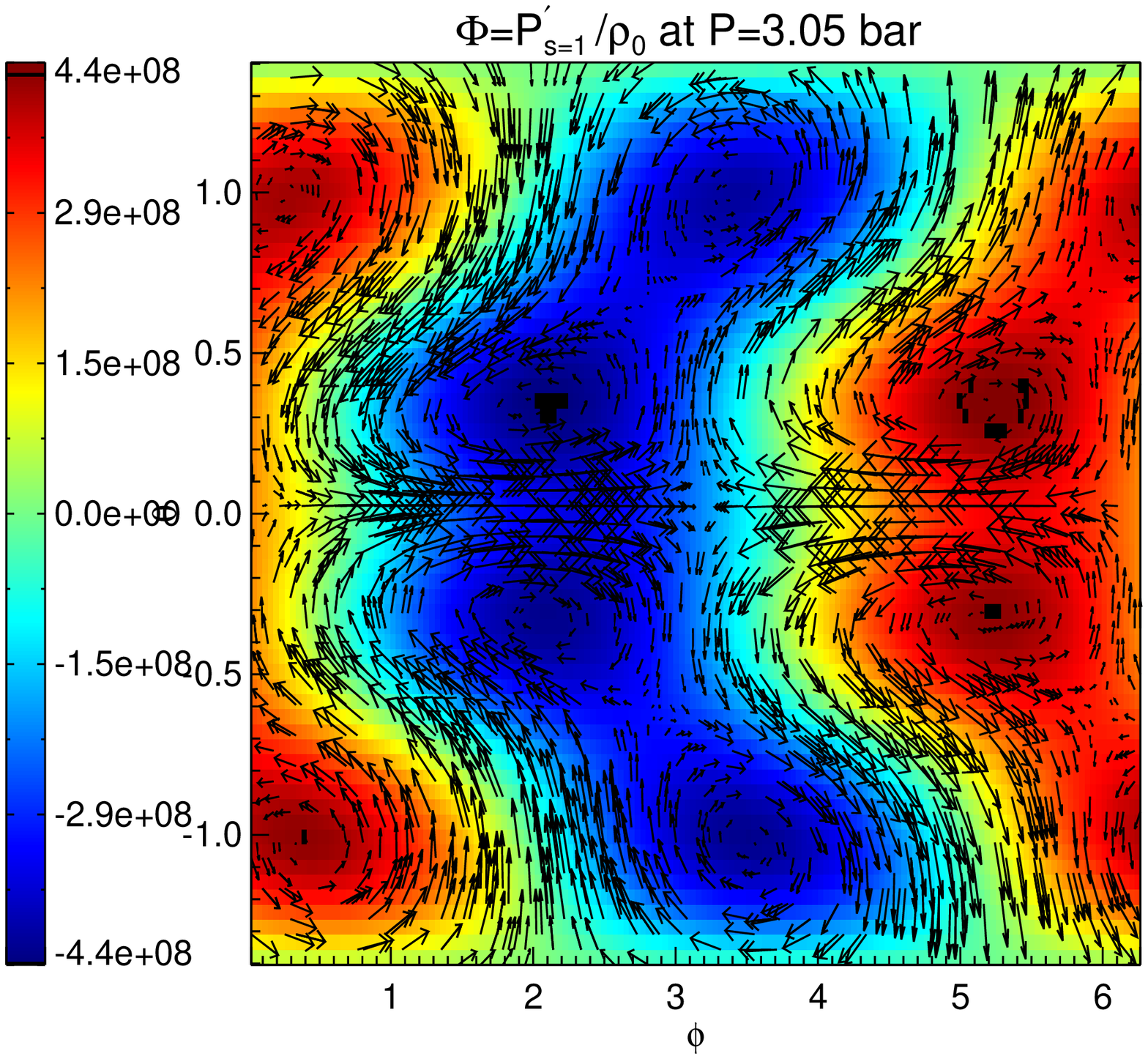}   \\
\includegraphics[height= 4cm]{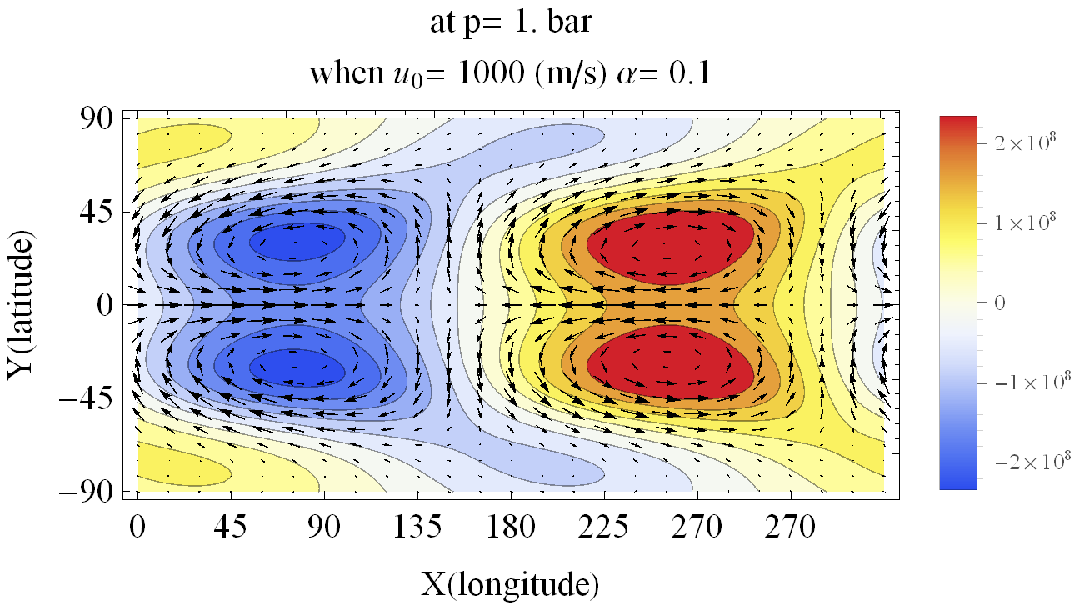}    &
\includegraphics[height= 3.3cm, width= 7cm]{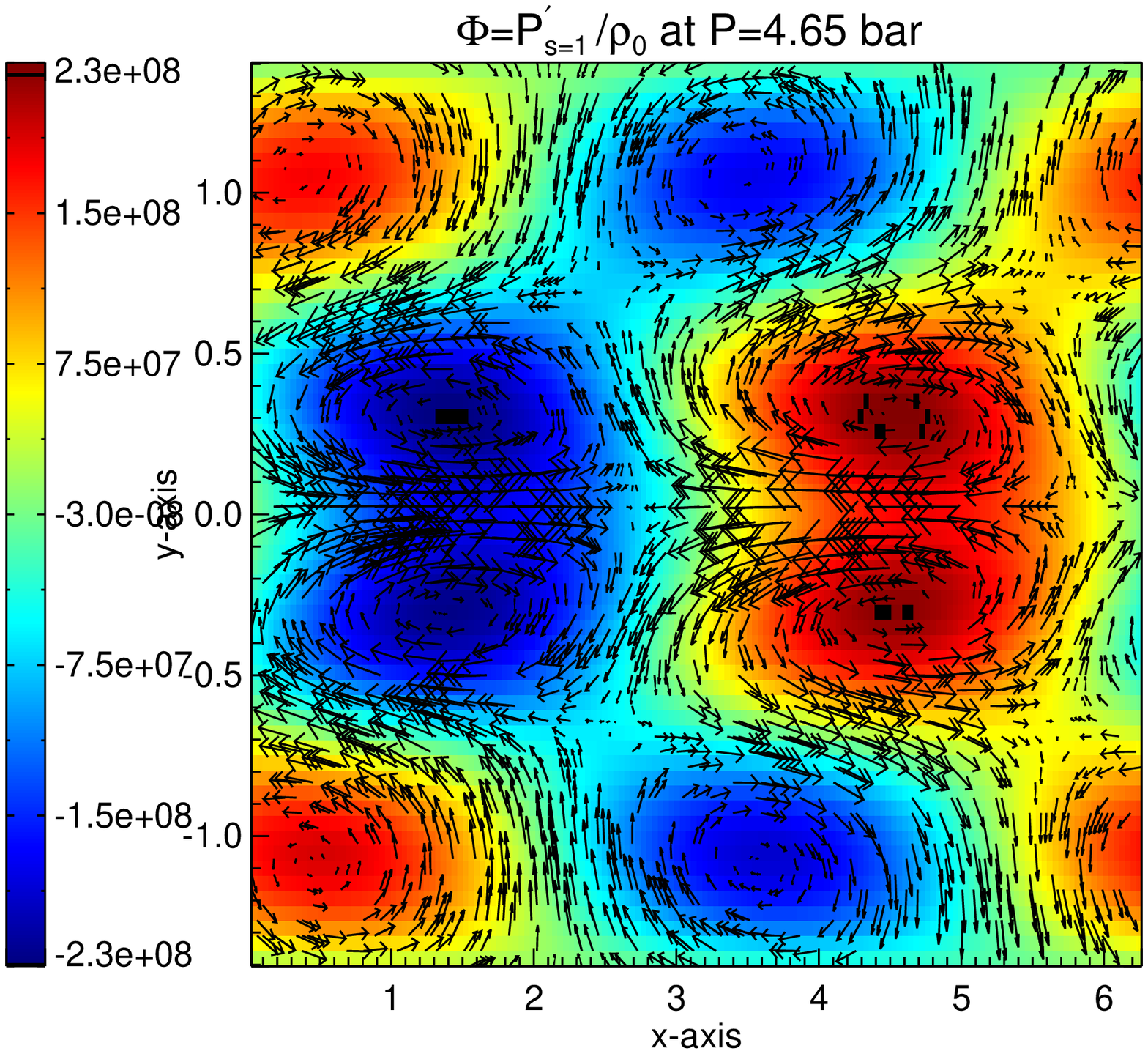}   \\
\end{tabular}
\end{center}
\caption{Wave velocities (arrows) and geopotentials (contours) for $u_{0}$ = 1000 m/s in the modest damping scenario at different pressure levels. The linear results (left column) are compared to the simulation results with $\nu=10^9$ cm$^2$/s (right column).
There are small-scale Rossby gyres westwardly separated from the larger scale structure at greater depths.}
\label{XZ plots u1000}
\end{figure}

\begin{figure}[ht]
\begin{center}
\begin{tabular}{cc}
\includegraphics[height= 4cm]{f6e.eps} &
\includegraphics[height= 3.3cm, width= 7cm]{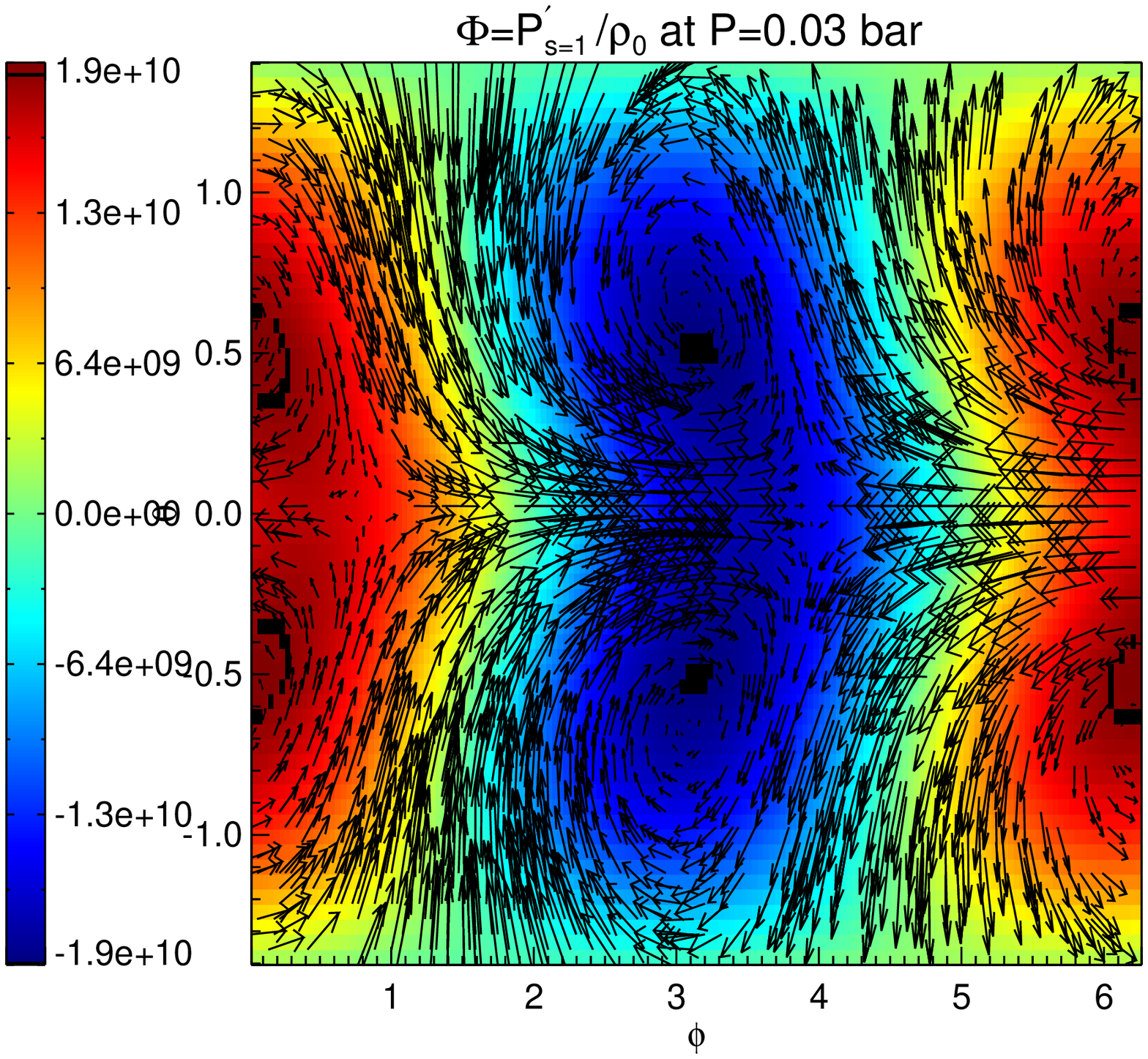}   \\
\includegraphics[height= 4cm]{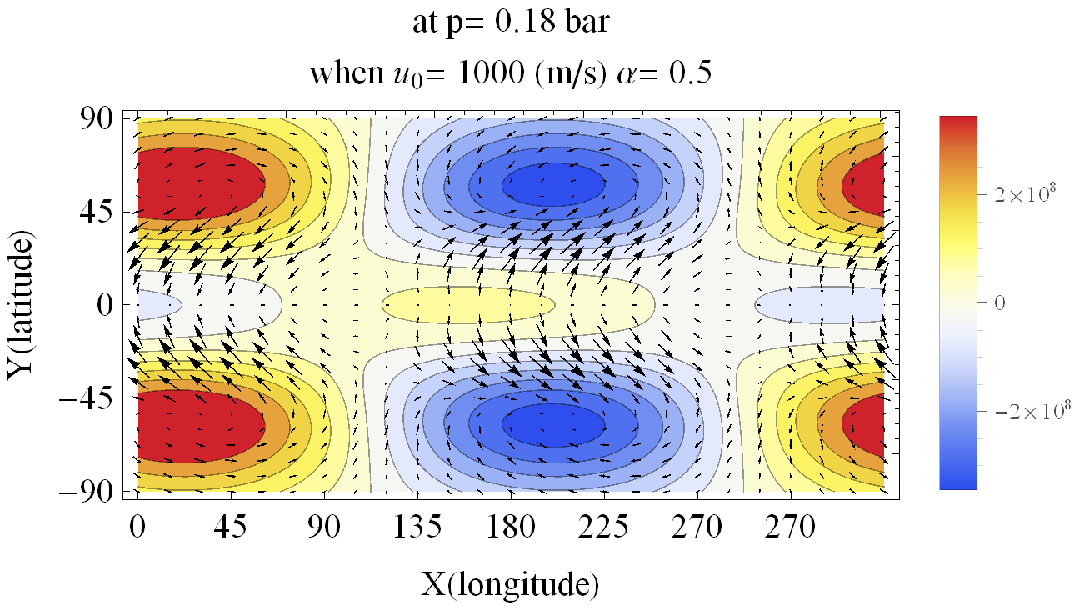}  &
\includegraphics[height= 3.3cm, width= 7cm]{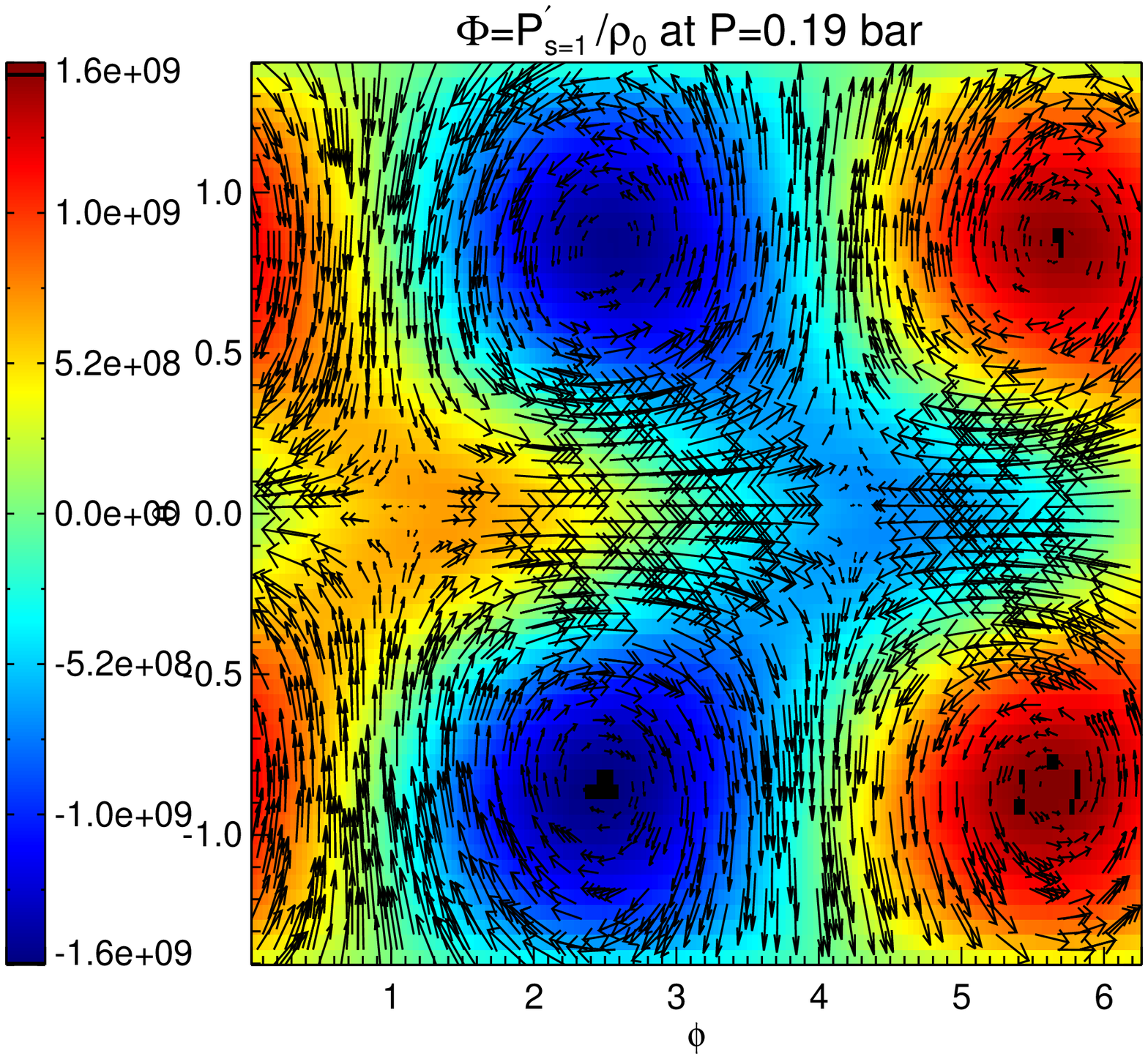}   \\
\includegraphics[height= 4cm]{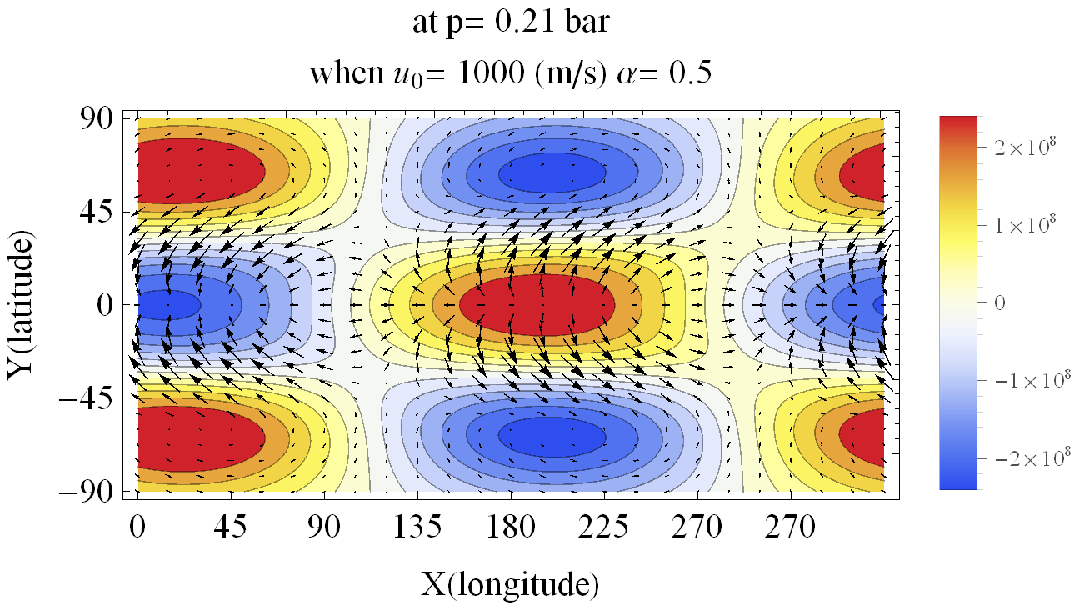}  &
\includegraphics[height= 3.3cm, width= 7cm]{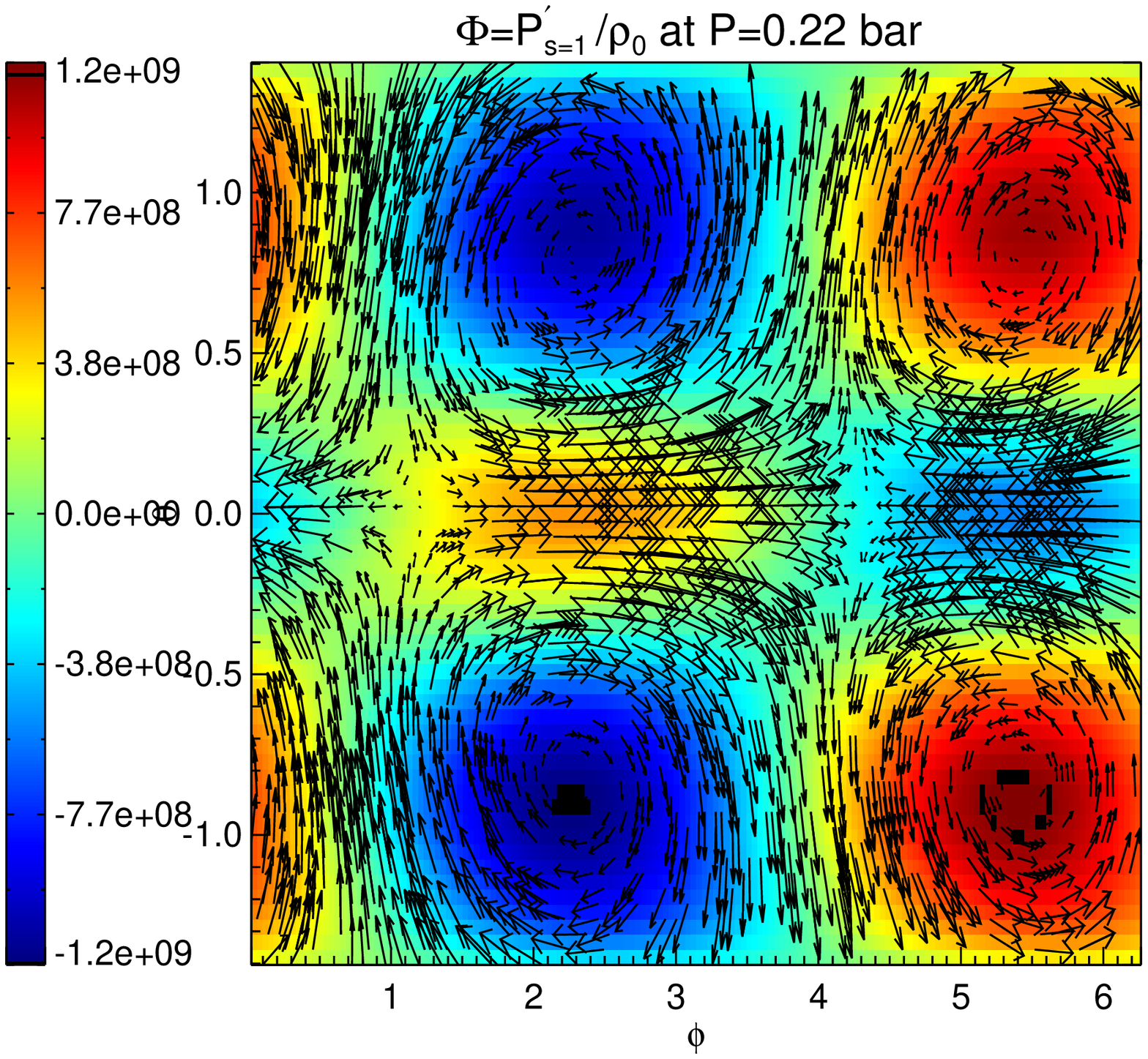}   \\
\includegraphics[height= 4cm]{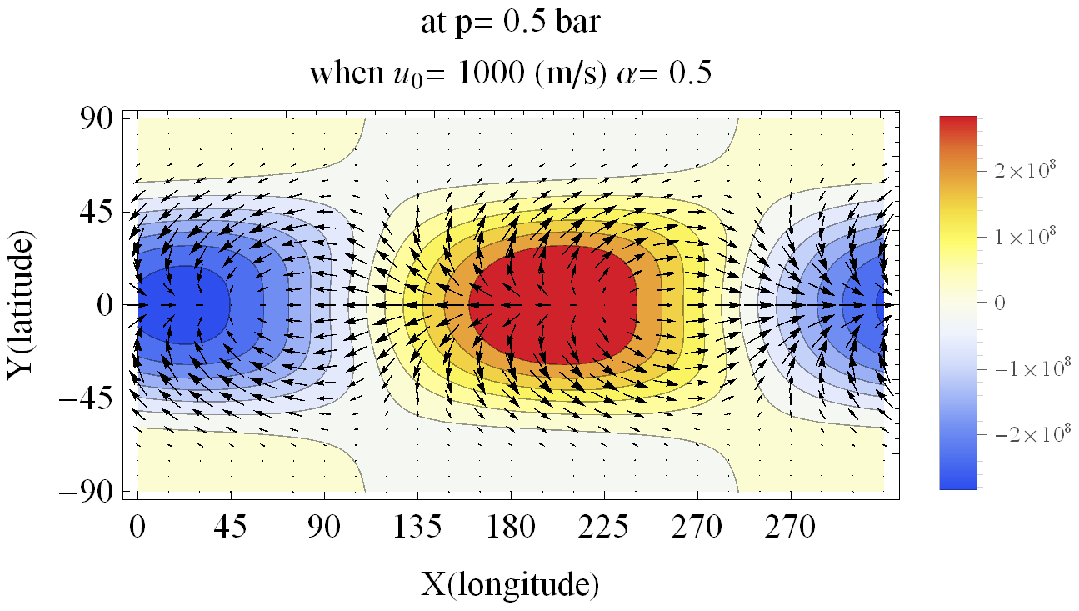}    &
\includegraphics[height= 3.3cm, width= 7cm]{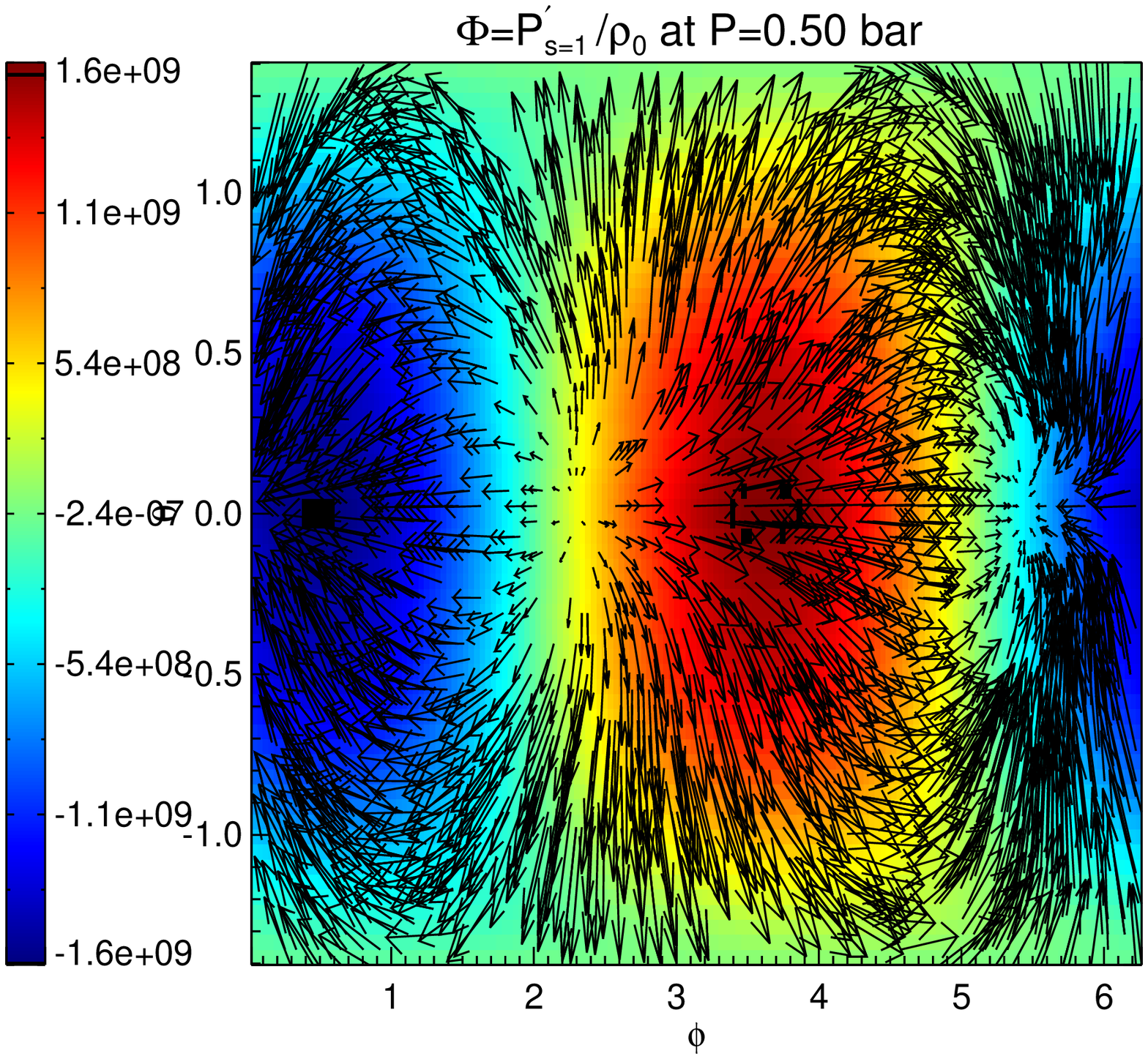}   \\
\end{tabular}
\end{center}
\caption{Same as Figure~\ref{XZ plots u1000} but for the strong damping scenario. $\nu=10^{10}$ cm$^2$/s is used for the simulation.
There are small-scale Kelvin modes  eastwardly separated from the larger scale structure
at greater depths.}
\label{fig:dispersion_alpha0.5}
\end{figure}

Figures~\ref{XZ plots u1000} and \ref{fig:dispersion_alpha0.5} provide another view to look at the tilted wavefronts, showing horizontal wave geopotentials overlaid with projected velocity fields at different depths obtained from the linear theory (left column) and from the numerical simulation (right column). The wave structures in the modest damping scenario are shown in Figure~\ref{XZ plots u1000} to compare to those in the strong damping scenario displayed in Figure~\ref{fig:dispersion_alpha0.5}.

As expected from the preceding subsection,
the Rossby gyres around the equator are gradually shifted westward at greater depths. We can see from Figure~\ref{XZ plots u1000} that on the wider horizontal scale, the Rossby gyres are also gradually tilted westward as going deeper, but interestingly there are smaller-scale gyres emerging from inside of the outer gyres around 1 bar, right below the bottom edge of the heating. The phenomena are seen in both the linear model and simulation.
This change of horizontal patterns at various depths is also present in Figure~6 of \citet{KH05} where the authors study zonal winds in the tropical upper troposphere of the Earth.

This particular pattern is a result of dispersion of a group of equatorial waves of different scales. As explained earlier in the paper, baroclinic Rossby modes dominate over baroclinic Kelvin modes in the modest damping scenario, which permits us to consider Rossby waves only to interpret the results in Figure~\ref{XZ plots u1000} without loss of generality.
Baroclinic Rossby waves with smaller vertical wavelengths have smaller equivalent depths $h_m$, and thus have smaller latitudinal scales (see Section \ref{sec:basic}). Rossby modes of smaller scales would tilt more horizontally compared to those of larger scales, meaning that small-scale Rossby gyres are shifted more westward than their large-scale counterparts at the same depth as the waves propagate downward from the level of excitation.
Consequently, smaller-scale Rossby gyres appear to be separated from the larger-scale counterparts over a certain range of depth, forming the ``double-gyre" pattern
as illustrated in Figure~\ref{XZ plots u1000}.

On the other hand, in the strong damping scenario shown in Figure~\ref{fig:dispersion_alpha0.5}, baroclinic Kelvin modes dominates over baroclinic Rossby modes. Hence we consider predominant Kelvin waves only in the strong damping scenario. By the same explanation for the modest damping case but with the opposite tilt for the wave front of Kelvin waves, small-scale Kelvin waves are shifted more eastward than their large-scale counterparts at the same depth as the waves propagate downward from the level of excitation. As a result, smaller-scale Kelvin waves appear to emerge from the larger-scale counterparts and shift more eastward at greater depths,
as illustrated in Figure~\ref{fig:dispersion_alpha0.5}. Note that the Kelvin modes do not possess any meridional velocities, and therefore the meridional velocity fields in the figure arise from the less-dominant Rossby modes in the strong damping scenario.

\begin{figure}
\plottwo{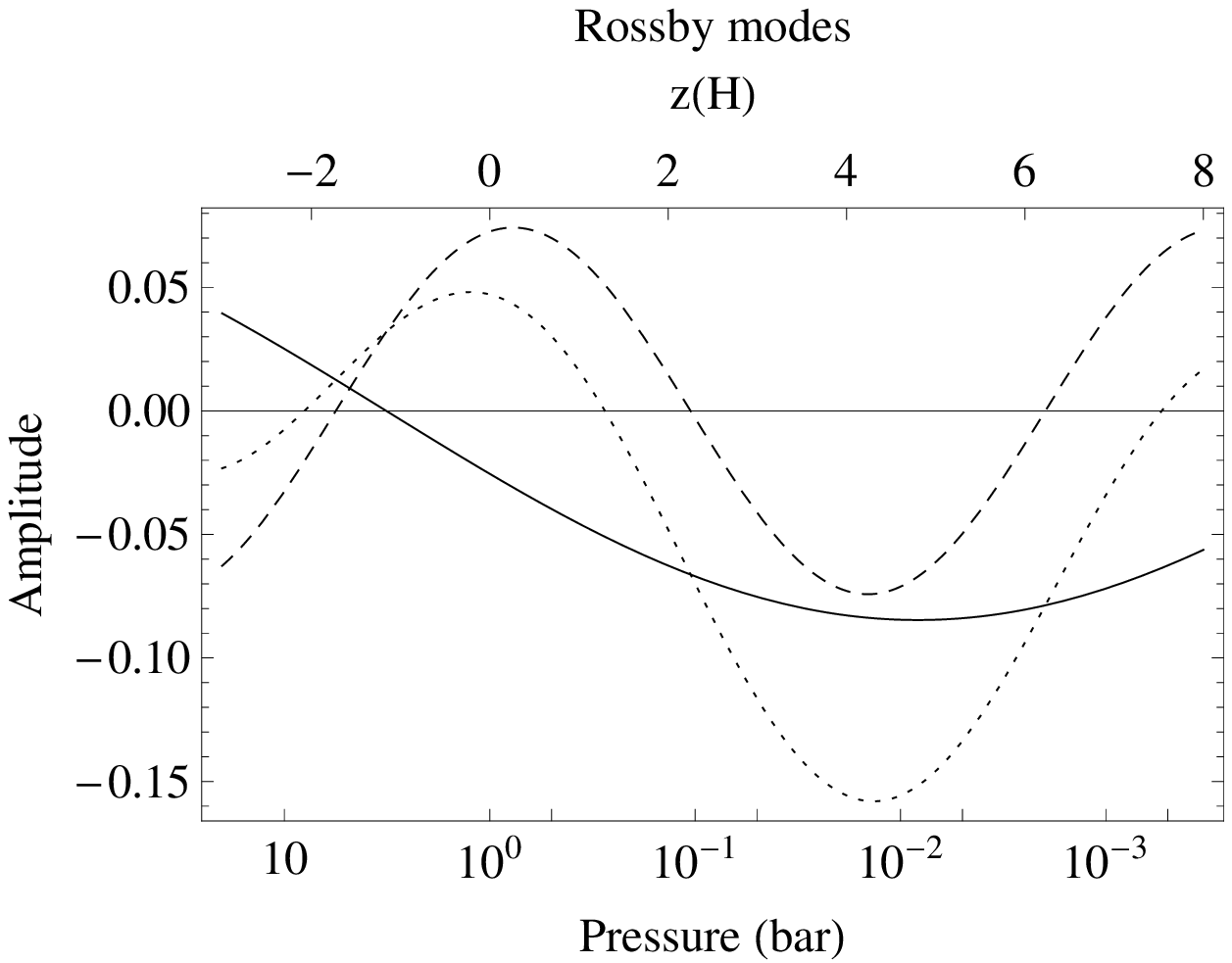}{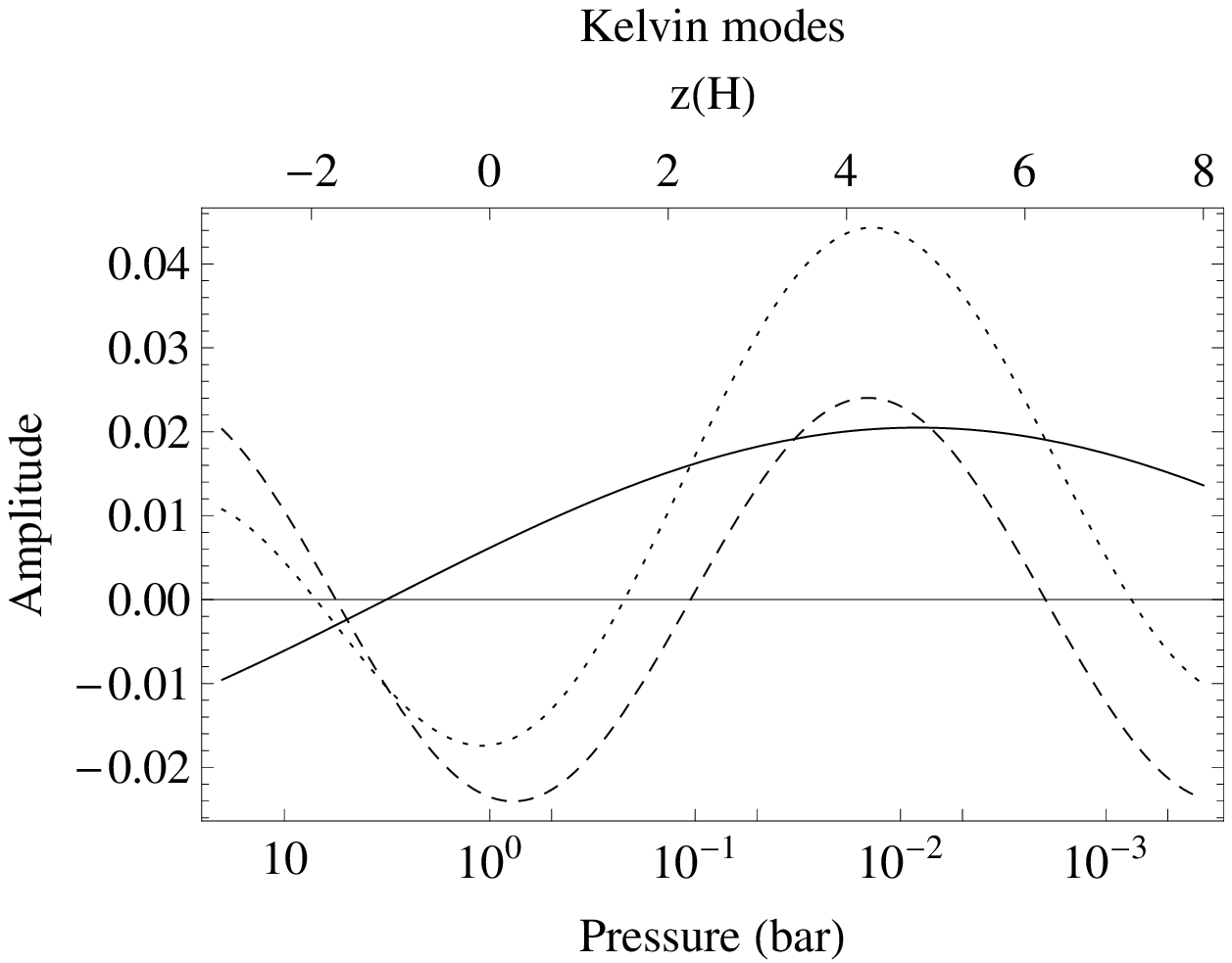}
\caption{Amplitudes of the two predominant vertical modes $m=0.26$ (solid) and $m=$0.79 (dashed) representative as large-scale and small-scale modes. The dotted curve is the sum of the two modes, showing the relation between the wave phase and the depth. The left panel illustrates the vertical modes associated with the Rossby component in the modest damping scenario, whereas the right panel displays those associated with the Kelvin component in the strong damping scenario.}
\label{m mode group plot}
\end{figure}

To be more specific, we further consider predominant vertical modes of the solution and inspect their amplitudes, defined by Re$[\sum_n a_{m,n,l}Z_m(z)]$, along the depth in the vertical domain. The left (right) panel of Figure~\ref{m mode group plot} shows the predominant vertical modes with $m =0.26$ and $m = 0.79$ related to the Rossby (Kelvin) component in the modest (strong) damping scenario\footnote{As studied above, the Kelvin (Rossby) component in the modest (strong) damping scenario is less dominant and thus is not shown here without loss of important physics.}. These two vertical modes
correspond to the equivalent depths about 150 km and 50 km respectively, selected from the power spectrum in Figure~\ref{fm2 plot},
referred to as the larger-scale mode (solid) and smaller-scale mode (dashed). In fact, the large excitation of these two modes results not only from a good match with the vertical heating profile but also from a strong coupling with the longitudinal heating profile.
The dotted line is the sum of the two modes and so the constructive or destructive phase interference can be seen. The two modes have the same phase and thus the maximum constructive effect around the heating level. This implies that they are simultaneously excited at the heating level. However, they have opposite signs and equal amplitude near 1 bar, implying that after traveling over a certain vertical distance the dispersive equatorial waves spatially split, leading to the dispersive wave patterns along the latitude.

Before closing the section, we would like to reiterate that as explained in Section 2.1,
the wave dispersion for Rossby waves can arise from baroclinic modes (denoted by $m$) in addition to horizontal modes (denoted by $n$, see Figure~\ref{n135 Rplot}) resulting from the shallow-water model. Generally, we find that modes of higher order $n$ are one order of magnitude smaller than the $n=1$ mode in amplitude, and therefore do not contribute to the double gyres. It is the baroclinic modes that are responsible for double gyres, a dispersive phenomenon unique to vertical stratification.

\subsection{Interaction between waves and mean zonal flows}
\label{sec: momentum flux}
We have studied the 3D structure and physical properties of equatorial waves in the presence of a zonal-mean zonal flow. We now proceed to the feedback of these waves to mean zonal flows.
We first compare the distribution of wave momentum fluxes from the linear theory  with the results from simulations. We then apply the linear model to explore the possible evolution of superrotating flows due to the feedback, i.e. the interaction between waves and mean zonal flows.

\subsubsection{Spatial distributions of momentum fluxes and mean-flow accelerations}
\label{sec:Distribution of momentum flux}

\begin{figure}[ht]
\begin{center}
\begin{tabular}{cc}
\includegraphics[height=5cm]{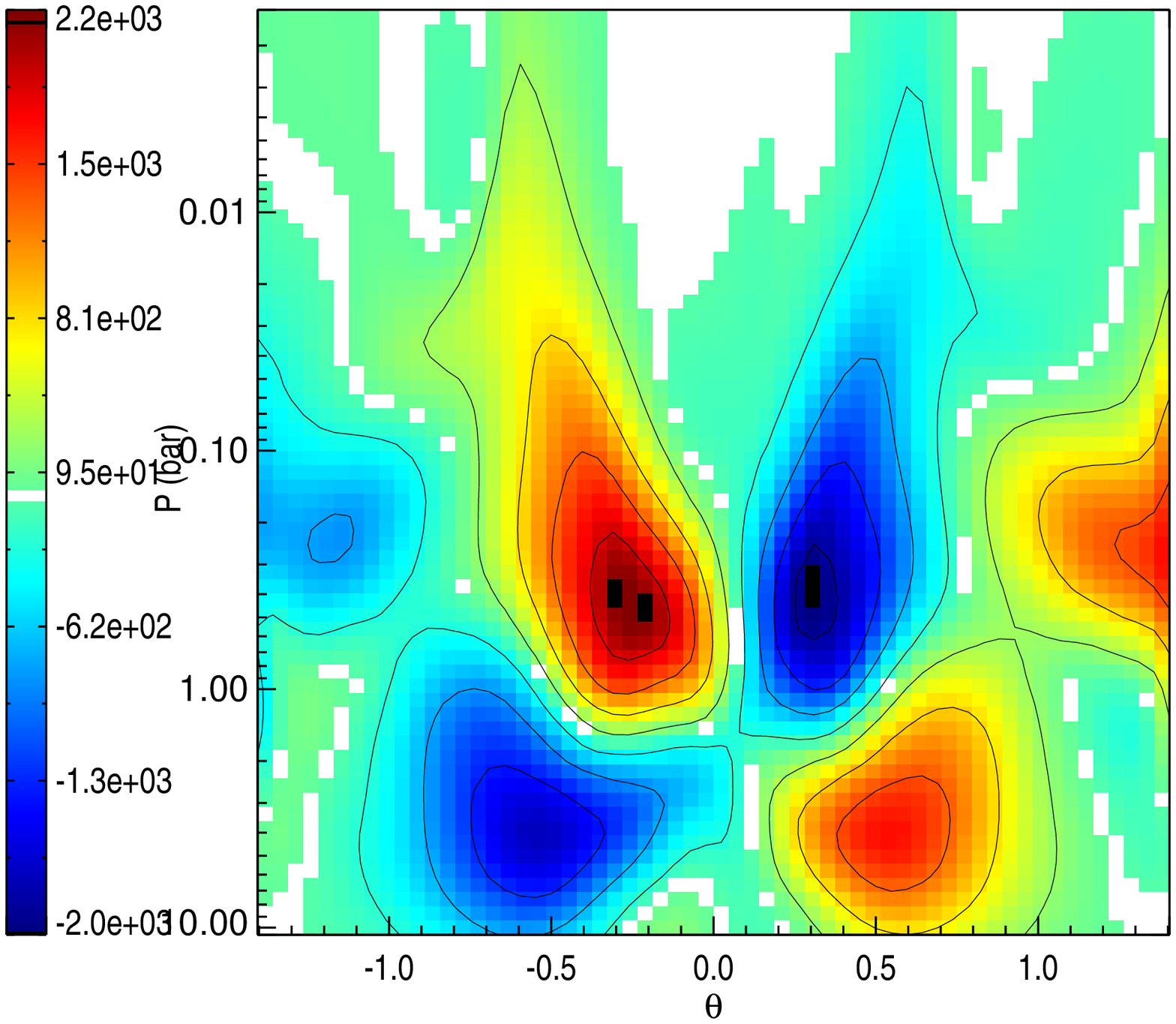} &
\includegraphics[height=5cm]{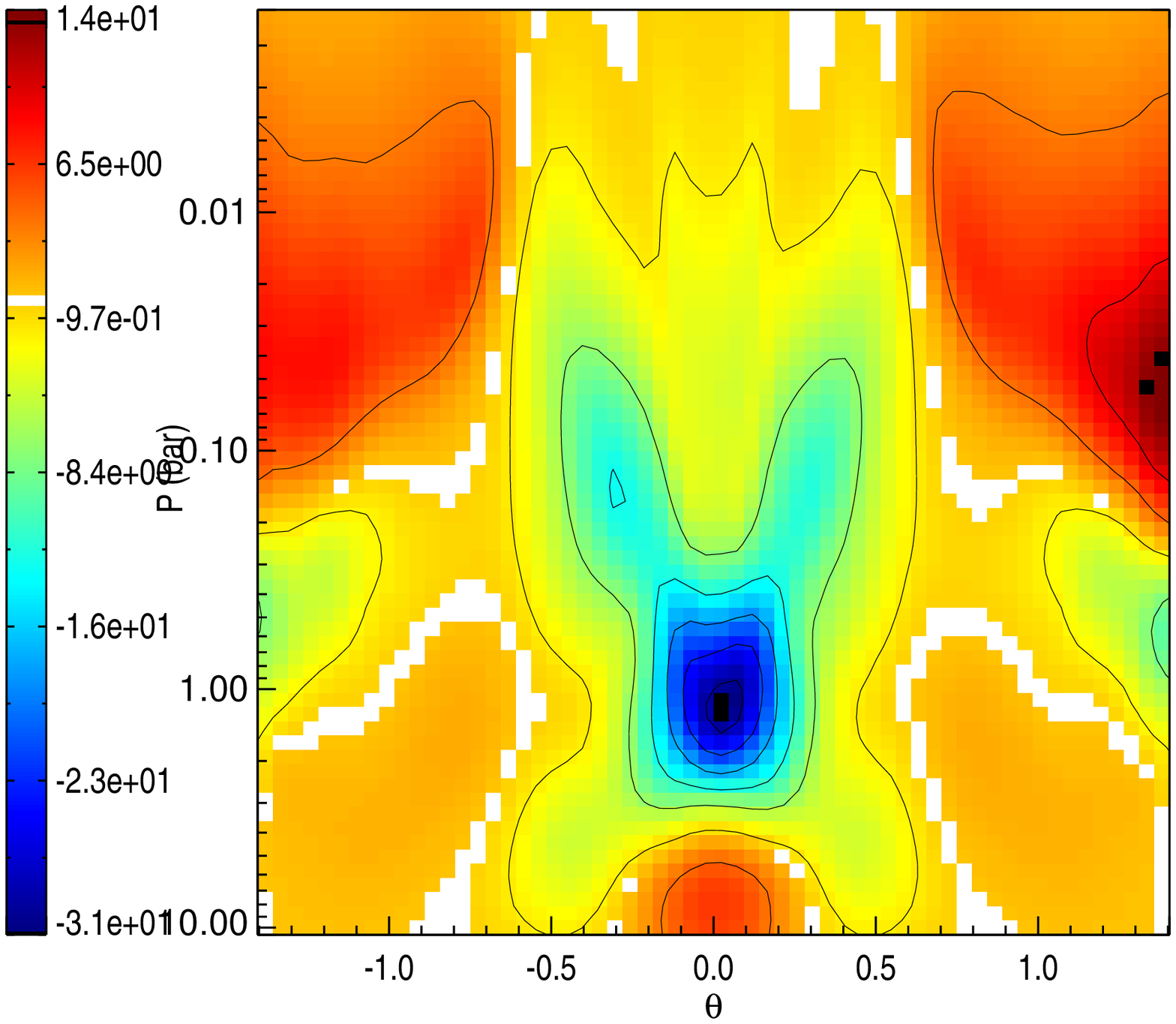} \\
\includegraphics[height=5cm]{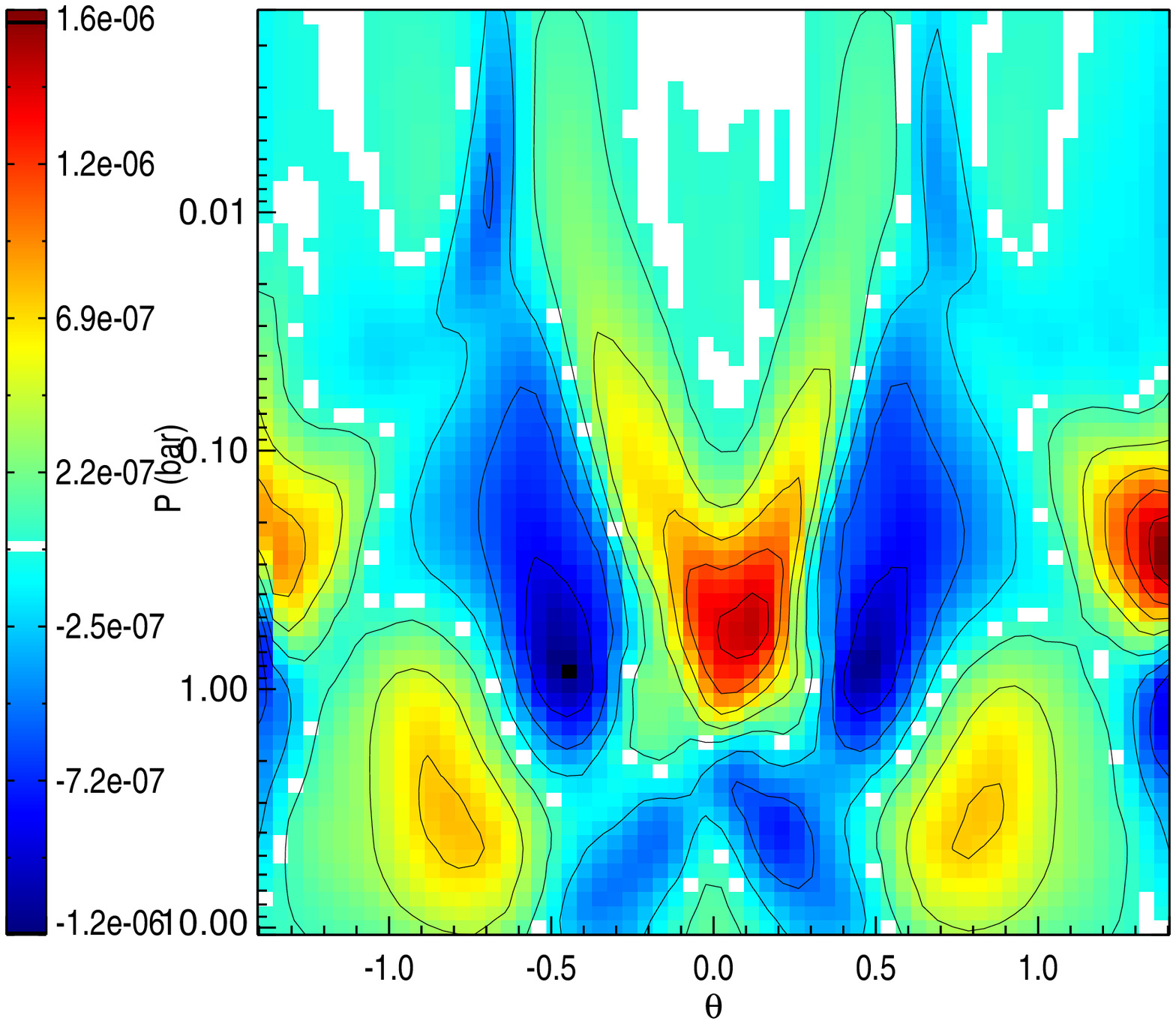} &
\includegraphics[height=5cm]{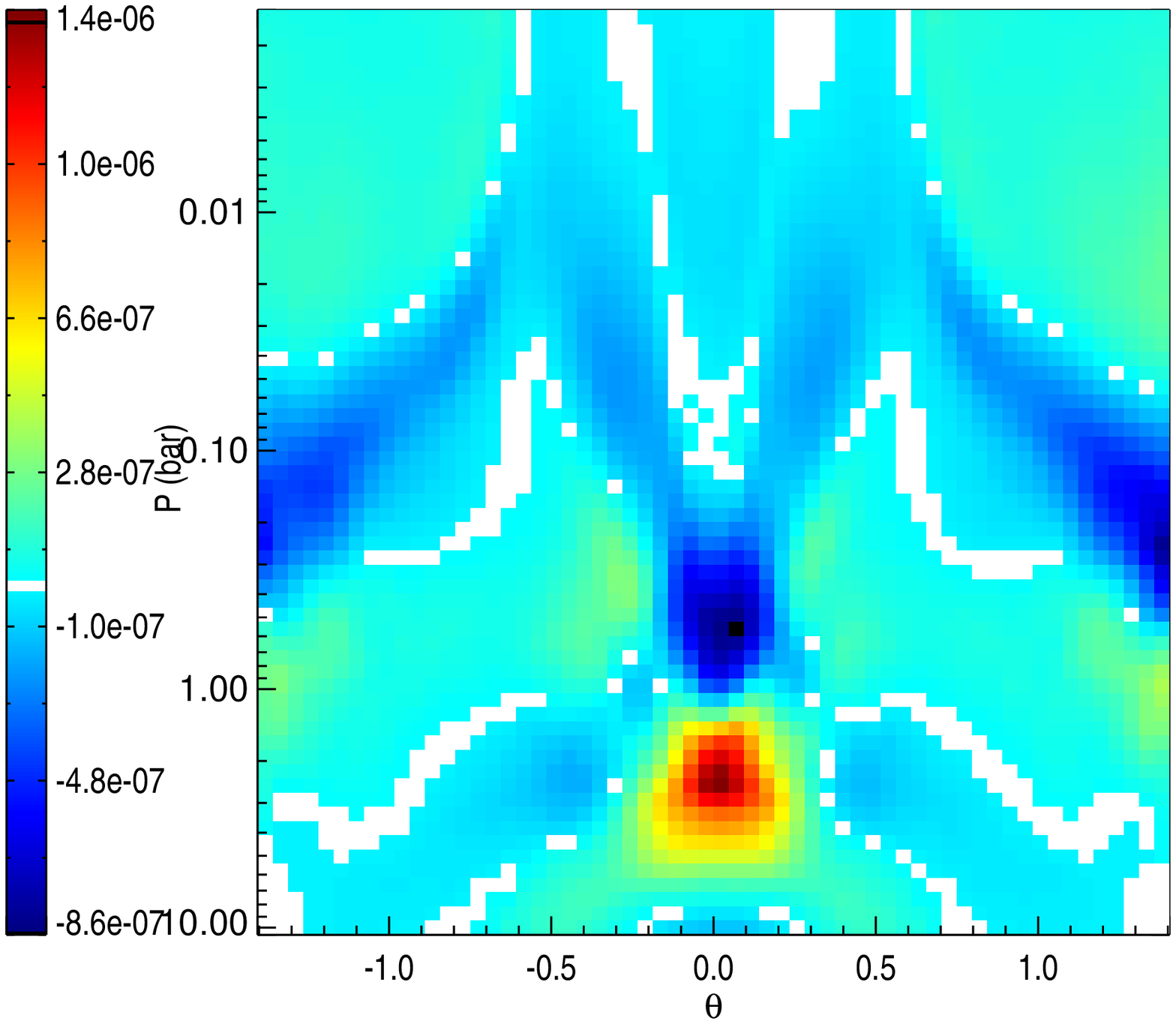} \\
\end{tabular}
\end{center}
\caption{Momentum fluxes $\rho_0 \overline{u'v'}$ (top left) and $\rho_0 \overline{u'w'}$ (top right) as well as their gradients $-(1/R\cos^2\theta )d/d\theta [\rho_0 \overline{u'v'}\cos^2 \theta]$ (bottom left) and $-d/dr[\rho_0 \overline{u'w'}]$ (bottom right) in the equilibrium state of the atmosphere from our numerical simulation for $\nu=10^{9}$ cm$^{2}$/s. Note that these are expressions in spherical coordinates $(r,\theta,\phi)$ used in the simulations.}
\label{Ian flux}
\end{figure}

\begin{figure}[ht]
\begin{center}
\begin{tabular}{cc}
\includegraphics[height= 5cm]{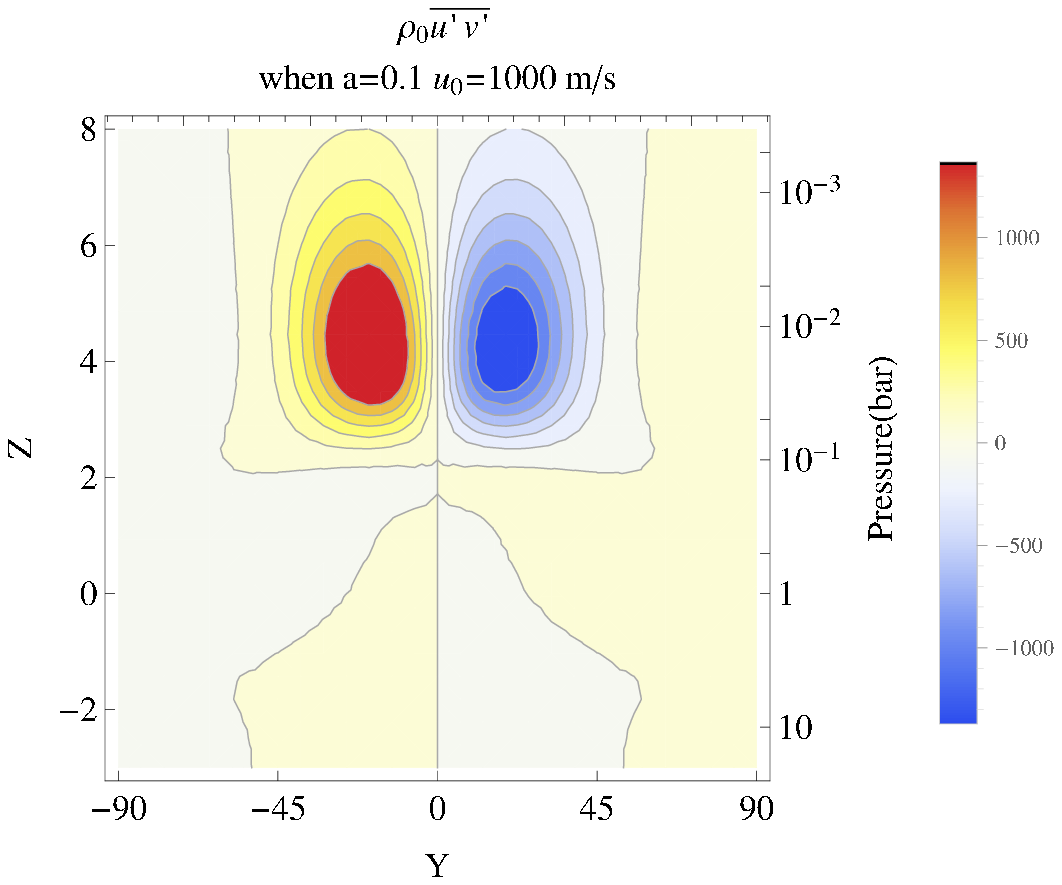} &
\includegraphics[height= 5cm]{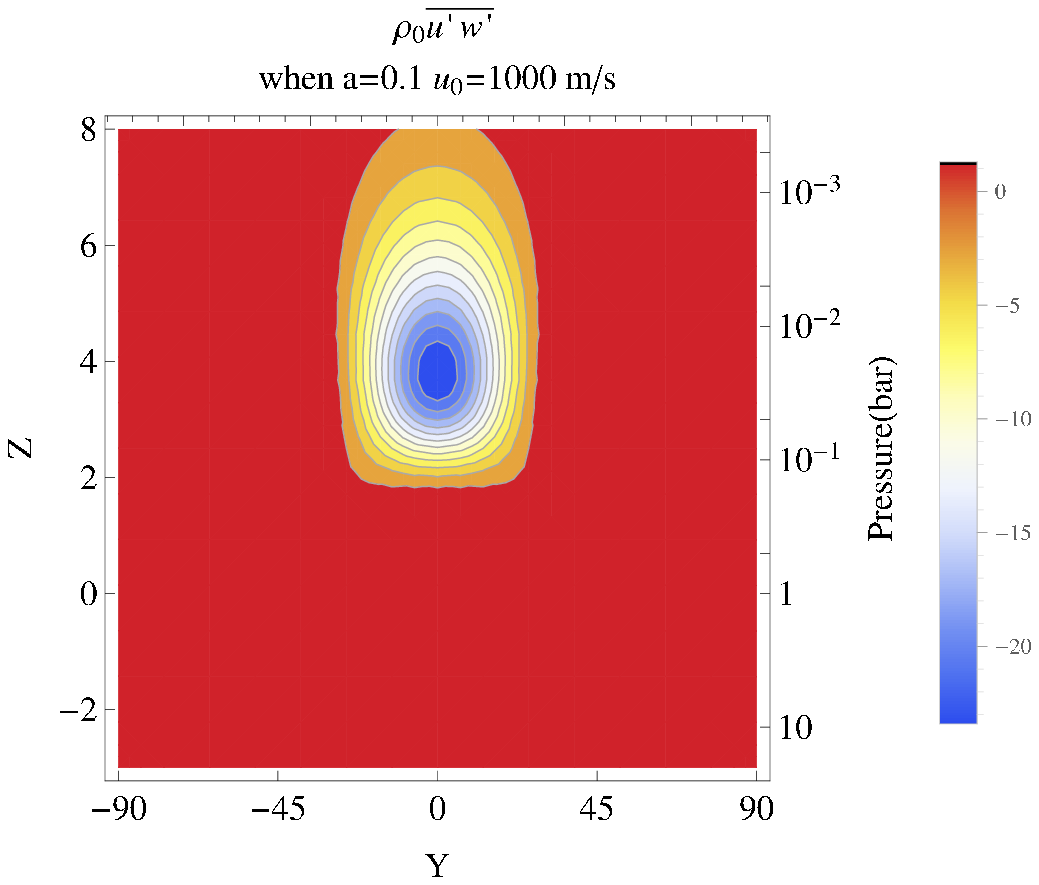} \\
\includegraphics[height= 5cm]{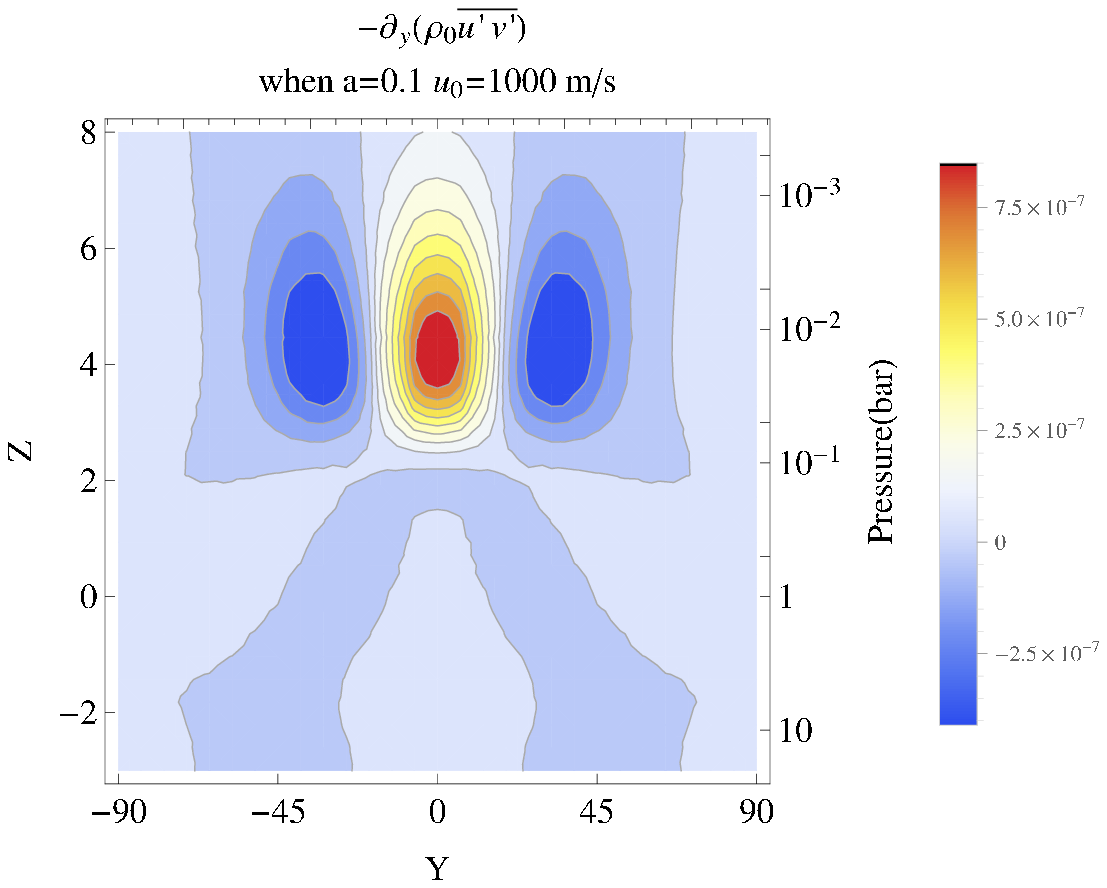} &
\includegraphics[height= 5cm]{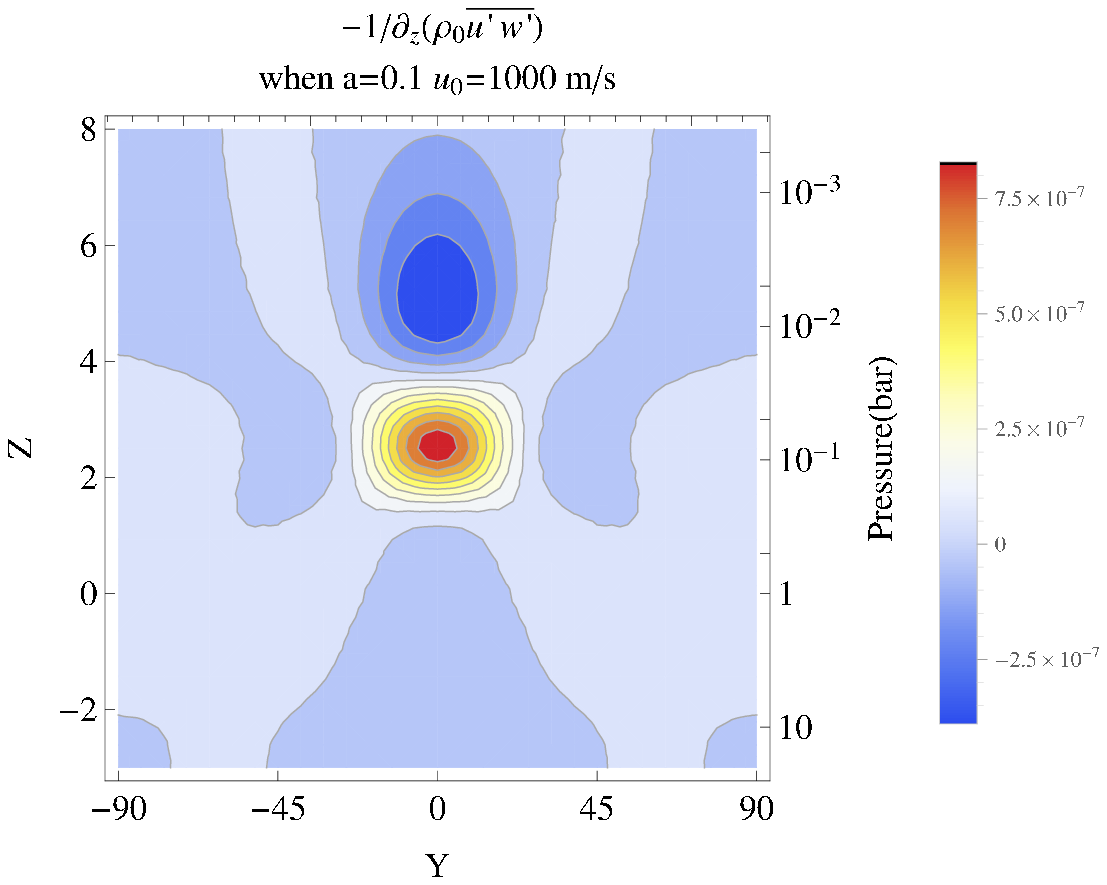}
\end{tabular}
\end{center}
\caption{Same as Figure~\ref{Ian flux} but calculated from the linear theory for the case with $u_{0} =$ 1000 m/s and $\alpha$ = 0.1.}
\label{flux u1000 plot}
\end{figure}

Figure~\ref{Ian flux} shows the spatial distributions of horizontal and vertical momentum fluxes as well as their gradients from our numerical simulation with $\nu=10^9$ cm$^2$/s in the final equilibrium state, which are compared with those from our linear model for $\alpha=0.1$ and $u_0=1000$ m/s illustrated in Figure~\ref{flux u1000 plot}. Owing to thermal forcing and dissipation, the gradients of wave momentum fluxes give the force acting on different parts of the atmosphere, as elaborated in Section 2.5.
The linear results for the momentum fluxes and their gradients are qualitatively consistent with the results from the simulation,
except that in the simulations the distributions of the momentum fluxes and corresponding gradients lie deeper in the atmosphere than those obtained from the linear model,
which might be due to different vertical density profiles for the background state of the atmosphere.
We can see from the figures that the horizontal momentum flux, $\rho_{0} \overline{u'v'}$, is negative in the northern hemisphere and positive in the southern hemisphere, creating the equatorward momentum transport to accelerate the equatorial jet around the heating level.  On the other hand, the vertical momentum flux $\rho_{0} \overline{u'w'}$ is negative around the heating level and slightly positive beneath it in the simulation, thereby decelerating the superrotating flow around the heating level.

The equatorward momentum flux has been discussed in Section~\ref{sec: 2D z3}. Here, we focus on the mechanism for the vertical momentum transport by studying velocity fields in our linear model.
As has been described in Section 3.2.2. for $u'$ and $w'$ at the equator,
upwelling (downwelling) flows primarily occur in
the horizontal westward (eastward) flows in Figure~\ref{XZ U1000 plots}.
Therefore in the vertical domain of the moderate damping case, the horizontal and vertical wave velocities have negative correlation $\overline{u'w'}<$ 0, implying an upward transport of westward momentum and a downward transport of eastward momentum, leading to the eastward acceleration at the bottom of heating region and westward acceleration (i.e. deceleration) above and around the heating level, in agreement with the result in the right panels of Figure~\ref{flux u1000 plot}.
Moreover, the mechanism of jet acceleration and deceleration in our 3D linear analysis generally agrees with the 3D numerical results performed by SP as well. The vertical (horizontal) momentum transport decelerates (accelerates) the superrotating jet.

\begin{figure}[ht]
\begin{center}
\begin{tabular}{cc}
\includegraphics[height= 5cm]{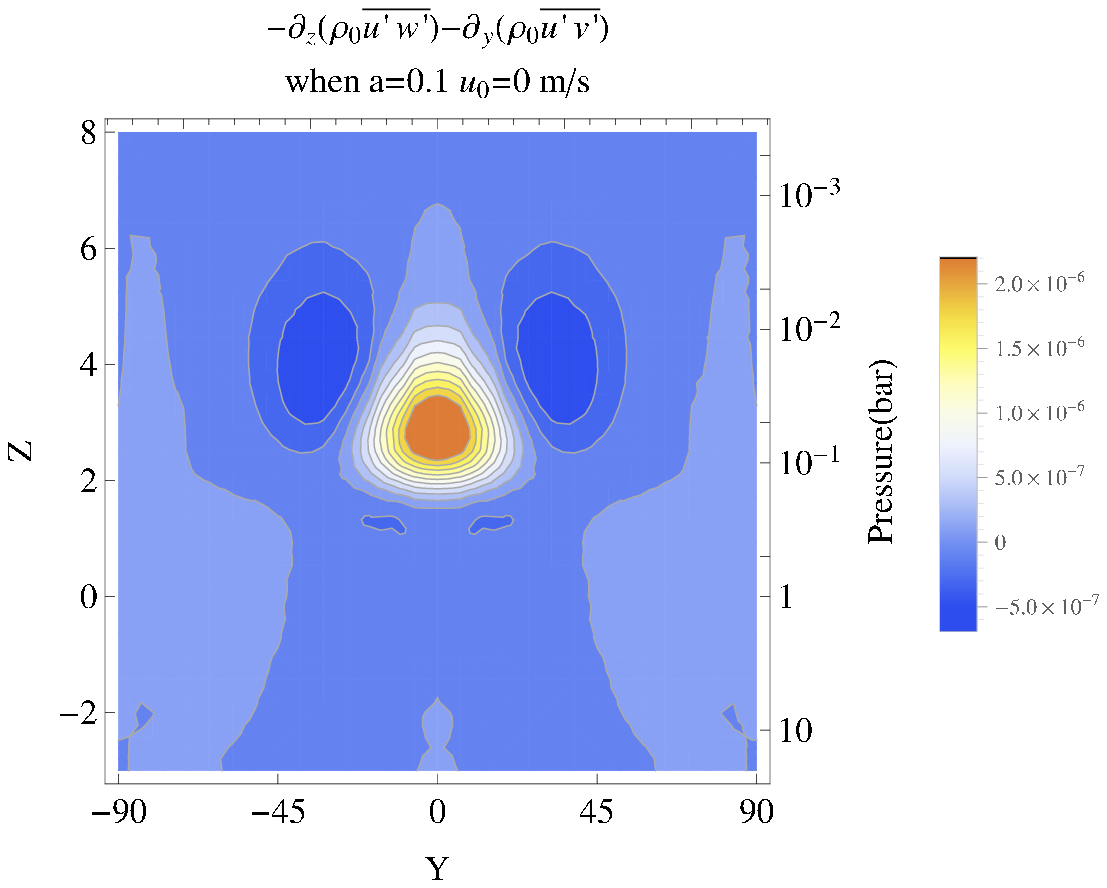} &
\includegraphics[height= 5cm]{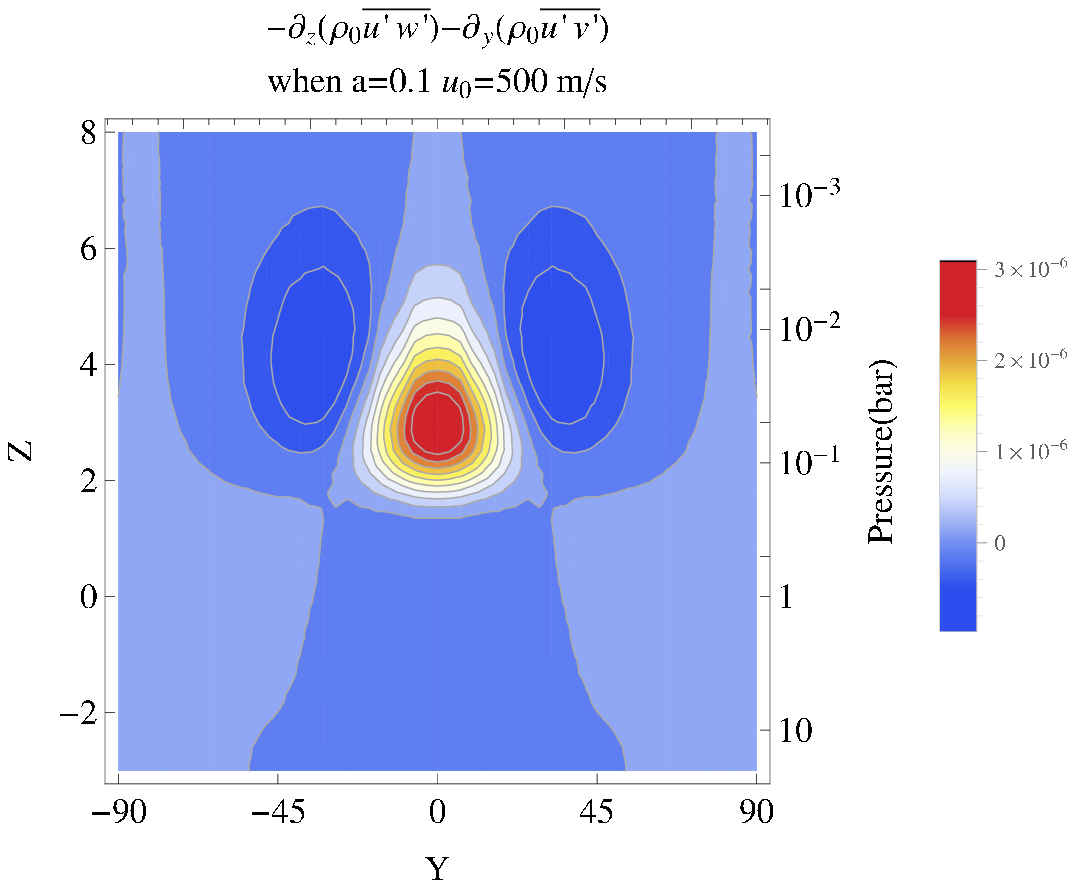} \\
\includegraphics[height= 5cm]{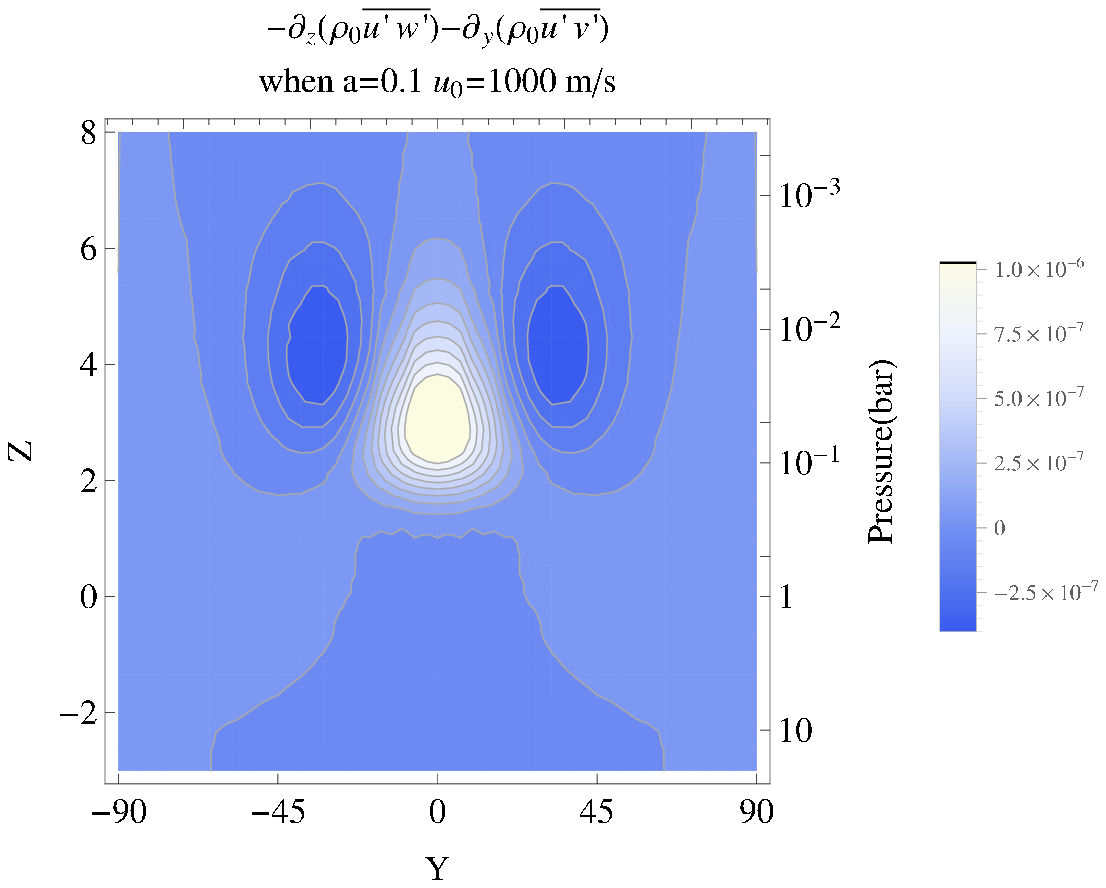} &
\includegraphics[height=5cm]{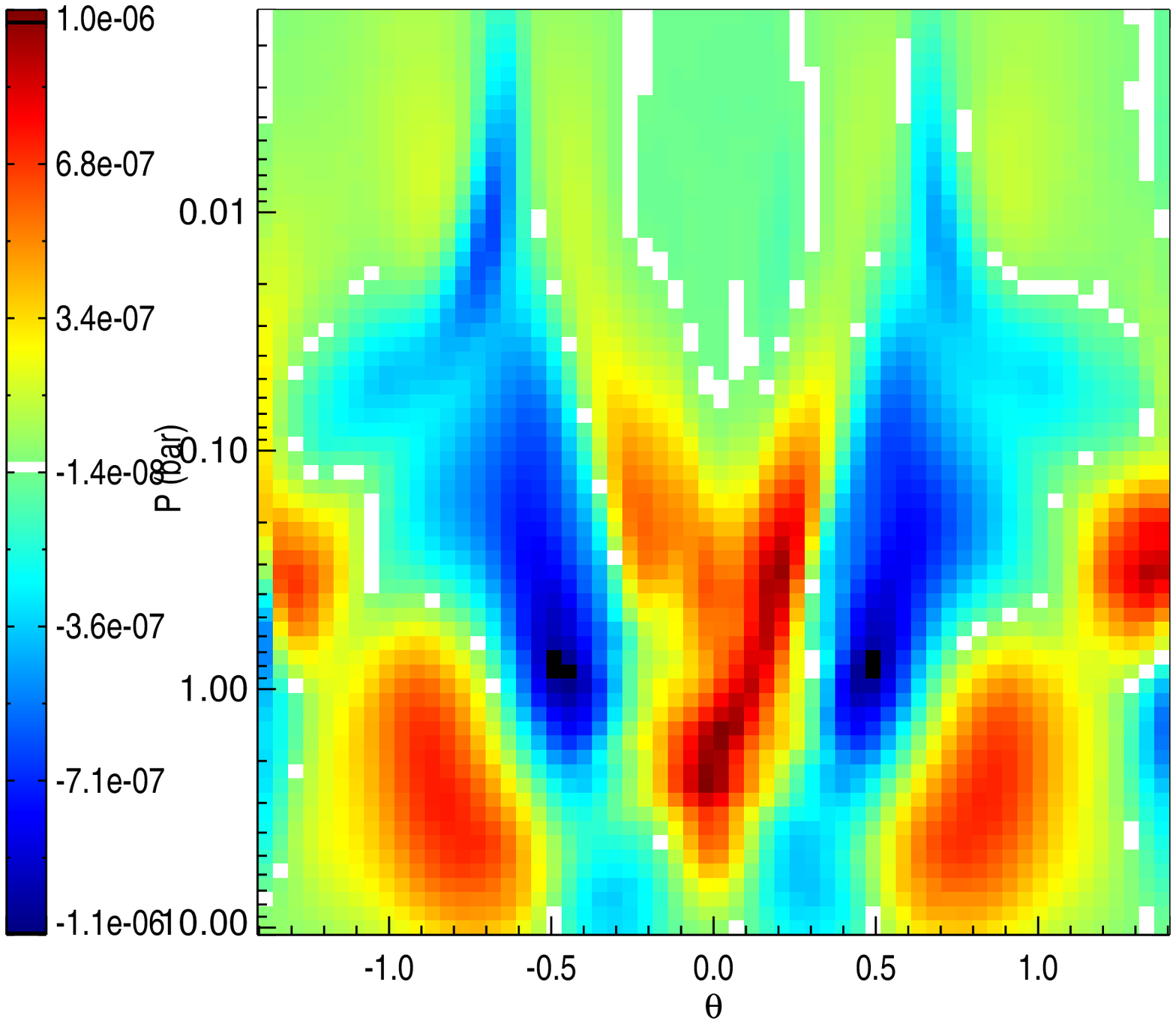}
\end{tabular}
\end{center}
\caption{Divergence of the total momentum flux obtained from the linear theory
for $u_0$ equal to 0 (top left panel), 500 (top right panel), and 1000 (bottom left panel) m/s. The same quantity
(i.e. $-d/dr[\rho_0 \overline{u'w'}]-(1/R\cos^2 \theta)d/d\theta [\rho_0 \overline{u'v'}\cos^2 \theta]$ in spherical coordinates) in the equilibrium state of the simulation is also displayed in the bottom right panel. The contours are color-coded in the same scale.}
\label{sum acc plot}
\end{figure}

The net force per unit volume acting on the superrotating jet as well as zonal flows away from the equator are further illustrated in Figure~\ref{sum acc plot}, which plots
the divergence of the total wave momentum flux for different $u_0$ in the linear theory and for the final equilibrium state in the simulation (see Equation \ref{zonal mean}).
The forces are strongest due to the Rossby resonance when $u_0=500$ m/s and subside when $u_0=1000$ m/s.
The results hint that the zonal-mean flow would be accelerated from rest, eastward around the equatorial region but westward at mid-latitudes and higher altitudes.
The result for $u_0=1000$ m/s from the linear theory (bottom left panel) resembles that from the simulation (bottom right panel), strongly suggesting that equatorial waves play a significant role in the formation and evolution of zonal flows observed in the simulation.
The implication of our linear wave theory to possible evolutions of the equatorial super-rotation flow will be explored in more detail in the next subsection.

\subsubsection{Implication to evolution of a superrotating flow}
\label{evo}

\begin{figure}[ht]
\begin{center}
\subfigure[]{\includegraphics[height= 4.5cm]{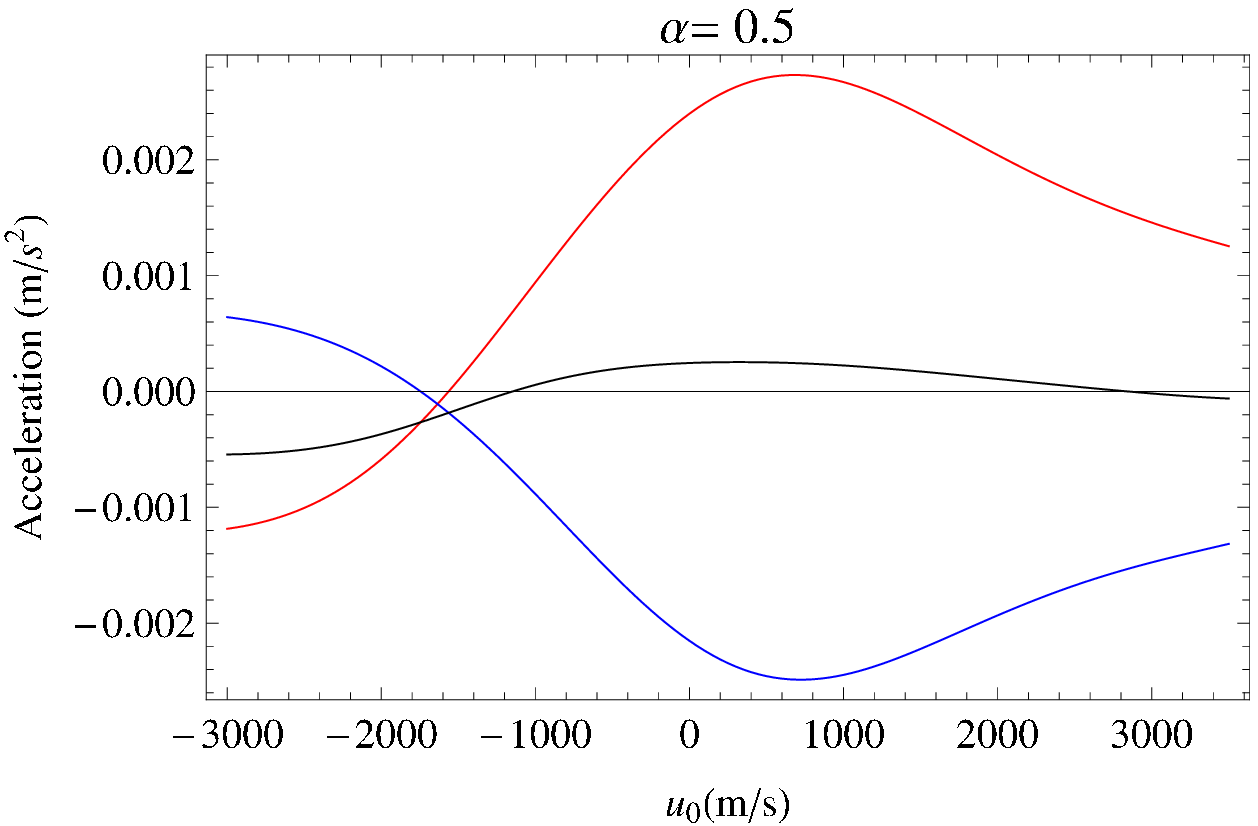}}
\subfigure[]{\includegraphics[height= 4.5cm]{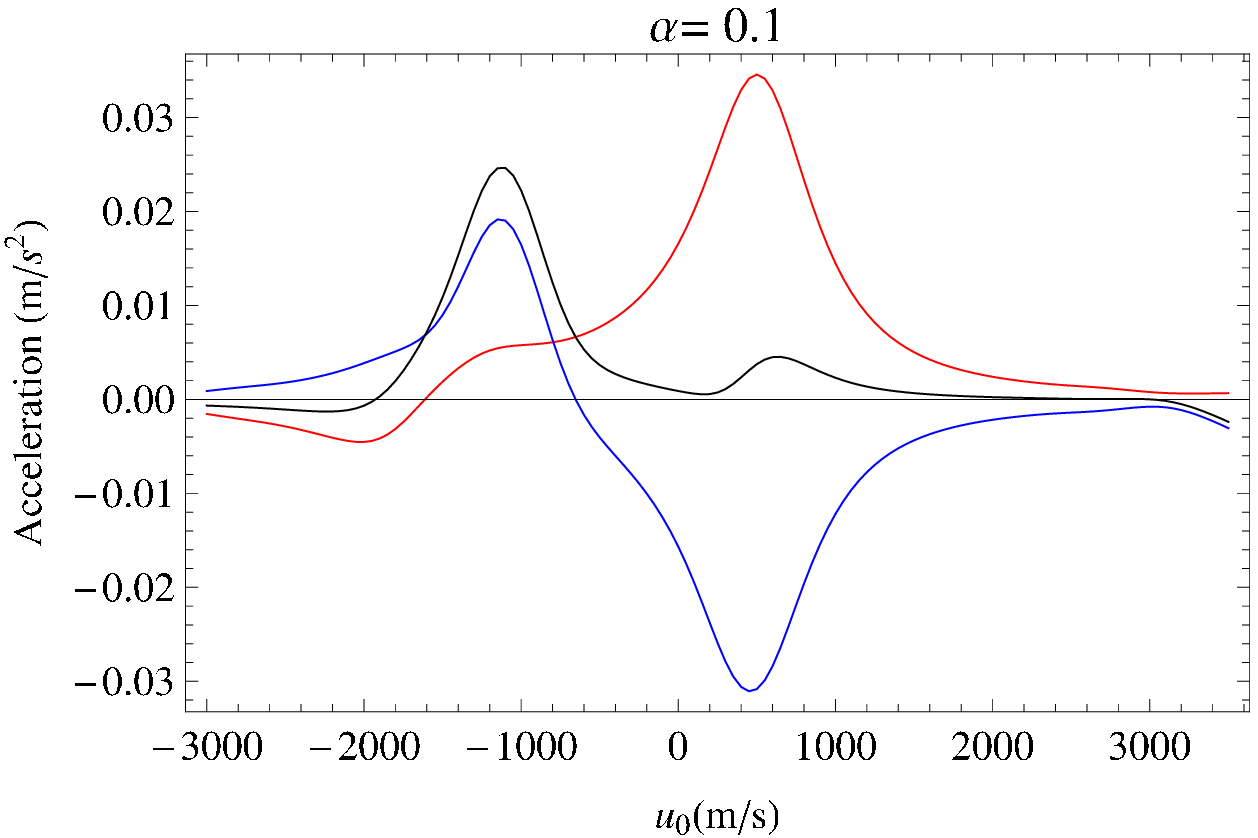}}
\subfigure[]{\includegraphics[height= 4.5cm]{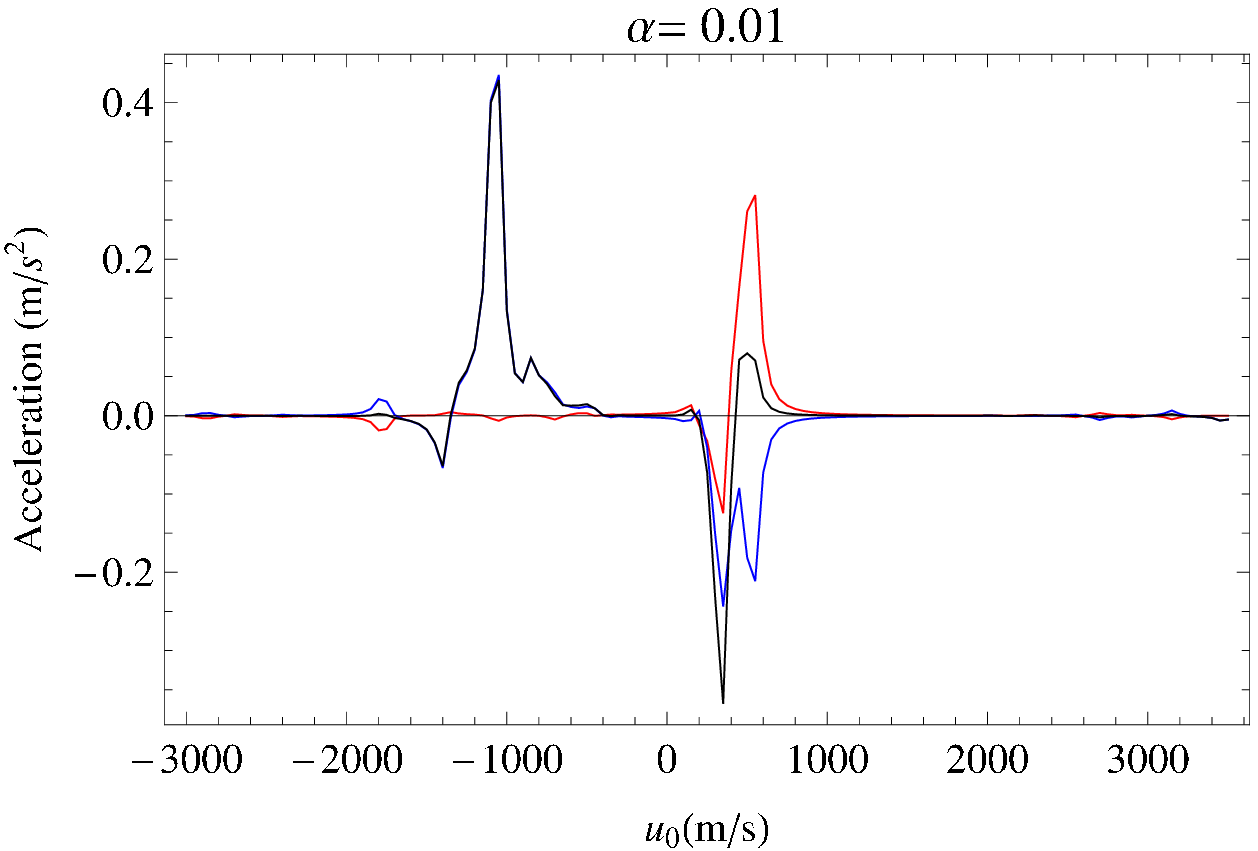}}
\subfigure[]{\includegraphics[height= 4.5cm]{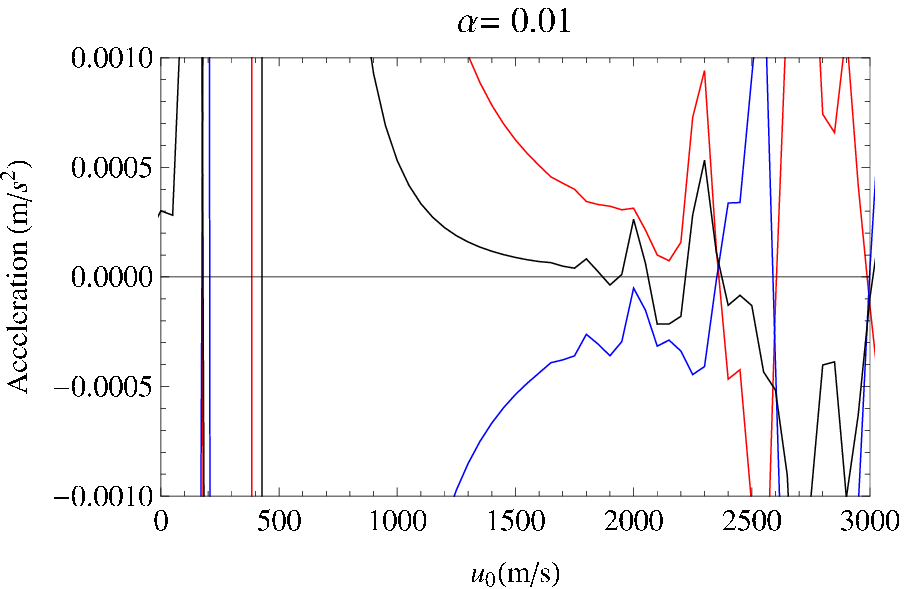}}
\end{center}
\caption{Mean equatorial accelerations to the zonal-mean flow from the horizontal and vertical eddy flux under varying mean-flow speed and tested for different damping rates. Red is $- \partial_{y} \overline{u' v'}$, blue is $- 1/\rho_{0} \partial_{z} \overline{\rho_{0} u' w'}$, and black is the sum of the two. The intersection points indicate the equilibrium $u_{0}$. The equilibrium point associated with a negative slope indicates a stable equilibrium.
}
\label{evo plot}
\end{figure}

As has been suggested by the 3D simulation of SP,
the evolution of the zonal flow mainly depends on the interaction with waves, and so the final equilibrium state occurs when the averaged accelerations from the waves vanishes. In our 3D linear analysis, we do not evolve linearized wave equations but solve for the solutions in a ``quasi-steady state" for any given $u_0$ as a free parameter, and then calculate the resulting acceleration. Here we relax the ``steady state" into ``quasi-steady state", to describe the possible evolution of $u_0$ by a sequence of intermediate quasi-steady states. As long as the acceleration is not too abrupt, $u_0$ may vary continuously from one quasi-steady state to the next. Figure~\ref{evo plot} shows the accelerations of the equatorial jet as a function of $u_0$ for the three damping scenarios derived from our linear model. The accelerations in the figure are
taken averaged across
the vertical domain roughly from 20 to $3 \times 10^{-4}$ bars, and from 20 degrees south of the equator to 20 degrees north of the equator, representing the latitudinal region of the equatorial jet as suggested by Figure~\ref{sum acc plot}. The red (blue) curve indicates the acceleration contributed from the horizontal (vertical) wave transport. The net acceleration is depicted by the black curve.

Figure~\ref{evo plot} can approximately describe the evolution of the zonal flow if each steady state for a given $u_0$ behaves as an intermediate steady state such that the entire evolution of $u_0$ can be considered as
a sequence of these intermediate steady states. There is no straightforward way to justify whether this approximation is reasonable unless a time evolution is performed to compare. Indeed, the approximation should be invalid if the evolution timescale of $u_0$ in the steady state is much shorter than the relaxation timescale for the entire atmosphere reaching an intermediate steady state. Hereafter, we just boldly assume that the approximation is applicable in order to explore the possible evolution of a superrotating flow when the relaxation time is shorter than the evolution time of $u_0$, estimated based on the steady-state results. As explained in Section 2.5, the evolution of $u_0$ resulting from the Reynolds stress is a second-order process in the linear theory and is therefore slower than the relaxation of the entire atmosphere in response to the change of $u_0$. However as we shall see, the linear theory breaks down when the dissipation is not strong enough to suppress the wave resonance.

The relaxation timescale is given by the longer of the dynamical and thermal timescales when planetary waves are not severely damped\footnote{When planetary waves are hard to propagate due to strong damping, it is meaningless to describe the dynamical timescale associated with planetary waves.}. We still apply the same estimate in our strong damping scenario with $\alpha=0.5$, in the sense that the damping rate is not much larger  than the dynamical time, which will be defined below. Because in our model the Newtonian cooling rate is set to equal the Rayleigh friction rate, the thermal timescale is just the damping time $1/\alpha$.
The dynamical timescale of the atmosphere is characterized by the wave travel time across the equatorial Rossby deformation radius \footnote{It would be more precise to use the equivalent depths $h_{m,n}$ of the Rossby and Kelvin modes instead of $H$ to estimate $T$ because we find that $h_{m,n}< H$. Nevertheless, the order-of-magnitude estimate in Equation(\ref{eq:T}) is sufficiently good for predominant modes.}:
\begin{equation}
T \sim \frac{L}{\sqrt{gH}} \sim 10^{4} sec
\label{eq:T}
\end{equation}
where $L$ is the equatorial Rossby radius, presenting the horizontal scale of planetary waves (for HD 189733b, $L = (\sqrt{gH}/\beta)^{1/2} \approx 5 \times 10^{7}$ m), and $\sqrt{gH}$ is the speed of the shallow-water gravity wave. Since $\alpha$ is normalized roughly by $1/T$, the thermal timescales corresponding to $\alpha=0.5$, 0.1, and 0.01 are about $2T$, $10T$, and $100T$, respectively. Obviously, the relaxation time is mainly governed by the thermal timescale in the damping scenarios that we consider.
On the other hand, the evolutionary timescale of the zonal flow $\tau_{jet}$ may be estimated from Figure~\ref{evo plot} by taking $u_0$ divided by the corresponding acceleration.
On average, for an eastward traveling jet (i.e. $u_0>0$),
$\tau_{jet}$ is $\gtrsim 10^{6}$ sec for $\alpha = 0.5$, $\gtrsim 10^{5}$ sec for $\alpha = 0.1$, and $\gtrsim 10^{4}$ sec for $\alpha = 0.01$. As a result, in the strong and moderate damping scenarios, the waves, although damped non-negligibly, are able to propagate and interact promptly to reach an intermediate steady state before $u_{0}$ has changed so significantly that the new $u_{0}$ would create a different wave state.
In the small damping scenario, $\tau_{jet}$ can be comparable to the relaxation time because the Rossby resonance is not severely suppressed, even leading to significant westward accelerations in the regime of $u_{0}$ = 200 -- 400 m/s.
Consequently, our linear model breaks down in the small damping scenario. We henceforth focus only on the evolutions in the strong and moderate damping scenarios.

In the moderate and strong damping scenarios, we can see that the net acceleration (black curves) due to the wave momentum flux continuously accelerates the mean flow from an initial rest state, and gradually declines to zero after $u_{0}$ exceeds 1000 m/s. The bump of the net acceleration in the moderate damping scenario results from the resonance. The time evolution process generally resembles that shown in Figure~11 of SP, despite different implementations of the damping processes. Note that the accelerations via horizontal and vertical transports in our model initially increase because of the Rossby resonance and then subside, which is also suggested by \citet{Arnold}.
It would be interesting to know whether
the decline
in the acceleration curves in SP starting from $u_0 \sim 500$ m/s is attributable to the Rossby resonance.
Also note that we ignore the drag acting on the zonal-mean flow, which would increase the rate that the acceleration curve approaches zero, and so a more realistic value of the final equilibrium velocity $u_0$ is expected to have a smaller value. Since the momentum flux is the product of wave velocities, the wave flux and the corresponding zonal acceleration scale quadratically with heating amplitude. Thus, our linear wave mechanism suggests that there is no threshold in the heating magnitude for generating superrotation (similar to SP). However, this is because the drag acting on the zonal-mean flow is ignored here, which may be responsible for inhibiting a superrotating jet when heating is too weak.

The intersections of the the net acceleration curve with the $x$-axis in Figure~\ref{evo plot} indicate the equilibrium states. The negative slope of the acceleration curve around an intersection point implies a stable equilibrium such as $u_0$ around 3000 m/s for $\alpha=0.1$ and around 2800 m/s for $\alpha=0.5$.
We may conclude from our linear model that when the damping rate is not too small, there is one unstable equilibrium state for negative $u_{0}$, and only one stable equilibrium state for positive $u_{0}$ over the tested range. The final equilibrium state $u_{0}$ should lie at the positive stable equilibrium point, or linearly runaway when the westward $u_{0}$ exceeds the unstable point, which occurs only when the initial state provides a very strong westward zonal-mean flow ($u_{0}=-1000$ m/s for $\alpha$ = 0.5 and $u_{0}=-2000$ m/s for $\alpha$ = 0.1). That is to say, the speed of the zonal-mean flow (or the total zonal-mean kinetic energy) in the final equilibrium state may have little dependence on the initial conditions according to our linear wave model with the strong and modest damping.

\section{Summary and discussions}
The superrotating equatorial flow in the atmosphere of a tidally-locked hot Jupiter is a special feature commonly found in radiative hydrodynamical simulations. The origin of the flow is explained by SP as a consequence of the momentum flux associated with equatorial Rossby and Kelvin waves excited by stationary stellar radiation in the 2D linear shallow-water model, which is then used as a guidance to interpret their 3D simulation results.  In this work, we extend their 2D wave model to 3D and additionally take into account a uniform zonal-mean zonal flow in the wave equations. We follow the linear analyses employed by \citet{matsuno} and \citet{Wu2000}, identifying the vertical modes with a shallow-water solution to construct a 3D structure for the $s=1$ equatorial trapped waves excited by stellar heating in an isothermal, vertically stratified atmosphere of a tidally locked hot Jupiter.
To clarify the distinct thermal responses from Rossby and Kelvin waves and their effects on the resulting momentum fluxes, we apply linear dampings in the thermal forcing problem and restrict ourselves to the simplified model with equal values of the Rayleigh friction and Newtonian cooling.
We take the atmosphere of the hot Jupiter HD 189733b as a fiducial example for the problem.
The results from the 3D linear theory are compared with the final equilibrium results for the $s=1$ Fourier component extracted from our 3D radiative hydrodynamical simulation based on DA. We show that the wave structures calculated by the linear model are qualitatively similar to those seen in the simulation. Our 3D linear model enables us to understand the vertical structures of the waves, which have not been analyzed previously. We find that the tilt of wavefronts is related to the strength of the Rossby-wave resonance and that the appearance of double-gyres below the heating level is an outcome indicative of dispersive equatorial waves due solely to vertical stratification.

We also compare our 3D linear results with the 2D linear results and 3D hydrodynamical numerical results in SP. The horizontal wave structures around the heating-center level in our linear model are generally consistent with the linear results in SP to demonstrate how a zonal flow is initiated by the Rossby and Kelvin waves driven by stationary stellar heating. The subsequent evolution of the zonal flow as seen in their 3D simulation can be qualitatively supported by our steady-state linear model.
As the superrotating flow speeds up, the evolution of the horizontal wave structure around the heating-center level can be explained by the eastward shift of the Rossby wave, whereas the Kelvin wave almost remains stationary. The resulting wave momentum fluxes therefore evolve,
with maximum values generally found when resonances occur.
The horizontal and vertical wave momentum fluxes
contribute together to give positive or negative feedback to the mean flow in order to achieve the final equilibrium state of the zonal-mean zonal flow in the atmosphere. The 3D distributions of momentum fluxes and their divergence derived from the linear theory resembles those in the final equilibrium state of the atmosphere in our radiative hydrodynamical simulation. Generally, the sum of the horizontal and vertical momentum fluxes have an eastward acceleration to the mean flow in the heating region and westward accelerations in higher altitudes and higher latitudes.
In addition, we argue that there exists only one single final equilibrium state for the zonal-mean flow with modest and strong damping according to our steady-state linear wave analysis with equal mechanical and thermal damping rates.

Because the atmospheric opacity is not included in our linear theory, the vertical heating profile in the linear theory is prescribed to mimic the stellar heating profile from our radiative hydrodynamic simulation, which gives rises to the numerical results in agreement with observed spectra and phasecurves particularly for HD 189733b. The fiducial model presents suggestive results regarding the interactions of the equatorial waves with superrotation. In view of the superrotation as a common phenomenon for the atmospheric circulation of hot Jupiters in both simulations (see Introduction) and observations \citep[e.g.][references therein]{FM14}, it would be of great interest in the future to apply our 3D linear theory to other hot Jupiters \citep[cf.][]{PS13}. Note that the vertical heating profile, the efficiency of re-radiated radiation energy, the background states, and the atmospheric opacity are intimately related \citep[e.g.][]{GO09,Guillot}. Consistent calculations built on our fiducial model should be conducted to provide a more complete picture that encompasses a large variety of physical conditions for atmospheric circulations of hot Jupiters.

Moreover, we have compared our linear results for hydrostatic waves with our simulation results in which the condition of hydrostatics in fact breaks down in a few locations of the dynamical atmosphere, such as flow subduction and shocks due to jet collisions. Nevertheless, these non-hydrostatic features are quite localized, which explains why our linear results still provide a verification and reasonable explanations for our simulation results. We note that shocks are not well resolved in our code and hence there are likely many more of them to exist. The role that the shocks play seems to be chiefly dissipation. A detailed study of these shocks in the simulation will be a subject for the future.

Our simple linear model is able to capture crucial structures of the atmosphere of a hot Juptier and possibly explain the evolution of superrotation, especially when the zonal-mean flow and the vertical structure are taken into consideration.
However, we remind readers of a few limitations in the linear model and envision potential improvements to the present work. All fluid quantities in the background state of the atmosphere are assumed to be uniform in the computation domain. In reality, a zonal flow away from the equator should be affected by the Coriolis force and shear, which are not considered in this work.  The linear damping employed in the linear model is over-simplified, namely, viscous, ohmic, and radiative damping are more realistic than the linear dissipation of waves\footnote{With regard to the Ohmic drag, its resulting Ohmic heating is expected to be negligible compared to stellar heating for a cooler hot Jupiter such as HD 189733b \citep{RM13}.}. In addition, there are a number of potentially important physical processes that have not been included in our model. These include (but are not limited to) self-consistent cloud/haze formation, MHD effects \citep{Batygin,RS13}, and non-equilibrium chemistry \citep{Knutson12}. All these processes may modify the resulting atmospheric flow patterns either through additional forcing of the flow or through modification of the radiative heating and cooling timescales.

Furthermore, the response of linear waves are intrinsically linear to the heating amplitude in our linear analysis. However, the amplitude of linear waves excited at the heating site, especially during the time near the Rossby resonance as the superrotating jet develops, could be so large that nonlinear dissipation may occur even in the modest damping scenario, which is out of the scope in the present work. Another issue relevant to nonlinearity is whether or not different initial conditions of the atmosphere can lead to different final equilibrium states.
Our linear analysis in the cases of modest and strong damping suggests that the final equilibrium state may not be sensitive to initial conditions. However,
nonlinear phenomena are well known to lead to potential stochastic outcomes. Our 3D numerical simulation without using the hyperviscosity (as adopted by SP) still indicates that turbulent flows are much weaker than the wave flows, suggesting that different initial conditions are likely to converge to the same final equilibrium state of the atmosphere. On the other hand, in the GCM by \citet{Arnold}, a superrotating flow suddenly driven by equatorial wave resonance can roughly sustain its speed even though the thermal forcing is reduced. This ``hysteresis" behavior implies contrarily that there are multiple equilibrium states developed by bifurcation due to nonlinear resonance.
More runs conducted in a systematic manner using our 3D radiative hydrodynamical simulation would be useful to explore the issue \citep[cf.][]{TC10,LS13}.

In addition, to compare with 3D hydrodynamical simulations, we apply the same/similar bottom boundary condition, i.e. the free-slip boundary, adopted in the simulations to the linear model, therefore preventing waves from continuing to travel downward. This limits us in that we cannot study the dynamical influence of thermal forcing to the entire planet, such as the quadrupole moment induced by thermal tides \citep{AS10}. We have shown via a simple test that wave dissipation may not be sensitive to the location of the bottom boundary. However, the downward propagating dissipated waves can still strive. In our qualitative studies, how much vertical wave energy flux can be brought down to deep layers of a hot Jupiter is not investigated, which could be one of the interesting mechanisms to inflate very hot Jupiters \citep{GS02,LCA11}. A more sophisticated linear model dealing with entire region of the radiative layer of hot Jupiter with a proper bottom boundary condition would be worthwhile to pursue this issue.
Lastly, while we calculate the mechanical feedback of waves to the zonal flow, we ignore the thermal feedback of waves (e.g., see Equation \ref{E flux eq}), which will be investigated in a future work.

\acknowledgements
We thank the anonymous referee for improving the paper.
This work is supported by a MOST grant in Taiwan through MOST 103-2112-M-001-027-. Numerical simulations were carried out utilizing NASA's High End Computing Program.


\clearpage

\begin{table}[h]
\caption{Parameters used for HD 189733b}
\begin{tabular}[t]{llc}
\hline
Parameter &     Symbol   &   Value   \\
\hline
$s=1$ mode of stellar flux & $F_{*,s=1}$ & $1.13\times 10^8$ erg/cm$^2$\\
Opacity parameter in Equation(\ref{eq heat})& $\kappa_e$ & 0.0175 cm$^2$/g\\
Planet radius   & $R_{p}$   & $80000$ km \\
Gravitational acceleration   & $g$   & $\rm 20\, m\, s^{-2}$  \\
Rotation rate   & $\Omega$  & $\rm 3 \times 10^{-5}\, sec^{-1}$ \\
Specific gas constant & $R$   & $\rm 3500\, J\, Kg^{-1}\,K^{-1}$ \\
Mean background temperature  &  $T_{0}$  & 1000 $K$\\
Specific heat at constant pressure & $c_p$ & 12500 K J/kg \\
\hline \label{parameter table}
\end{tabular}
\end{table}

\clearpage

\clearpage

\clearpage

\clearpage

\clearpage

\clearpage

\clearpage

\clearpage

\clearpage

\clearpage

\clearpage

\clearpage

\clearpage

\clearpage

\clearpage

\clearpage

\clearpage

\clearpage

\clearpage

\end{document}